\documentclass[structabstract]{aa}

\usepackage[utf8]{inputenc}
\usepackage{graphicx}
\usepackage{natbib}
\usepackage{xcolor}
\usepackage{hyperref}
\usepackage{txfonts}

\renewcommand{\d}{{\rm d}}

\hypersetup{hidelinks}

\definecolor{pantone}{RGB}{204,171,171}
\definecolor{darkblue}{RGB}{0,0,102}
\definecolor{brown}{RGB}{128,0,0}
\definecolor{yellow}{RGB}{255,200,0}

\title{Source regions of carbonaceous meteorites and NEOs}

\author{
    M.~Bro\v{z}\inst{\ref{prague}}              \and 
    P.~Vernazza\inst{\ref{lam}}                 \and 
    M.~Marsset\inst{\ref{eso}}                  \and 
    R.P.~Binzel\inst{\ref{mit}}                 \and 
    F.~DeMeo\inst{\ref{mit}}                    \and 
    M.~Birlan\inst{\ref{imcce},\ref{bucharest}} \and 
    F.~Colas\inst{\ref{imcce}}                  \and 
    S.~Anghel\inst{\ref{imcce},\ref{bucharest}} \and 
    S.~Bouley\inst{\ref{geops},\ref{imcce}}     \and 
    C.~Blanpain\inst{\ref{pyth}}                \and 
    J.~Gattacceca\inst{\ref{cerege}}            \and 
    S.~Jeanne\inst{\ref{imcce}}                 \and 
    L.~Jorda\inst{\ref{lam}}                    \and 
    J.~Lecubin\inst{\ref{pyth}}                 \and 
    A.~Malgoyre\inst{\ref{pyth}}                \and 
    A. Steinhausser\inst{\ref{impmc}}           \and 
    J.~Vaubaillon\inst{\ref{imcce}}             \and 
    B.~Zanda\inst{\ref{impmc},\ref{imcce}}           
}

\institute{
     Charles University, Faculty of Mathematics and Physics, Institute of Astronomy, V~Hole{\v s}ovi{\v c}k{\'a}ch 2, 18000 Prague, Czech Republic%
     \label{prague}%
     \and 
     Aix-Marseille University, CNRS, LAM, Laboratoire d'Astrophysique de Marseille, 38 rue Frederic Joliot Curie, 13388 Marseille, France%
     \label{lam}
     \and 
     European Southern Observatory (ESO), Alonso de Cordova 3107, 1900 Casilla Vitacura, Santiago, Chile
     \label{eso}
     \and 
     Department of Earth, Atmospheric and Planetary Sciences, MIT, 77 Massachusetts Avenue, Cambridge, MA 02139, USA
     \label{mit}
     \and 
     IMCCE, Observatoire de Paris, PSL Research University, CNRS UMR 8028, Sorbonne Université, Université de Lille, 77 av. Denfert-Rochereau, 75014 Paris, France
     \label{imcce}
     \and 
     Astronomical Institute of the Romanian Academy, Cutitul de Argint 5, 040557, Bucharest, Romania
     \label{bucharest}
     \and 
     GEOPS-Géosciences, CNRS, Université Paris-Saclay, 91405 Orsay, France
     \label{geops}
     \and 
     Service Informatique Pythéas, CNRS, OSU Institut Pythéas, UMS 3470, Marseille, France
     \label{pyth}
     \and 
     Aix-Marseille University, CNRS, IRD, Coll France, INRA, CEREGE, Aix-en-Provence, France
     \label{cerege}
     \and 
     Institut de Minéralogie, Physique des Matériaux et Cosmochimie (IMPMC), Muséum National d’Histoire Naturelle, CNRS UMR 7590, Sorbonne Université, 75005 Paris, France
     \label{impmc}
}

\date{Received x-x-2024 / Accepted x-x-2024}

\abstract
      {
The source regions of ordinary chondrites ($\sim$80\% of all falls)
and large S-type near-Earth objects ($\sim$30\% of all NEOs)
have recently been identified (Brož et al., Marsset et al.),
with three young asteroid families (Karin, Koronis, Massalia)
being at the origin of most ordinary chondrite falls.
}
   {
The present work is a continuation of these two studies
and aims to determine the source regions of the remaining meteorite and NEO classes
with an emphasis on carbonaceous chondrites
(CM, CI, CO, CV, CK, CR, CH, CB, or C-ungrouped).
}
   {
We studied 38 individual asteroid families, including young and old ones,
and determined their contributions to the NEO populations at metre and kilometre sizes
using collisional and orbital models.
Our models are in agreement with
spectroscopic observations of NEOs,
cosmic-ray exposure ages of meteorites,
statistics of bolides,
infrared emission from dust bands,
composition of interplanetary dust particles (IDPs), or
abundance of extraterrestrial helium-3.
}    
  {
We identified the Veritas, Polana and Eos families
as the primary sources of CM/CR, CI and CO/CV/CK chondrites, respectively.
Substantial contributions are also expected from
CM-like K\"onig and CI-like Clarissa, Misa and Hoffmeister families.
The source regions of kilometre-sized bodies are generally different.
The Adeona family is by far the main source of CM-like NEOs,
whereas the Polana (low-i) and Euphrosyne (high-i) families are at the origin of most CI-like NEOs.
The Polana family is the likely source of both Ryugu and Bennu.
We were able to link spectroscopically and dynamically several NEOs to the Baptistina family.
Finally, it appears that the pre-atmospheric flux of carbonaceous chondrites at metre sizes
is about the same as that of ordinary chondrites.
Given the difference in fall statistics between the two groups (80\% versus 4.4\%),
this implies either substantial atmospheric fragmentation of carbonaceous bodies
at the level of ${\sim}0.5\,{\rm MPa}$,
or destruction by thermal cracking and water desorption.
}
   {
The source regions of most meteorites and kilometre-sized NEOs have now been determined;
including some minor classes like
enstatite chondrites and achondrites (Nysa, Hungaria),
acapulcoites/lodranites (Iannini).
Future work should focus on the few remaining classes (essentially, iron meteorites, pallasites, ureilites).
}
  

\begin{document}

\maketitle


\begin{table*}
\caption{
Overview of carbonaceous chondrites classes.
}
\label{tab1}
\centering
\begin{tabular}{@{}l@{\kern3pt}llrl@{}}
    & class   & example          & falls   & notes \\
\hline
\vrule width0pt height9pt
{}* & CI1     & Ivuna (TZ)       & 0.4\%   & no chondrules, no CAIs, abundant water, IDP-like, solar-like \\
    & CY1     & Yamato (An)      & ?       & aqueous + thermal alteration, \cite{King_2019ChEG...79l5531K} \\
{}* & CM1-2   & Mighei (UA)      & 1.8\%   & 0.3-mm chondrules, 20\% of chondrules $\leftrightarrow$ 80\% of matrix \\
    & CR1-3   & Renazzo (IT)     & 0.2\%   & 0.7-mm, 55\% $\leftrightarrow$ 45\% \\
    & CH2-3   & high-metal       & 0.05\%? & 0.02-mm, 70\% $\leftrightarrow$ 30\%, high-metal \\
{}* & CO3     & Ornans (FR)      & 0.5\%   & 0.15-mm, 50\% $\leftrightarrow$ 50\% \\
{}* & CV3     & Vigarano (IT)    & 0.6\%   & 1-mm, 45\% $\leftrightarrow$ 55\% \\
    & CK3-6   & Karoonda (AU)    & 0.2\%   & 1-mm, 45\% $\leftrightarrow$ 55\% \\
    & CB3     & Bencubin (AU)    & 0.05\%? & up to 10-mm, 30\% $\leftrightarrow$ 70\%, high-metal \\
    & CL4     & Loongana (AU)    & ?       & 1-mm, 80\% $\leftrightarrow$ 20\%, \cite{Metzler_2021GeCoA.304....1M} \\
    & C-ungr. & Tagish Lake (CA) & 0.5\%   & distinct, e.g., fragile, low density, \dots \\
\hline
\vrule width0pt height9pt
    &         &                  & $\Sigma$4.9\% & \\
\end{tabular}
\tablefoot{
A typical range of petrologic types (1-6) is shown, corresponding to
aqueous alteration: 1$\,\leftarrow\,$3 (CI$\,\leftarrow\,$CH); or
thermal alteration: 4$\,\rightarrow\,$6 (CO$\,\rightarrow\,$CL).
Adapted from \citet{Cobb_2014ApJ...783..140C}.
The percentages of falls (out of 36 falls) are from
\url{https://www.lpi.usra.edu/meteor/} 
\citep{Gattacceca_2022M&PS...57.2102G}.
The mean bulk densities are:
CI $1.6$,
CM $2.3$,
CO $3.0$,
CV $3.1\,{\rm g}\,{\rm cm}^{-3}$
\citep{Colsolmagno_2008ChEG...68....1C,Macke_2011M&PS...46..311M}.
}
\end{table*}


\section{Introduction}

Both telescopic observations and dynamical studies of solar system small bodies imply
that most meteorites and near-Earth objects (NEOs) originate from the main asteroid belt
\citep{Ceplecha_1961BAICz..12...21C,Wisdom_1983Icar...56...51W,Binzel_1996Sci...273..946B,Farinella_1998Icar..132..378F,Morbidelli_2002aste.book..409M,Granvik_2017A&A...598A..52G,Nesvorny_2023AJ....166...55N}.
Until recently, however, it has been an unsuccessful quest to unambiguously identify
the origin of even the better-studied meteorite groups
with the exception of Lunar, Martian and Vestian (HEDs) meteorites.
The reason is pretty simple: asteroids as opposed to planets are not unique in terms of composition
and spectral properties (Vesta being an exception).
It follows that there are several possible sources in the asteroid belt
for a given meteorite class and for each compositional group of NEOs.

\begin{figure*}
\centering
\includegraphics[width=18cm]{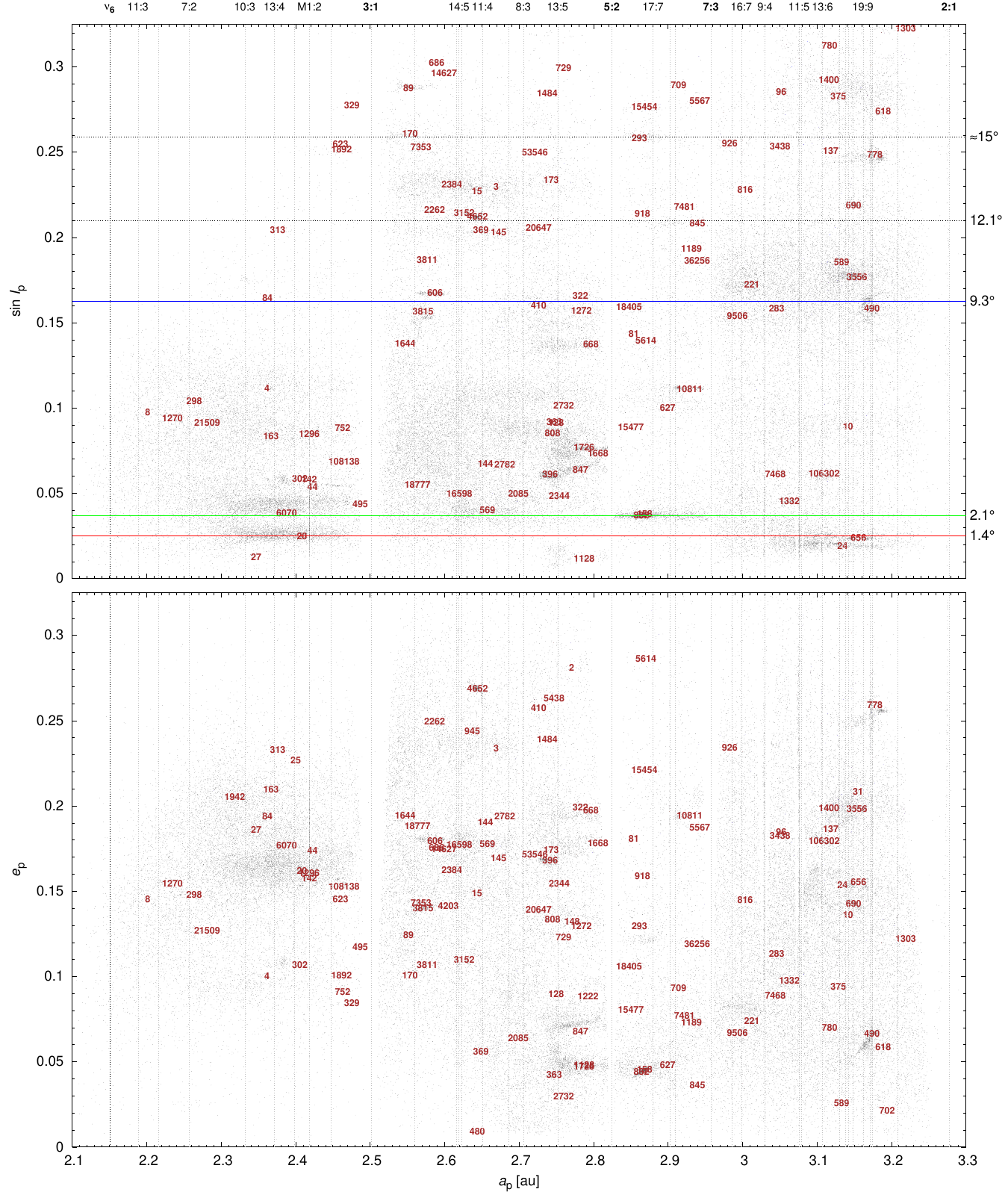}
\caption{
"Faint Main Belt",
showing only bodies with the absolute magnitude
close to the limit of the Catalina Sky Survey,
i.e., $H \gtrsim 19.25 + 5(\log(2.2(1-0.1)) + \log(2.2-1))) - 5(\log(a(1-e)) + \log(a-1)))$.
The proper semimajor axis~$a_{\rm p}$
versus eccentricity~$e_{\rm p}$ (bottom)
versus inclination~$\sin i_{\rm p}$ (top) are plotted;
together with locations of the mean-motion resonances (vertical lines),
IRAS dust bands (horizontal lines), and
known asteroid families \citep{Nesvorny_2015aste.book..297N} (labels).
Big and old ones are almost invisible here
(e.g., ``4'' Vesta).
Small and young ones --having a steep SFD-- are prominent.
The distribution of faint bodies is surprisingly irregular.
The concentrations are directly related to the sources of meteorites:
L          .. ``20'' Massalia \citep{Marsset_2023},
H          .. ``158'' Koronis \citep{Broz_2023},
CM         .. ``490'' Veritas,
CI         .. ``142'' Polana,
CO/CV/CK   .. ``221'' Eos,
M          .. ``293'' Brasilia, or
M          .. ``606'' Brang\"ane (this work).
}
\label{aei7_sykes}
\end{figure*}

Two recent studies \citep{Broz_2023,Marsset_2023} have provided a breath of fresh air to this quest,
identifying the sources of the two main meteorite groups,
H and L chondrites, representing $\sim\,$70\% of all falls.
Specifically, H chondrites originate from two breakups related to (832)~Karin and (158)~Koronis,
which occurred 5.7 and 7.6\,My ago,
whereas L chondrites originate from a cratering or reaccumulative event on (20)~Massalia,
$\sim\,$40\,My ago.
We summarize the implications of these studies that can be generalized to other classes of meteorites and NEOs hereafter:

\begin{enumerate}
\item Faint ($H \gtrsim 18.5\,{\rm mag}$) asteroids are not distributed evenly in the main belt,
instead, they are concentrated strongly around few specific asteroid families.
This is seen nicely in Fig.~\ref{aei7_sykes}.

\item The youngest (${\lesssim}\,40\,{\rm My}$) asteroid families
are at the origin of the prominent dust bands
observed at $1.4^\circ$ (Massalia) and $2.1^\circ$ (Karin, Koronis),
highlighting that their size-frequency distribution (SFD) is steep
and continuous from sub-km asteroids to ${\sim}100$-$\mu$m dust grains.

\item Consequently, a sizeable breakup temporarily `overshoots'
the whole main belt population also at metre sizes.

\item The background population must be a negligible source of meteorites,
because both H and L chondrites exhibit unique radiometric features
\citep{Eugster_2006mess.book..829E,Swindle_2014GSLSP.378..333S}
for about half of all finds.

\item The intervals of iridium or helium-3 excess found in terrestrial strata
\citep{Schmitz_1997Sci...278...88S,Farley_1998Sci...280.1250F},
which have a relatively short duration,
confirm the transient nature of the extraterrestrial dust and meteoroid populations,
created during breakups and sustained by a collisional cascade.

\item Old families (Vesta, Flora, Eunomia, etc.) contribute significantly less
than young ones to the meteorite flux,
because their SFD is 'bent' at sub-km sizes
due to a collisional cascade.
They are, however, the primary source of $D\geq 1\,{\rm km}$ NEOs.
This difference is at the origin of the longtime meteorite--NEO conundrum
\citep{Vernazza_2008Natur.454..858V}.

\item The Phocaea, Juno and Flora families are the primary sources of H-, L- and LL-like NEOs,
respectively \citep{Broz_2023}.

\end{enumerate}

In this work, we present a follow-up study
focusing on the source regions of carbonaceous chondrites (CCs) and CC-like NEOs.
CCs represent 4.4\% of all falls
(\citealt{Gattacceca_2022M&PS...57.2102G}; see Tab.~\ref{tab1}),
whereas their parent bodies
(C-complex asteroids along with K-type asteroids);
represent more than 50\% of the mass of the asteroid belt
\citep{DeMeo_2013Icar..226..723D}.
The discrepancy between CC falls and their abundance in the asteroid belt
is at first order a direct consequence of their friable nature,
making it more difficult for those types of materials to survive atmospheric entry
\citep{Borovicka_2019M&PS...54.1024B}.
One can not exclude that most CCs actually reach the surface of the Earth
in the form of micrometeorites or interplanetary dust particles
(IDPs; \citealt{Vernazza_2015ApJ...806..204V}).

\section{Family identification}\label{identification}

\begin{figure*}
\centering
\begin{tabular}{c@{\kern0.1cm}c@{\kern0.1cm}c}
\kern1cm CM-chondrite &
\kern1cm CI-chondrite, steep &
\kern1cm CI-chondrite, shallow \\
\includegraphics[width=5.9cm]{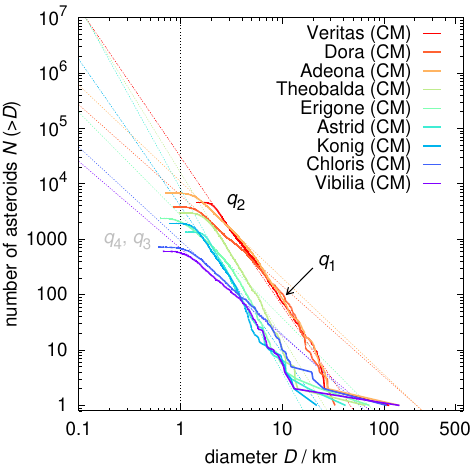} &
\includegraphics[width=5.9cm]{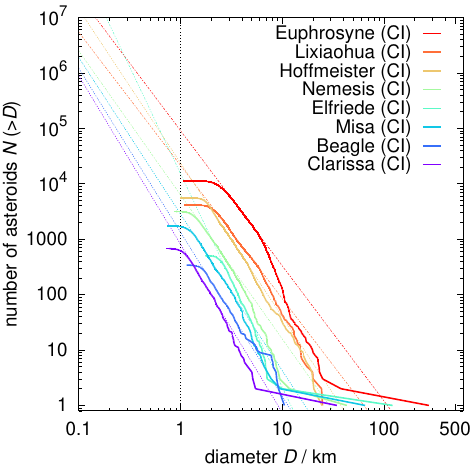} &
\includegraphics[width=5.9cm]{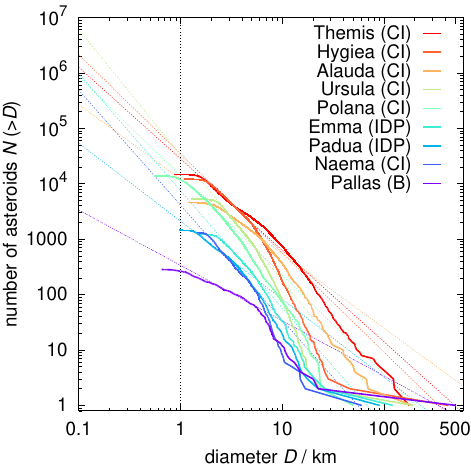} \\
\kern1cm CO/CV/CK-type &
\kern1cm M-type &
\kern1cm unknown \\
\includegraphics[width=5.9cm]{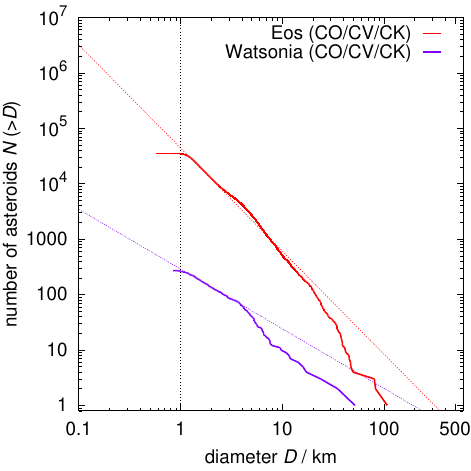} &
\includegraphics[width=5.9cm]{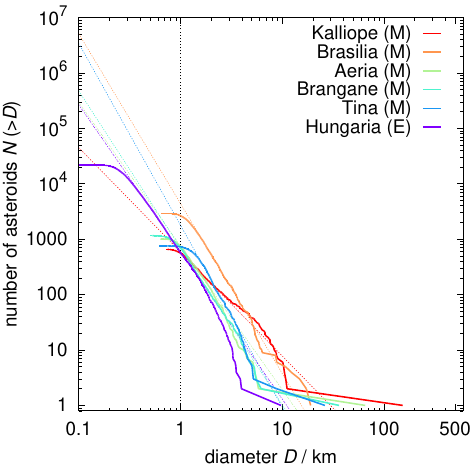} &
\includegraphics[width=5.9cm]{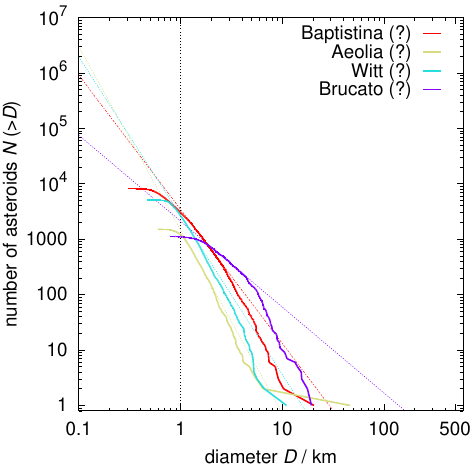} \\
\end{tabular}
\caption{
Observed SFDs of C-type asteroid families.
The SFDs usually exhibit a steep slope~$q_1$ at sizes $D\gtrsim 5\,{\rm km}$
and a shallow slope~$q_2$ at $D\lesssim 5\,{\rm km}$.
This break is often created in the course of collisonal evolution.
An observational bias affects the SFDs at even smaller sizes (0.5-3\,km),
depending on the distance and albedo of the respective populations.
Additionally, families of P, B and M types were studied
(together with 4 of unknown taxonomical classification).
}
\label{CM_chondrite}
\end{figure*}

The first step of our work consisted in the identification and characterization
of CC-like families selected according to the following criteria:
(i)~abundance of small, {\em faint\/} bodies as these are the most promising sources of meteorites (Fig.~\ref{aei7_sykes}),
(ii)~abundance of large ($D\geq1\,{\rm km}$) bodies as these are the most promising sources of large NEOs,
(iii)~the slope of the SFD (Tab.~\ref{tab:q}),
(iv)~the presence/absence of an associated dust band \citep{Sykes_1990Icar...85..267S,Reach_1997Icar..127..461R,Nesvorny_2006Icar..181..107N,Ade_2014A&A...571A..14P}.
We identified families using recent catalogues:
Astorb \citep{Moskovitz_2019EPSC...13..644M},
AFP \citep{Knezevic_Milani_2003A&A...403.1165K,Novakovic_2019EPSC...13.1671N},
Wise \citep{Nugent_2015ApJ...814..117N},
Akari \citep{Usui_2011PASJ...63.1117U}, and
SDSS \citep{Parker_2008Icar..198..138P};
versions as of July 2023.
They contained 1298281 orbits in total. We applied a hierarchical clustering
(HCM; \citealt{Zappala_1995Icar..116..291Z})
with a removal of interlopers.
Specifically, family members must fulfill several criteria.
For C-type families,
we assumed a geometric albedo of $p_V \in (0; 0.125)$ and
an SDSS colour index of $a^\star \in (-1; 0.1)$.
We further used a magnitude criterion \citep{Vokrouhlicky_2006Icar..182..118V}:
\begin{equation}
H > 5\log_{10}\left({|a-a_c|\over C}\right)\,,
\end{equation}
where
$a_c$ is the centre and $C$ the extent of a family.
Some families also contain sub-families
(e.g., Beagle within Themis; Fig.~\ref{24_Themis_ei}),
to be studied separately.
For K-type families,
we assumed a geometric albedo of $p_V > 0.075$ and
an SDSS colour index of $a^\star \in (-0.5; 0.5)$.
For each family, we obtained
the median albedo,
diameters of all bodies,
and observed SFDs (Fig.~\ref{CM_chondrite}).
Their well-defined slopes are provided in Tab.~\ref{tab:q}.
Some of the families are steep down to the observational limit
(e.g., Veritas).
On the other hand, none of the previously proposed old families
(``old Polana", ``Eulalia", ``Athor'')
is relevant here as they have too shallow (too depleted) SFDs.

\begin{figure*}
\centering
\includegraphics[width=16cm]{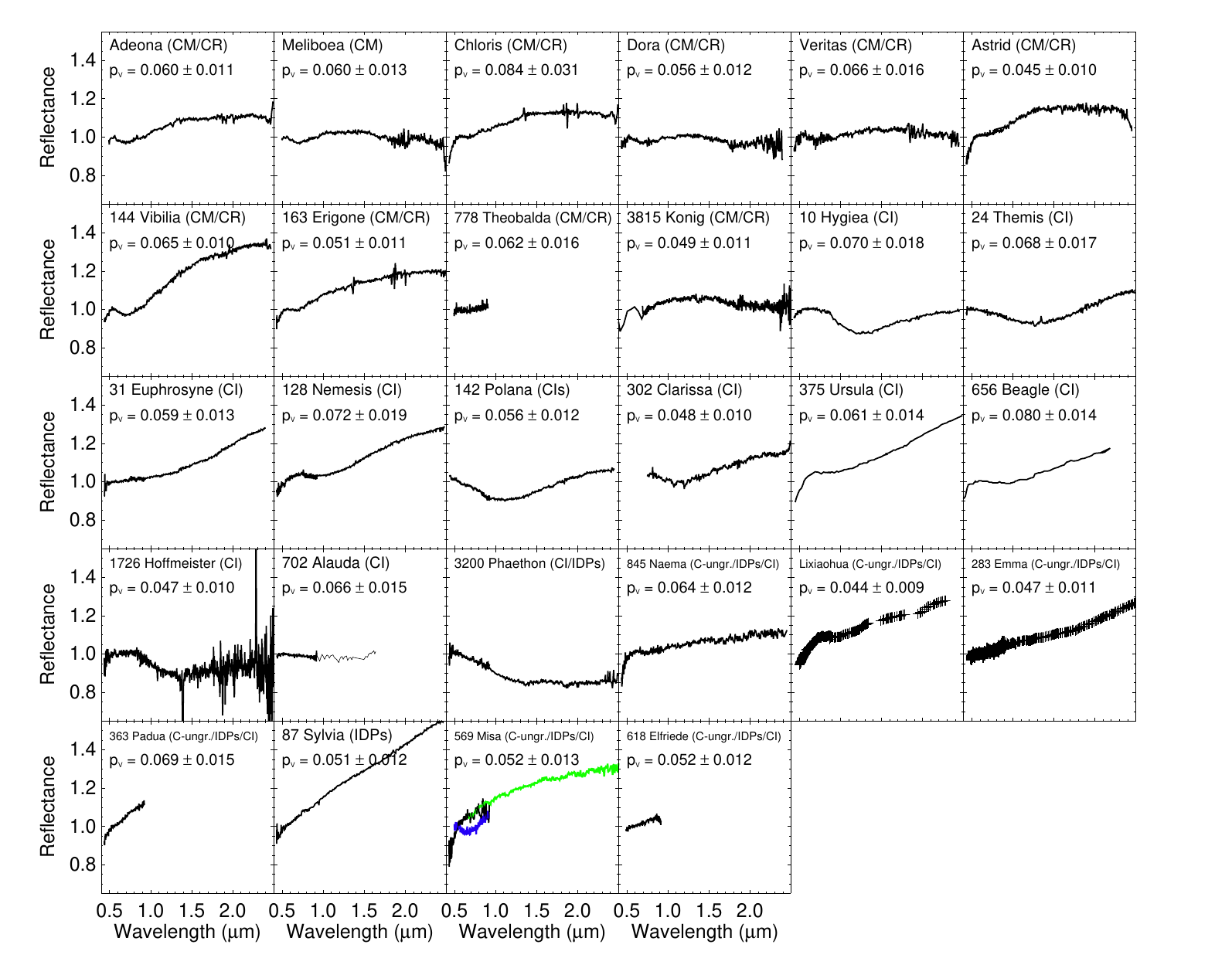}
\includegraphics[width=16cm]{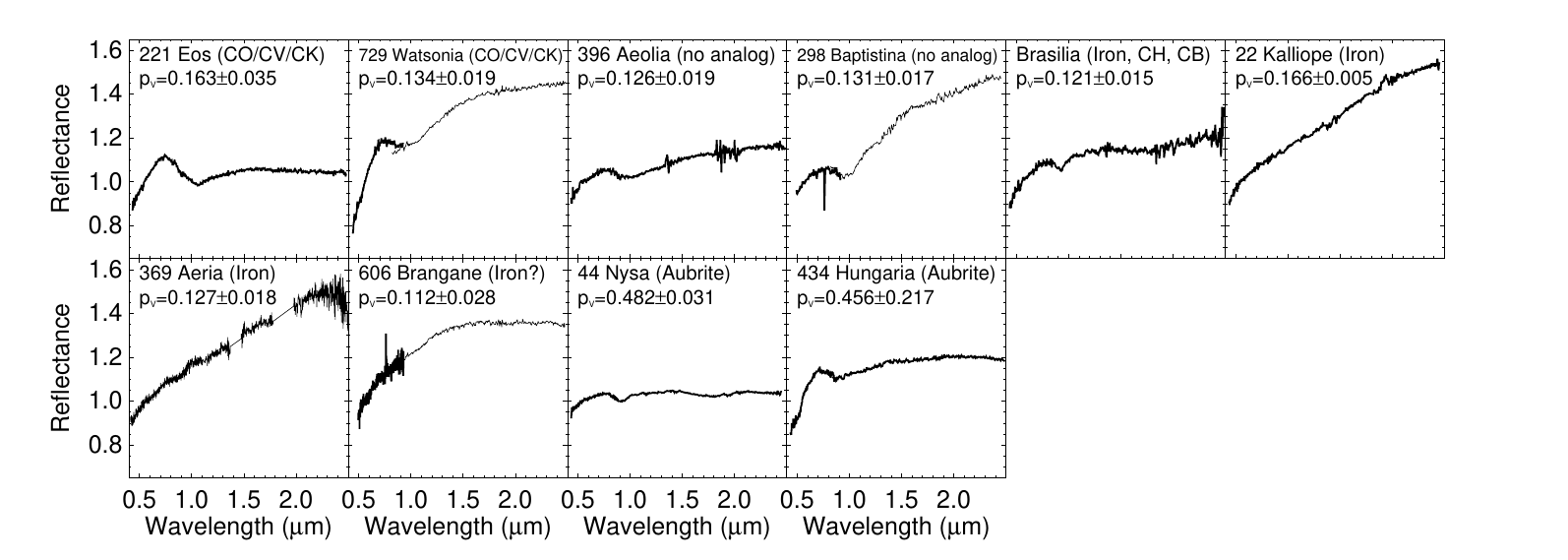}
\caption{
The mean spectra of asteroid families along with their classification.
If only one spectrum (for the largest member) was available,
the designation includes its number.
If more spectra were available, but the mean was unreliable,
spectra were plotted separately (\textcolor{blue}{blue}, \textcolor{green}{green}).
}
\label{CtypeFamilies}
\label{EKMPtypeFamilies}
\end{figure*}


\section{Taxonomic classification}\label{classification}

As a next step, we determined the meteoritic analog of these CC-like
main belt families as well as those of CC-like NEOs. We used publicly available
colors in the optical domain (SDSS survey; \citealt{Parker_2008Icar..198..138P}),
visible spectra (SMASS and S3OS2 surveys;
\citealt{Bus_2002Icar..158..106B,Bus_2002Icar..158..146B,Lazzaro_2004Icar..172..179L}),
near-infrared spectra from the MITHNEOS survey
\citep{DeMeo_2009Icar..202..160D,Binzel_2019Icar..324...41B,Marsset_2022AJ....163..165M},
other sources from the literature
\citep{Fornasier_2010Icar..210..655F,Fornasier_2016Icar..269....1F,Masiero_2015ApJ...809..179M,Depra_2020A&A...643A.102D}
as well as our own unpublished near-infrared spectra. 
We also used albedo information from WISE \citep{Nugent_2015ApJ...814..117N}
to distinguish chondrule-poor (CI/CM/C ungrouped/IDPs) from chondrule-rich (CO/CV/CK) carbonaceous materials
and to ascertain our classification based on available spectrophotometric measurements.
Typically, chondrule-poor materials possess a reflectance at 0.55 microns
that falls in the 0.03-0.09 range,
whereas chondrule-rich materials possess a reflectance at 0.55 microns
that essentially falls in the 0.1-0.2 range.
This dichotomy is well supported by albedo measurements
($p_V\leq0.1$ and $p_V\geq0.1$)
of their carbonaceous parent bodies,
for which a similar trend is observed \citep{Vernazza_2021A&A...654A..56V}.

\subsection{Families with $p_V\leq0.1$}

The vast majority of the families studied here possess low ($p_V\leq0.1$) albedos 
\citep{Masiero_2015aste.book..323M}. From a reflectance/albedo perspective, these families are suitable sources for at least four carbonaceous chondrite classes (CI, CM, CR, C ungrouped) as well as for IDPs. 

Spectroscopy in the visible was particularly useful to distinguish CM as well as CR-like families from CI/C ungrouped/IDP-like ones. Indeed, all CM chondrites as well as a sizeable fraction of CR chondrites (e.g., \citealt{Prestgard_2023M&PS...58.1117P}) possess spectral properties that are identical to those of Ch- and Cgh-type asteroids with the presence of an absorption feature at $\sim$0.7 microns (see \citealt{Vernazza_2022vcid.book....3V} for a detailed review). CM/CR-like families comprise Adeona, Chloris, Dora, Veritas, Astrid, Vibilia, Erigone, Theobalda and K\"onig.

Spectroscopy in the near-infrared was particularly useful to identify plausible sources of CI chondrites. Both the present study (see discussion) as well as previous studies \citep{Bottke_2020AJ....160...14B} have linked CI-like NEOs Ryugu and Bennu with the Polana family. 
Given that the spectrum of the Polana family
\citep{Pinilla_2016Icar..274..231P} is typical of those of C, Cb, Cg-type spectra, displaying a broad convex band centered around 1.1-1.3 micron, we assigned a CI-like analog to all families with this convex feature (Hygiea, Themis, Euphrosyne, Nemesis, Polana, Clarissa, Ursula, Beagle, Hoffmeister). 

At this stage, we were left with one B-type family (Alauda) and several P-type families (Misa, Elfriede, Naema, Lixiaohua, Emma, Padua, Sylvia). Given the B-type spectrum of NEO Bennu and its similarity to Ryugu, we also assigned a CI-like analog to Alauda. P-types appear as natural sources for spectrally red and featureless meterorites such as Tagish Lake \citep{Vernazza_2013Icar..225..517V}, a member of the C ungrouped class. It is interesting to note that other meteorites of the C ungrouped class appear spectrally similar to Tagish Lake \citep{Applin_2022LPICo2678.2218A}. Meteorites such as Tagish Lake are, however, an unlikely analog for many P-types, including Sylvia. Indeed, measurements in the mid-infrared suggest that many P and D-types are spectrally similar to anhydrous IDPs instead of aqueously altered meteorites such as Tagish Lake \citep{Vernazza_2013Icar..225..517V,Vernazza_2015ApJ...806..204V,Vernazza_2022vcid.book....3V}. 
In the absence of further measurements of P-type families in both the critical 3 micron and 7-25 micron regions (aside from Sylvia; \citealt{Usui_2019PASJ...71....1U,Vernazza_2015ApJ...806..204V}), and laboratory spectroscopic measurements in the visible and near-infrared range (0.4-4 microns) for IDPs (e.g., \citealt{Bradley_1996M&PS...31..394B,Maupin_2020PSJ.....1...62M}),  we cannot unambiguously identify their analog. We consequently leave all possibilities open (C ungrouped, IDPs, CIs). Note that some prominent CI-like families such as Themis and Euphrosyne \citep{Marsset_2016A&A...586A..15M,Yang_2020A&A...643A..38Y} show a mix of C and P-types, with objects filling the continuum between the two classes. Deciphering properly this spectral heterogeneity requires further measurements in the critical 3 micron and mid-infrared regions. 

\subsection{Families with $p_V\geq0.1$}

It has now been established for a long time that K-type asteroids (alike (221) Eos and its family) comprise the parent bodies of CV, CO, and CK meteorites
(e.g., \citealt{Bell_1988Metic..23..256B,Burbine_2001M&PS...36..245B,Clark_2009Icar..202..119C,Mahlke_2023A&A...676A..94M}). Recently, it has been demonstrated that Barbarians L-type asteroids alike (729) Watsonia are also plausible parent bodies for CO and CV chondrites \citep{Mahlke_2023A&A...676A..94M}.

For the Aeolia and Baptistina families, we were unable to identify a satisfactory match with any meteorite group. The spectra of both largest members (298 and 396) are characterized by the presence of a 1-micron feature and the absence of a 2-micron feature. Whereas this is very consistent with olivine being a substantial surface component, the band~I centers are shifted towards lower values with respect to olivine (${\approx}1.00\,\mu$m for asteroids, 1.05$\,\mu$m for olivine). It follows that CO, CV or CK chondrites are unlikely analogs for both families. We also considered CR chondrites but these meteorites possess either a CM-like spectrum \citep{Prestgard_2023M&PS...58.1117P} or a spectrum with a 1-micron band located close to $0.9\,\mu{\rm m}$.
Similarly, ureilites are not convincing analogs for these two bodies.


We identified four families as plausible source of iron and/or metal-rich meteorites (Brasilia, Kalliope, A\"eria, Br\"angane). 
The mean spectrum of two Brasilia family members (3985, 4461) displays a narrow $\sim$0.9$\mu$m feature alike that seen in many M-type spectra although significantly deeper \citep{Hardersen_2005Icar..175..141H,Hardersen_2011M&PS...46.1910H,Neeley_2014Icar..238...37N} as well as a $\sim$1.9$\mu$m feature, indicating the presence of low-Ca pyroxene. This spectrum appears consistent with the one of metal-rich chondrites (e.g., CH, CB; \citealt{Dibb_2022M&PS...57.1570D}). In the case of (22) Kalliope, the asteroid with the highest density so far \citep{Ferrais_2022A&A...662A..71F} and a differentiated metal-rich interior \citep{Broz_2022A&A...664A..69B}, radar measurements indicate a metal-poor surface \citep{Shepard_2015Icar..245...38S}.


The mean spectra of asteroid families or of the largest family member are shown in Fig.~\ref{CtypeFamilies}.


\section{Models}\label{modeling}

Unfortunately, observations of multi-km asteroids alone are not sufficient
to identify the sources of meteorites.
It is necessary to understand SFDs down to metre sizes,
as well as the ages of families,
because SFDs evolve considerably in the course of time.
Eventually, meteoroid fluxes must be determined
by transporting bodies from the main belt to the NEO space
and by estimating collisional probabilities with the Earth.

\subsection{Collisional model}\label{collisional}

In order to characterize the family SFDs,
we used a statistical collisional model,
described in detail in \citet{Broz_2023}.
It is based on 'Boulder' \citep{Morbidelli_2009Icar..204..558M},
with a number of subsequent improvements
\citep{Cibulkova_2016A&A...596A..57C,Sevecek_2017Icar..296..239S,Vernazza_2018A&A...618A.154V}.
The number of collisions is computed as:
\begin{equation}
n_{ijkl} = p_{ij} f_{\rm g} (D_{ik} + D_{jl})^2 \d N(D_{ik}) \d N(D_{jl}) \Delta t\,,
\end{equation}
where
$i$, $j$ denote populations,
$k$, $l$ corresponding size bins,
$p_{ij}$ is the collisional probability (see Tab.~\ref{tabc1}),
$f_{\rm g}$, the gravitational focussing factor,
$D_{ik}$, $D_{jl}$, corresponding sizes,
$\d N(D)$, the differential distributions, and
$\Delta t$, the time step.

Fragmentation is determined by the specific energy of impact:
\begin{equation}
Q = {0.5 M_{ik} v_{ij}^2 \over M_{ik}+M_{jl}}\,,
\end{equation}
where
$M_{ik}$ denotes the projectile mass,
$M_{jl}$, the target mass,
$v_{ij}$, the relative speed.
We assumed a size-strength scaling law:
\begin{equation}
Q^* = 9.0\cdot 10^7\,{\rm erg}\,{\rm g}^{-1}\left({D\over 2\,{\rm cm}}\right)^{-0.53}\!\!\! + \,0.5\,{\rm erg}\,{\rm cm}^{-3}\rho\left({D\over 2\,{\rm cm}}\right)^{1.36}\!,
\end{equation}
which involves a strength (small $D$) and a gravity regimes (large $D$).
Hereinafter, we assume a single law for the main belt population,
even though S- and C-types might be different in terms of their rheology.
Further relations for
$M_{\rm lr}(Q)$, largest remnant mass,
$M_{\rm lf}(Q)$, largest fragment mass,
$q(Q)$, slope of SFD of fragments,
constrain the outcome of collisions.

For the decay time scale in the main belt,
describing orbital dynamics, not collisions,
we assumed the following analytical prescription
($\tau$ in My, $D$ in km):
\begin{equation}
\tau_{\rm mb} = \cases{
30                      & for $D < 10^{-7}$ \cr
30\,(D/10^{-7})^{0.38}  & $D < 0.001$ \cr
1000                    & $D < 0.003$ \cr
1000\,(D/0.003)^{-0.58} & $D < 0.01$ \cr
500                     & $D < 0.02$ \cr
500\,(D/0.02)^{0.86}    & $D < 0.1$ \cr
2000\,(D/0.1)^{0.30}    & $D < 1$ \cr
4000\,D^{0.41}          & $D < 20$ \cr
13800\,(D/20)^{1.05}    & $D < 100$ \cr
75000\,(D/100)^{1.30}   & otherwise\,,\cr
}
\end{equation}
which contains contributions from
the diurnal Yarkovsky effect,
the seasonal Yarkovsky effect,
the YORP effect,
spin-size dependence,
conductivity-size dependence, or
the Poynting--Robertson effect.

For the average lifetime in the NEO population ($\tau$ in My), we assumed:
\begin{equation}
\tau_{\rm neo} = 3\,.
\end{equation}
It is shorter compared to \cite{Granvik_2018Icar..312..181G},
but closer to an average over all families.
The two time scales cannot be easily separated from each other.
The absolute NEO population is derived from the main-belt population by the ratio of
$\tau_{\rm neo}/\tau_{\rm mb}$,
assuming an equilibrium.

Our model was calibrated by:
  (i)~the observed SFD of the asteroid belt \citep{Bottke_2015aste.book..701B} (Fig.~\ref{sfd_1100});
 (ii)~the observed SFD of NEOs \citep{Harris_2015aste.book..835H} (Fig.~\ref{sfd_1100_NEA});
(iii)~the Vesta family SFD;
 (iv)~the Rheasylvia basin age \citep{OBrien_2014P&SS..103..131O};
  (v)~(4) Vesta's cratering record \citep{Marchi_2012Sci...336..690M}.

Initial conditions for the Vesta family were as simple as possible,
a single power-law $N({>}D) = CD^q$ down to sub-km sizes,
with the initial slope $q$.
The break at 4\,km is created in the course of evolution
by a collisional cascade.
After approximately 1100\,My,
the slope $q_1 = q$ for $D > 4\,{\rm km}$ remains steep,
but the slope $|q_2| < |q_1|$ for $D < 4\,{\rm km}$ becomes shallow,
which is in accord with the observed slopes
(see again Fig.~\ref{sfd_1100}).
The break at 0.5\,km, attributed to observational incompleteness,
does not affect this calibration.

A similar approach was used for all other families,
which led to an independent estimate of their ages
(Tab.~\ref{tab2}).
It is based on the assumption that the break of the SFD at about 3 to 5\,km
has been created by the collisional cascade.

\begin{table}
\caption{Ages of the C-type families estimated from our collisional model.}
\label{tab2}
\centering
\begin{tabular}{lr}
\vrule width 0pt depth 4pt
family & age \\
--     & My \\
\hline
\vrule width 0pt height 9pt
\\
\\
\\
\\
Adeona (CM)             & $ 350\pm  50$!\\
{\bf Aeolia (?)}        & $\lesssim 50$ \\
A\"eria (M)             & $ 350\pm  50$ \\
Alauda (CI)             & $2500\pm 500$ \\
Astrid (CM)             & $\lesssim 50$ \\
{\bf Baptistina (?)}    & $ 300\pm  50$ \\
Beagle (CI)             & $\lesssim100$ \\
{\bf Brang\"ane (M)}    & $\lesssim 50$ \\
Brasilia (M)            & $\lesssim100$ \\
Brucato (?)             & $ 500\pm 100$ \\
Chloris (CM)            & $1100\pm 300$ \\
Clarissa (CI)           & $ 150\pm  50$ \\
Dora (CM)               & $ 250\pm 100$ \\
Elfriede (CI)           & $\lesssim100$ \\
Emma (IDP)              & $1200\pm 200$ \\
Eos (CO/CV/CK)          & $1800\pm 300$ \\
Erigone (CM)            & $ 500\pm 100$ \\
{\bf Euphrosyne (CI)}   & $ 800\pm 100$ \\
Hoffmeister (CI)        & $\lesssim 50$ \\
Hungaria (E)            & $ 150\pm  50$ \\
Hygiea (CI)             & $2500\pm 300$ \\
Iannini (Aca/Lod)       & $\lesssim 50$ \\
Kalliope (M)            & $ 900\pm 100$ \\
{\bf K\"onig (CM)}      & $\lesssim 50$ \\
Lixiaohua (CI)          & $1200\pm 200$ \\
Misa (CI)               & $ 200\pm  50$ \\
Naema (CI)              & $1500\pm 200$ \\
Nemesis (CI)            & $1300\pm 200$ \\
Padua (IDP)             & $ 900\pm 150$ \\
Pallas (B)              & $2000\pm 500$ \\
{\bf Polana (CI)}       & $ 200\pm  50$!\\
Sylvia (P)              & $1200\pm 200$ \\
Themis (CI)             & $2000\pm 500$ \\
Theobalda (CM)          & $\lesssim100$ \\
Tina (M)                & $ 800\pm 200$ \\
Ursula (CI)             & $1800\pm 200$ \\
{\bf Veritas (CM)}      & $\lesssim200$ \\
Vesta (HED)             & $1100\pm 100$ \\
Vibilia (CM)            & $ 700\pm 200$ \\
Watsonia (CO/CV/CK)     & $2500\pm 500$ \\
Witt (?)                & $\lesssim100$ \\
\hline
\end{tabular}
\end{table}


\subsection{Backward integrations}\label{backward}

For some of the `promising' families,
we tried to determine their ages independently
by backward integrations of the orbits
\citep{Nesvorny_2002Natur.417..720N,Nesvorny_2003ApJ...591..486N}.
Our orbital model is based on a modified version of SWIFT
\citep{Levison_Duncan_1994Icar..108...18L,Broz_2011MNRAS.414.2716B},
the symplectic integrator MVS2.
It involves 5 massive bodies (Sun, Jupiter to Neptune),
100 mass-less asteroids,
the Yarkovsky effect, and
the YORP effect.
The sampling time step of osculating elements was 1\,y.
On these, we applied filters A, A, B of \citet{Quinn_1991AJ....101.2287Q},
with the decimation factors 10, 10, 3,
so that the output time step was 300\,y,
which is sufficient to sample secular oscillations.

The thermal parameters were set up as follows:
density $\rho = 1.3\,{\rm g}\,{\rm cm}^{-3}$,
conductivity $K = 10^{-3}\,{\rm W}\,{\rm m}^{-1}\,{\rm K}^{-1}$,
capacity $C = 680\,{\rm J}\,{\rm kg}^{-1}\,{\rm K}^{-1}$,
the Bond albedo $A = 0.1$,
emissivity $\varepsilon = 0.9$.
For each asteroid, we created 20 clones 
with a uniform sampling of the obliquity ($\cos\gamma$),
in order to explore different Yarkovsky drift rates.

In the post-processing,
we computed a differential precession $\Delta\Omega$
with respect to the 1st body
(unless its orbit is chaotic, e.g., (490) Veritas).
At each time step and for each asteroid, we selected the best clone,
that is, the clone with the smallest $\Delta\Omega$.
The percentage of interlopers we removed from the output population
varied from 0 to 50\%, depending on the surrounding background population.

We determined either a precise age,
when the orbits exhibit a convergence and then a divergence of $\Delta\Omega$,
or a lower limit of the age $t_\downarrow$,
when the orbits only converge, but then do not diverge,
because there is always a clone which can reach zero $\Delta\Omega$.

The upper limit $t_\uparrow$ on the family age is given by the Yarkovsky semimajor axis drift
\citep{Nesvorny_2015aste.book..297N}:
\begin{equation}
t_{\uparrow} = 1\,{\rm Gy}\, {C\over 10^{-4}\,{\rm au}} \left({a_{\rm c}\over 2.5\,{\rm au}}\right)^2 {\rho\over 2.5\,{\rm g}\,{\rm cm}^{-3}} \left({0.2\over p_V}\right)^{1/2}\,.
\end{equation}
However, even $t_{\uparrow}$ is not a 'hard' limit,
if the Yarkovsky drift is not nominal
(e.g., due to different thermal parameters)
or is modified
(e.g., due to the stochastic YORP related to shape variations;
\citealt{Statler_2009Icar..202..502S,Bottke_2015Icar..247..191B}).

A summary of our results is presented in Tab.~\ref{tab3}.
It is important to focus on the rows ``young $\Omega$''.
From the point of convergence,
the most promising C-type families seem to be:
Aeolia,
Misa,
Elfriede,
Hoffmeister,
K\"onig,
and, of course, Veritas
\citep{Carruba_2017MNRAS.469.4400C}.

\begin{figure}
\centering
\includegraphics[width=8.5cm]{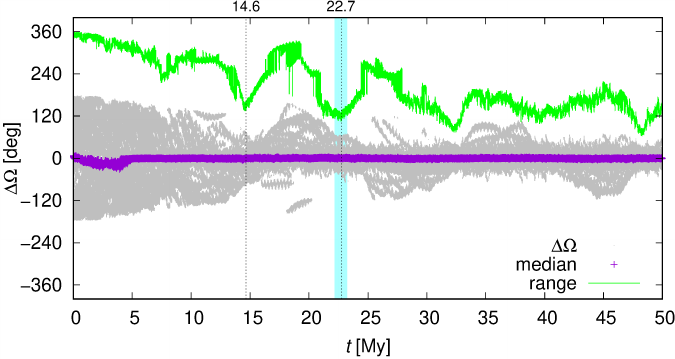}
\includegraphics[width=8.5cm]{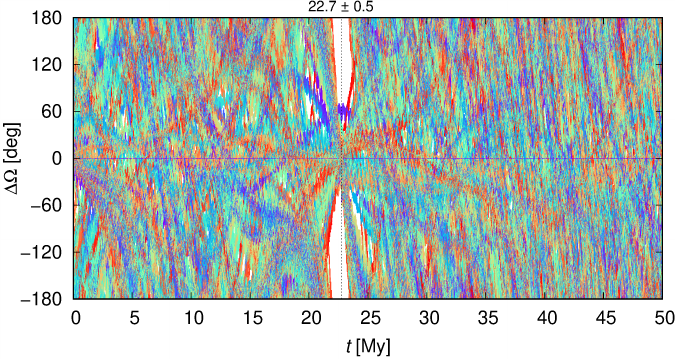}
\caption{
K\"onig family exhibits a convergence of nodes.
The range of $\Delta\Omega$ with respect to (3815) K\"onig (\textcolor{green}{green}; top)
is plotted together with individual values (\textcolor{gray}{gray}).
For each body, we used a set of 20 clones with different Yarkovsky effect
and we selected the best clone at each time step.
Not all bodies converge, because some of them are interlopers.
The temporal dependence $\Delta\Omega(t)$ exhibits local minima,
in particular, 14.6\,My and 22.7\,My seem to be possible ages.
For 22.7\,My, the converging orbits of the best clones
were plotted (coloured; bottom).
For 14.6\,My, a different set of clones would have to be selected.
The orbits often exhibit more than one differential-precession cycle.
See also Fig.~\ref{konig-1_0003.200}.
}
\label{konig-1_converg_20_11_0.00}
\end{figure}

An example of the K\"onig family is shown in
Fig.~\ref{konig-1_converg_20_11_0.00}.
A clear convergence of $\Omega$ occurs
either at 14.6, or 22.7\,My,
with a local uncertainty of 1\,My.
We will identify the correct solution soon (Sec.~\ref{forward}).

An independent argument is based on statistics.
If there are breakups between 0-10\,My
(cf. Karin, Koronis$_2$, Veritas),
there must be also some between 10-20\,My,
consequently, some of the convergences must be real.

\subsection{Forward integrations}\label{forward}

Again, for some of the `promising' families
we tried to determine their ages independently
by forward integrations.
Our orbital model is again based on a modified version of SWIFT
\citep{Levison_Duncan_1994Icar..108...18L,Broz_2011MNRAS.414.2716B},
the symplectic integrator RMVS3.
Hereinafter, we included 10 massive bodies (Sun, Mercury to Neptune, Ceres),
about 1000 mass-less asteroids,
the Yarkovsky effect,
the YORP effect, and also
collisional reorientations.
The sampling time step was 1\,y.
Mean elements were computed
with the filters A, A, A, B,
the factors 10, 10, 10, 3,
the intermediate time step was 3000\,y;
unless it was necessary to sample short-period oscillations.
Proper elements were computed by the Fourier transform
\citep{Sidlichovsky_1996CeMDA..65..137S},
out of 1024 samples,
with an exclusion of planetary frequencies ($g$, $s$).
The output time step was 0.1\,My.
Thermal parameters were the same as above;
the efficiency of YORP $c = 0.33$.

In the case of forward integrations,
we needed a synthetic family,
representing an initial configuration before dispersal.
The geometry of individual impacts is presented in Tab.~\ref{tab:synthetic}.
While the velocity field was isotropic,
the distribution of $a$, $e$, $i$ elements was determined
by the true anomaly~$f$ and the argument of perihelion $\omega$.
For example, $f = 0^\circ$ (or $180^\circ$) corresponds to
a narrow diagonal ellipse.
We tested several combinations of $f$, $\omega$ and selected suitable geometries.
The uncertainty is typically of the order of $20^\circ$.
A long-term evolution is chaotic, diffusive, i.e., irreversible,
so that we must compare synthetic vs. observed distributions
of $a$, $e$, $i$.

Perturbations, which actually allowed us to estimate the ages, included
secular resonances,
chaotic diffusion,
(1)~Ceres, or
the YORP effect,
i.e., a depletion of small bodies in the centre of the family.
Again, a summary of our results is presented in Tab.~\ref{tab3}.
It is important to focus on the rows denoted ``young $e$'' or ``young $i$''.
The most promising C-type families are:
Aeolia,
Beagle,
Brang\"ane,
K\"onig, and
Theobalda.

\begin{figure}
\centering
\begin{tabular}{c}
\kern1cm 3.2\,My \\
\includegraphics[width=8.5cm]{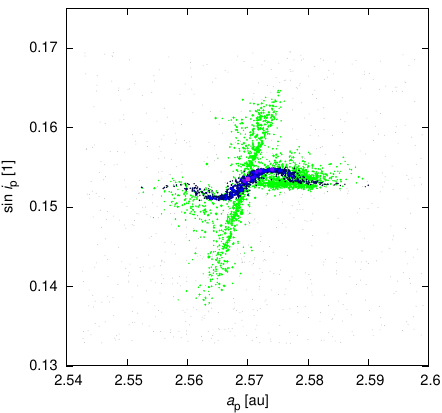} \\
\kern1cm 22\,My \\
\includegraphics[width=8.5cm]{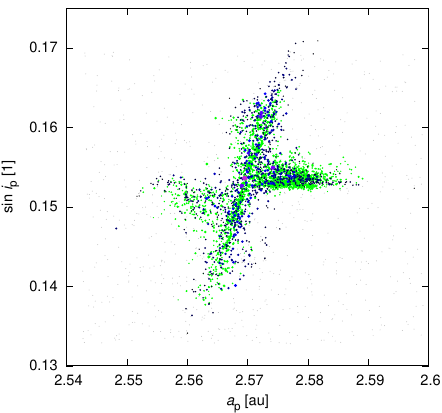} \\
\end{tabular}
\caption{
K\"onig family is 22\,My old according to orbital evolution.
The observed family (\textcolor{green}{green}) has a very peculiar structure
in the proper semimajor axis~$a_{\rm p}$ versus the proper inclination~$\sin i_{\rm p}$.
The synthetic family (\textcolor{blue}{blue}; top) is initially distributed
in a simple 'ellipse', created by an isotropic impact.
The geometry of the impact was
$f = 85^\circ$,
$\omega = 0^\circ$.
In the course of evolution,
the distribution is modified by the $g-g_5+s-s_6$ secular resonance.
At the time $(22\pm2)\,{\rm My}$ (bottom),
it exactly corresponds to the observed distribution.
Such a young age indicates that the SFD can still be steep.
}
\label{konig-1_0003.200}
\end{figure}

For example, the K\"onig family is perturbed by
the $g-g_5+s-s_6$ secular resonance (Fig.~\ref{konig-1_0003.200}).
It evolves along the libration centre
and subsequently, away from it.
Starting from a simple isotropic field (ellipsoidal)
ends up with a complex non-isotropic ('criss-cross').
The best-fit age is $(20\pm 5)\,{\rm My}$.
Combined with the convergence of $\Omega$ (Sec.~\ref{backward}),
the true age is $22\pm 2\,{\rm My}$.

Taking the Astrid family as another example, it is affected by
the $g-g_{\rm C}$ secular resonance with Ceres
\citep{Novakovic_2016IAUS..318...46N,Tsirvoulis_2016Icar..280..300T}.
This perturbation increases the spread of inclinations at $a = 2.763\,{\rm au}$;
there is no spread without Ceres.
According to Fig.~\ref{astrid-1_CERES_0003.200},
the lower limit $t_\downarrow$ is 150\,My.
Importantly, the perturbations from Ceres allowed us to rule out the existence of
a steep `tail' of the SFD for Astrid,
as well as for
A\"eria,
Brasilia, or
Hoffmeister.
The argument is exactly the same as for Vesta (Sec.~\ref{collisional}).

Other interesting examples of perturbations are shown in
Figs.~\ref{misa-2_ai_0003.200}, \ref{brangane-1_VESTA_ae_obs}.

\begin{table}
\caption{Ages of the C-type families estimated from our orbital model.}
\label{tab3}
\centering
\begin{tabular}{l@{}rrrl}
\vrule width 0pt depth 4pt
family & $t_\downarrow$ & age & $t_\uparrow$ & notes \\
-- & My & My & My \\
\hline
\vrule width 0pt height 9pt
Karin (H)               &            & $5.7\pm 0.1$     & 50             & young $\Omega$, $\omega$ \\
Koronis$_2$ (H)         & 7.6        & $7.6\pm 0.2$     & 30             & young $\Omega$           \\  
Massalia (L)            & 18.7       &                  & 250            &                          \\  
\\                       
Adeona (CM)             &            &                  & 1000           &                          \\
{\bf Aeolia (?)}        & 12         & $25\pm5$         & 45!            & young $e$                \\
A\"eria (M)             &            & ${>}\,200$       & 500            & Ceres                    \\
Alauda (CI)             &            &                  & ${\sim}\,4000$ &                          \\
Astrid (CM)             & 15         & ${>}\,150$       & 230$^{\rm c}$  & Ceres                    \\
{\bf Baptistina (?)}    &            &                  & 400            &                          \\
Beagle (CI)             & 8          & $30\pm10$        & 90             & young $i$, YORP          \\
{\bf Brang\"ane (M)}    & 20         & $45\pm15$        & 70             & young $a$, $e$           \\
Brasilia (M)            & 16.5       & ${>}\,100$       & 400$^{\rm c}$  & Ceres                    \\
Brucato (?)             &            &                  & 700            &                          \\
Chloris (CM)            &            &                  & 900            & YORP                     \\
Clarissa (CI)           & 14         & $50\pm10$        & 60!            & old $e$                  \\
Dora (CM)               &            &                  & 650            &                          \\
Elfriede (CI)           & 35         & $55\pm20$        & 230            & cf. $e$                  \\
Emma (IDP)              &            &                  & 750            & old $e$, $i$             \\
Eos (CO/CV/CK)          &            &                  & 3000           & old $e$, $i$             \\
Erigone (CM)            &            & $200\pm 30$      & 250            & YORP                     \\
{\bf Euphrosyne (CI)}   &            &                  & 1900           & old $e$, $i$             \\
Hoffmeister (CI)        & 23         & ${>}\,100$       & 350$^{\rm c}$  & Ceres                    \\
Hungaria (E)            &            &                  & 150            &                          \\
Hygiea (CI)             &            &                  & 3400           &                          \\
Iannini (Aca/Lod)       & 4          & $6\pm 2$         & 20             & young $\Omega$, $e$, $i$ \\
Kalliope (M)            &            &                  & 2700           & depleted                 \\
{\bf K\"onig (CM)}      & 14         & $22\pm2$         & 100            & young $\Omega$, $i$      \\
Lixiaohua (CI)          &            & ${>}\,150$       & 600            & old $i$                  \\
Misa (CI)               & 9.5        & ${\gtrsim}50$    & 150            & old $e$, $i$, YORP       \\
Naema (CI)              & 18         & $150\pm50$       & 350$^{\rm C}$  & old $e$                  \\
Nemesis (CI)            &            & ${>}\,100$       & 700            & old $a$, $e$             \\
Padua (IDP)             &            &                  & 700            &                          \\
Pallas (B)              &            &                  & 1800           & depleted                 \\
{\bf Polana (CI)}       &            &                  & 2400           &                          \\
Sylvia (P)              &            &                  & 1600           &                          \\
Themis (CI)             &            &                  & ${\sim}\,4000$ &                          \\
Theobalda (CM)          & 5          & $20\pm 5$        & 300            & young $i$                \\
Tina (M)                & 15         & $200\pm 50$      & 250            & old $i$                  \\
Ursula (CI)             &            &                  & 2000           &                          \\
{\bf Veritas (CM)}      &            & $8.3\pm 0.1$     & 270            & young $\Omega$, $e$      \\
Vesta (HED)             &            &                  & 1400           &                          \\
Vibilia (CM)            &            &                  & 700            & YORP                     \\
Watsonia (CO/CV/CK)     &            &                  & 1900           & depleted                 \\
Witt (?)                &            &                  & 700            & young $\Omega$?          \\
\hline
\end{tabular}
\tablefoot{
$^{\rm c}$ collisional age is significantly shorter,
$^{\rm C}$ collisional age is significantly longer.
For comparison,
Astrid is $(140\pm 30)\,{\rm My}$ old according to \cite{Carruba_2016MNRAS.461.1605C},
Clarissa $(56\pm 6)\,{\rm My}$ \citep{Lowry_2020AJ....160..127L},
Eos $(1700\pm 200)\,{\rm My}$ \citep{Broz_2013Icar..223..844B},
Erigone $(280\pm 100)\,{\rm My}$ \citep{Vokrouhlicky_2006Icar..182..118V},
Euphrosyne $(280\pm 100)\,{\rm My}$ \citep{Yang_2020A&A...643A..38Y},
Hoffmeister $(220\pm 50)\,{\rm My}$ \citep{Carruba_2017MNRAS.465.4099C},
Iannini $(6\pm 1)\,{\rm My}$ \citep{Carruba_2018MNRAS.477.1308C},
Kalliope $(900\pm 100)\,{\rm My}$ \citep{Broz_2022A&A...664A..69B},
Lixiaohua $(155\pm 36)\,{\rm My}$ \citep{Novakovic_2010MNRAS.402.1263N},
Pallas $(1700\pm 300)\,{\rm My}$ \citep{Marsset_2020NatAs...4..569M},
Sylvia $>1000\,{\rm My}$ \citep{Vokrouhlicky_2010AJ....139.2148V},
Theobalda $(6.9\pm 2.3)\,{\rm My}$ \citep{Novakovic_2010MNRAS.407.1477N},
where the value corresponds to our lower limit, and
Tina $(170\pm 30)\,{\rm My}$ \citep{Carruba_2011MNRAS.412.2040C}.
}
\end{table}


\subsection{Transport model}\label{transport}

We constrained the transport from the family location to the near-Earth space
by another set of orbital simulations.
This time, we included 11 massive bodies
(Sun, Mercury to Neptune, Ceres, Vesta),
and about 1000 mass-less particles,
which were initially located at the observed positions of the family members.
The temporal decay of the populations is shown in Fig.~\ref{nbody_observed_decay_CM},
the lifetime in the NEO space in Fig.~\ref{nbody_observed_nea3_CM}.

Assuming steady state (cf.~the discussion in \citealt{Broz_2023}),
the NEO population can be estimated as:
\begin{equation}
N_{\rm neo} = {\bar\tau_{\rm neo}\over\tau_{\rm mb}}N_{\rm mb}\,,\label{N_neo}
\end{equation}
where
$\tau_{\rm mb}$ is the decay time scale in the main belt,
$\bar\tau_{\rm neo}$, the {\em mean\/} lifetime in the NEO region, 
and $N_{\rm mb}$, the population in the main belt.
A substantial number of families is too young or too small
to contribute to the population of 1-km NEOs.
Especially, if the families are located far from major resonances,
there is no way how to transport bodies to the near-Earth space.

Regarding the transport of metre-sized bodies,
we changed two important parameters,
the conductivity $K = 1\,{\rm W}\,{\rm m}^{-1}\,{\rm K}^{-1}$,
corresponding to monoliths,
and the tensile strength according to \citet{Holsapple_2007Icar..187..500H},
which allows for fast-rotating bodies.
This substantially suppresses the diurnal Yarkovsky effect \citep{Vokrouhlicky_1998A&A...335.1093V}.
The outcomes are shown in
Figs.~\ref{nbody_metresized_decay_CM}, \ref{nbody_metresized_nea3_CM}.

Using Eq.~(\ref{N_neo}) is not sufficient though,
because C-type families are found at very different locations.
One has to compute collisional probabilities~$p$ with the Earth
(Tab.~\ref{tabc2}),
and estimate the meteoroid flux
(in $10^{-9}\,{\rm km}^{-2}\,{\rm y}^{-1}$)
as:
\begin{equation}
\Phi = p N_{\rm neo}\,.
\end{equation}
The populations and fluxes from individual sources are listed in Tabs.~\ref{tab:1km} and~\ref{tab:1m}.


\section{Results}\label{mb_1km}

The SFDs of individual families (Fig.~\ref{CM_chondrite}) exhibit substantial differences
in terms of observational bias, which occurs for diameters in the 1--3\,km range.
To make them comparable, we extrapolated SFDs from a multi-kilometer population consistently to 1\,km.
Most of the time, the extrapolation was straightforward
(a straight line on a log-log plot). 
Sometimes, however, we had to use a collisional model (Sec.~\ref{collisional}),
because it turned out that the SFD was 'bent' already at multi-km sizes (e.g., Euphrosyne).

To further extrapolate these SFDs to metre sizes,
the use of a collisional model was absolutely necessary (Fig.~\ref{sfds}).
Moreover, SFDs evolve in the course of time and it is necessary to determine their current state.
Regarding the transport from the main belt to the NEO space,
our model is similar to \cite{Broz_2023},
with only a few exceptions (see Sec.~\ref{transport}).

A notable result is that the meteorite--NEO conundrum \citep{Vernazza_2008Natur.454..858V, Broz_2023}
is well present among carbonaceous bodies.
Prominent sources for kilometre-sized bodies turn out in most cases to be irrelevant
to modest sources for meteorites and {\em vice versa\/}.
Hereafter, we describe in more detail the main respective source of kilometre- and metre-sized NEOs
for the main CC compositional groups (CI/IDP/C-ungr., CM/CR, CO/CV/CK).

\begin{figure*}
\centering
\begin{tabular}{ccc}
synthetic main belt 1-km &
synthetic NEO 1-km &
observed NEO 1-km \\
\includegraphics[width=6cm]{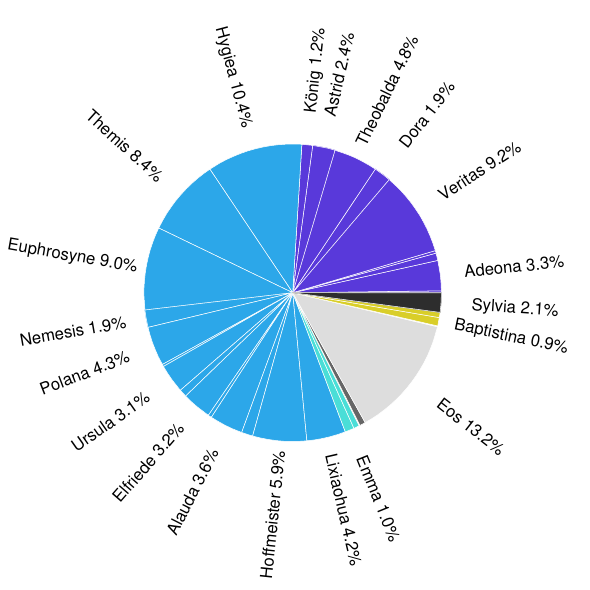} &
\includegraphics[width=6cm]{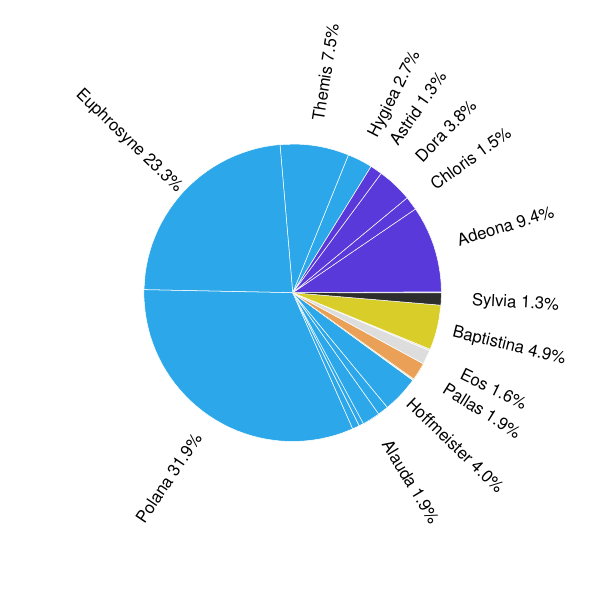} &
\includegraphics[width=6cm]{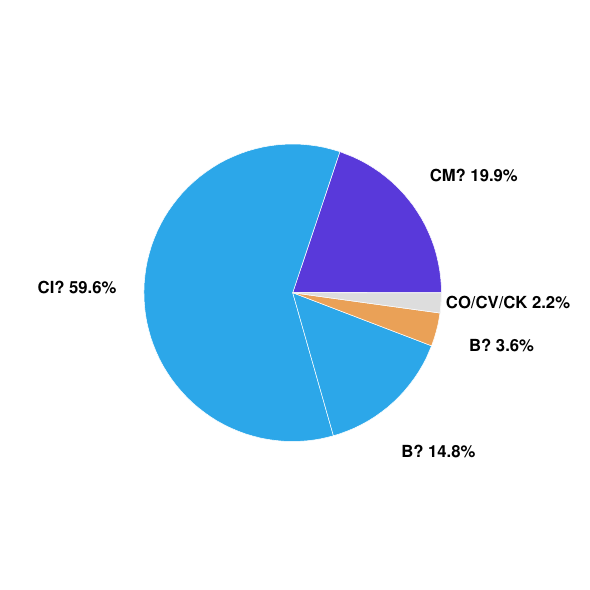} \\
synthetic main belt 1-m &
synthetic meteoroids 1-m (flux) &
observed meteorites 1-m (no irons) \\
\includegraphics[width=6cm]{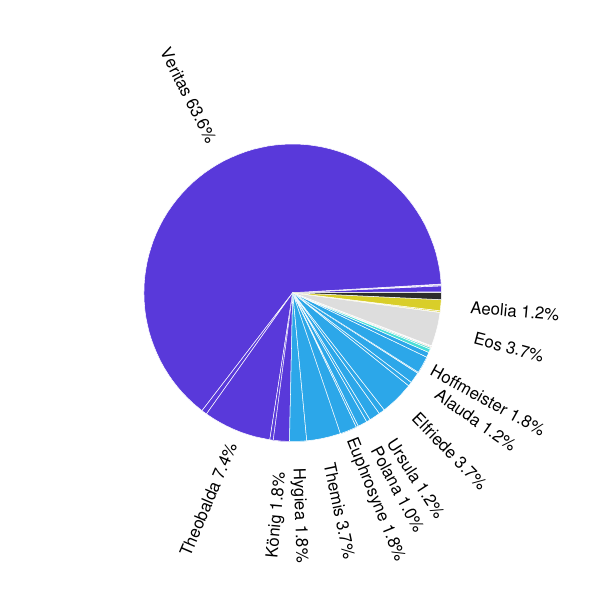} &
\includegraphics[width=6cm]{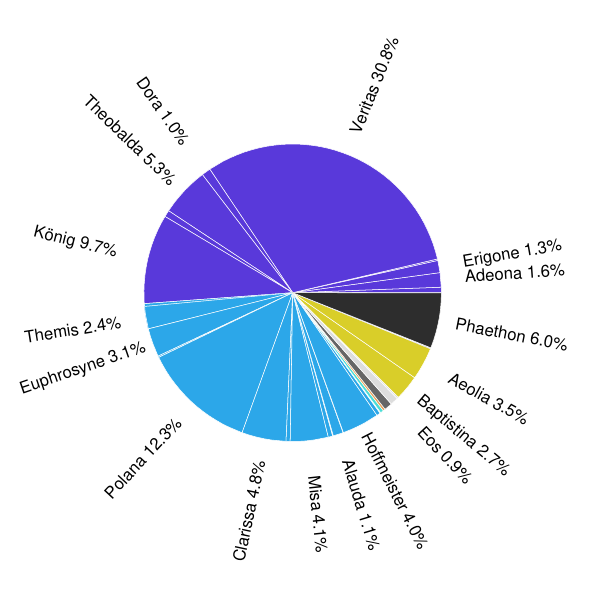} &
\includegraphics[width=6cm]{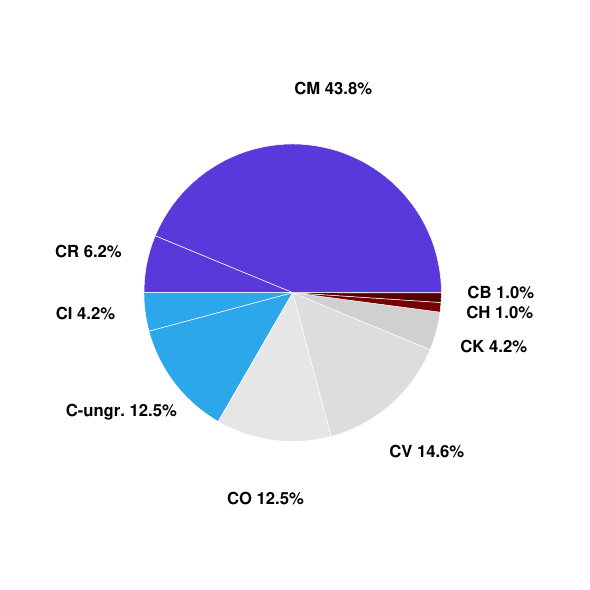} \\
\end{tabular}
\caption{
Percentages of
CM/CR- (\color{blue}blue\color{black}),
CI/IDP- (\color{cyan}cyan\color{black}),
CO/CV/CK- (\color{gray}gray\color{black})
and other carbonaceous-chondrite-like bodies
of the synthetic main belt (left),
of the synthetic NEO (middle)
and the observed NEO (right) populations.
These percentages are different for
kilometre-sized (top) {\em versus\/}
metre-sized (bottom)
bodies, due to substantial differences in the SFDs of the source populations.
The contributions of individual asteroid families are indicated within the respective pie charts.
For comparison, the observed percentages of meteorite falls from
\url{https://www.lpi.usra.edu/meteor/},
with respect to {\em all\/} classes, are as follows:
HED~6.0\%,
OC~80\%,
{\bf CC~4.4\%\/},
irons~4.1\%,
enstatite chondrites~1.5\%,
aubrites~0.9\%,
mesosiderites~0.6\%,
ureilites~0.5\%,
pallasites~0.3\%,
acapulcoites~0.1\%,
lodranites~0.1\%,
respectively.
}
\label{pie}
\end{figure*}

\subsection{Source regions of kilometre-sized CC-like NEOs}

The vast majority of kilometre-sized CI-like NEOs originate from three families:
Polana (approximately 58 bodies),
Euphrosyne (39), and also
Themis (10)
(Tab.~\ref{tab:1km}, or percentages in Fig.~\ref{pie}).
Regarding CM-like NEOs, the Adeona family is by far the main source (22),
followed at a respectable distance by the Dora family (7).
The Eos family alone is a sufficient source of CO/CV/CK NEOs.
A subset of NEOs with a characteristic shallow 1.00-micron band is related to Baptistina, respectively.

Overall, CI-like NEOs are the dominant population among CC-like NEOs
and they represent the vast majority (${\approx}80\,\%$) of low-albedo ($p_V\leq 0.1$) bodies
that are either CI- or CM-like.
In this context, the CI-like composition of Ryugu and Bennu (see discussion; \citealt{Yokoyama_2023Sci...379.7850Y}),
both with low ($p_V\leq 0.06$) albedos,
is very consistent with our findings and actually not surprising.
Finally, the relative abundance of the different compositional types predicted by our model
are in excellent agreement with the observed ones (Fig.~\ref{pie}).

\subsection{Source regions of carbonaceous chondrites}

\begin{figure}
\centering
\begin{tabular}{c}
\kern0.5cm {\bf Polana (CI)}, 200\,My \\
\includegraphics[width=8.5cm]{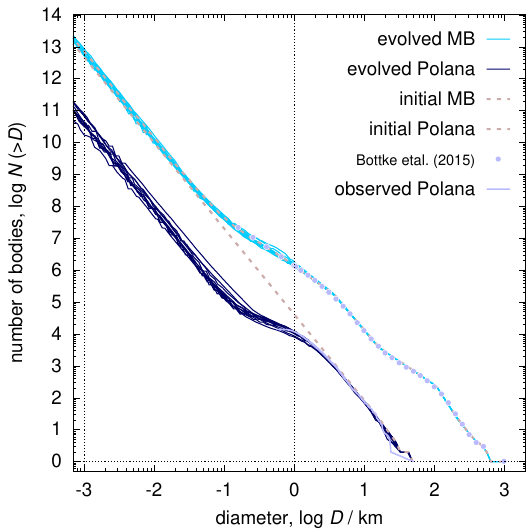} \\
\end{tabular}
\caption{
Synthetic SFD of the Polana family derived from our collisional model.
Initially, it was steep (\textcolor{pantone}{Pantone 7633C}, dashed),
but since then it has evolved and it is now shallow (\textcolor{darkblue}{blue}),
in agreement with observations at kilometre sizes (\textcolor{gray}{gray}, solid).
For comparison, the main belt distribution is also plotted (\textcolor{cyan}{cyan}),
as well as corresponding observations \citep{Bottke_2015aste.book..701B}.
Each model was run 10 times to account for stochasticity.
}
\label{sfd_0200.000}
\end{figure}

There are five families (Polana, Euphrosyne, Clarissa, Misa, Hoffmeister)
that contribute to at least 10\% of the incoming flux of CI chondrites (Tab.~\ref{tab:1m}),
with the Polana family being the most prominent source
(approximately $80\times 10^{-9}\,{\rm km}^{-2}\,{\rm y}^{-1}$).
Its SFD is evolved and shallow at sub-km sizes (Fig.~\ref{sfd_0200.000}).

\begin{figure}
\centering
\begin{tabular}{c}
\kern0.5cm {\bf Veritas (CM)}, 8.3\,My \\
\includegraphics[width=8.5cm]{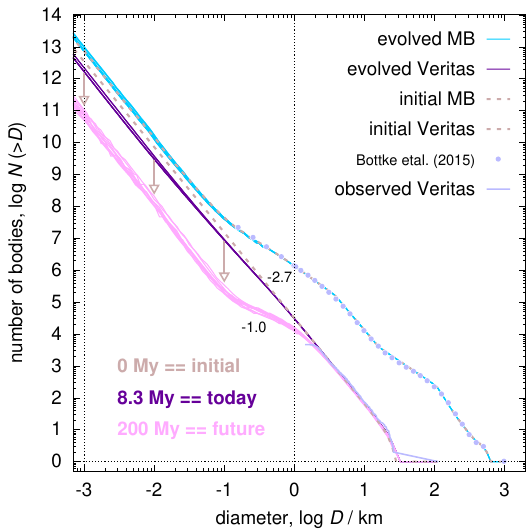} \\
\end{tabular}
\caption{
Same as Fig.~\ref{sfd_0200.000} for the Veritas family.
For a family as young as Veritas, however,
the initially steep SFD is still preserved.
In the future, it will become as shallow as Polana.
}
\label{sfd_0008.300}
\end{figure}

On the contrary,
the 8.3-My-old Veritas family \citep{Nesvorny_2003ApJ...591..486N}
still has a steep SFD (Fig.~\ref{sfd_0008.300}).
It is by far the main source of CM chondrites (200 in the same units),
followed by the K\"onig family (50).
Overall, CM-like metre-sized bodies seem to be the dominant population over CI-like bodies,
essentially because of the massive flux originating from the young Veritas family.

Of great interest, we identify Phaethon as the fourth most important source of CC meteorites,
representing about $30\times 10^{-9}\,{\rm km}^{-2}\,{\rm y}^{-1}$ of the incoming flux,
provided its SFD is continuous from sub-km asteroids to dust.

The Baptistina and Aeolia families, respectively, should also be sources of some meteorites,
even though we were unable to identify an analogue.
Their combined flux (${\approx}\,30$) is non-negligible,
in comparison to other C-type families.

Finally, the Eos family is a viable source of CO/CV/CK chondrites.
Its flux $5\times 10^{-9}\,{\rm km}^{-2}\,{\rm y}^{-1}$ is sufficient,
in comparison to the calibration,
i.e., the Vesta family and HED meteorites ($40\times 10^{-9}\,{\rm km}^{-2}\,{\rm y}^{-1}$).
This suggests that the denser (CO/CV/CK) carbonaceous bodies
are much less affected by atmospheric or pre-atmospheric bias
compared to less dense (CI/CM) bodies (see discussion).
Quantitatively, CO/CV/CK-like meteoroids represent only about 1\% of the incoming CC flux in our simulations,
whereas they should represent more than 30\% of all CC falls.
Such a discrepancy between synthetic and observed CI/CM compositions versus CO/CV/CK ones
is {\em not\/} observed at large kilometre sizes
(see again Fig.~\ref{pie}),
strengthening the likelihood of some sort of bias at small metre sizes.

\subsection{Source regions of other types of compositions}

The majority of metal-rich (M-type) kilometre-sized NEOs originate from
the Brasilia family (13 bodies), followed by the Tina family (4).
The Kalliope, A\"eria and Br\"angane families are negligible source at kilometre sizes.
The situation is more even at metre sizes,
with all families contributing to the meteorite flux
and the young Brang\"ane family being the most prominent source
(about $10\times 10^{-9}\,{\rm km}^{-2}\,{\rm y}^{-1}$).

\vskip\baselineskip
We further identified the Iannini family as the source of acapulcoite and lodranite meteorites.
The match in terms of family age versus CRE ages of these meteorites is excellent,
and the same can be said in terms of composition (pyroxene-rich).
Note that the flux predicted from this family
--in tandem with the flux predicted from the Vesta family--
is consistent with meteorite falls statistics,
acapulcoite and lodranite falls amounting to ${\approx}\,0.2\%$
and HED $6\%$ of all falls (Tab.~\ref{tab:1m}).

\vskip\baselineskip
Finally, there is only one dedicated source of P-type kilometre NEOs, namely the Sylvia family.
Nevertheless, P-type asteroids seem to be present also in
the Euphrosyne and Themis families,
at the levels of 20\% \citep{Yang_2020A&A...641A..80Y}
and 50\% \citep{Marsset_2016A&A...586A..15M,Fornasier_2016Icar..269....1F}, respectively.
These families therefore appear as the main sources of P-types.

\clearpage

\begin{table*}
\centering
\caption{
Dynamical time scales and cumulative numbers of 1-km asteroids
in the main belt (mb) and the near-Earth region (neo).
}
\label{tab:1km}
    
\tiny
\renewcommand{\arraystretch}{0.7}
\def\s#1{\color{gray}#1\color{black}}

\begin{tabular}{rllrrrrrrrrl}
&
&
&
&
1-km &
1-km &
&
1-km &
1-km &
1-km &
1-km &
\\
num. &
family &
res. &
$\tau_{\rm g18}$ &
$\tau_{\rm neo}$ &
$\tau_{\rm mb}$ &
$\rho$ &
$N_{\rm mb}$ &
$N_{\rm neo}$ &
obs. &
obs. &
notes \\
\vrule width 0pt depth 4pt
-- &
-- &
-- &
My &
My &
My &
${\rm g}\,{\rm cm}^{-3}$ &
$10^3$ &
1 &
1 &
\% &
\\
\hline
\vrule width 0pt height 9pt
   4 & Vesta (HED)         & $\nu_6$ & 6.98 & 4.39     & 1713     & 2.5! &  11.4     & 29.2       &           &            & \\ 
\\                                                                                                 
  25 & Phocaea (H)         & $\nu_6$ & 6.98 & 5.91     &  796     & 2.5  &   2.7     & 20.0!      &           &            & \\ 
 170 & Maria (H)           & 3:1     & 1.83 & 0.95$^4$ & 1533     & 3.0  &   5.5     &  3.4       &           &            & \\ 
 808 & Merxia (H)          & 5:2     & 0.68 & 0.24     &  866     & 2.5  &   2.0     &  0.6       &           &            & \\
 847 & Agnia (H)           & 5:2     & 0.68 & 0.19     & 1004     & 2.5  &   3.1     &  0.6       &           &            & \\
 158 & Koronis (H)         & 5:2     & 0.68 & 0.82$^4$ & 1438     & 3.0  &   9.2     &  5.2       &           &            & \\
 158'& Koronis$_2$ (H)     & 5:2     & 0.68 & $^a$     & $^a$     & 3.0  &   0.4     &  0.0       &           &            & \\
 832 & Karin (H)           & 5:2     & 0.68 & $^a$     & $^a$     & 2.5  &   1.1     &  0.0       &           &            & \\
\\                                                                                                                         
  20 & Massalia (L)        & 3:1     & 1.83 & 0.45     & 1140     & 2.5  &   1.3?    &  0.5?      &           &            & \\ 
  20'& Massalia$_2$ (L)    & 3:1     & 1.83 & 0.45     & 1140     & 2.5  &   1.3?    &  0.5?      &           &            & \\ 
1272 & Gefion (L)          & 5:2     & 0.68 & 0.69     &  749     & 2.5  &   3.8     &  3.5       &           &            & \\ 
   3 & Juno (L)            & 8:3     & 1.70 & 2.55     &  519     & 2.5  &   4.2     & 20.6       &           &            & \\
\\                                                                                                                         
   8 & Flora (LL)          & $\nu_6$ & 6.98 & 11.93    &  722     & 2.5  &   7.2     & 119.5      &           &            & \\ 
  15 & Eunomia (LL)        & 3:1     & 1.83 & 4.48     & 3078     & 3.54 &   7.0     & 10.2       &           &            & \\ 
  44 & Nysa (LL)           & 3:1     & 1.83 & 4.04     &  789     & 2.5  &   2.9     & 14.6       &           &            & \\
\\                                                                                                                         
 144 & Vibilia (CM)        & 8:3     & 1.70 & 0.60$^u$ & 2412$^u$ & 1.7  &   0.7     &  0.2       &           &            & all CM are CM/CR \\
 145 & Adeona (CM)         & 8:3     & 1.70 & 1.76     &  864     & 1.7  &  10.9     & 22.2       &           &            & \\
 163 & Erigone (CM)        & 3:1     & 1.83 & 4.64$^a$ &  932$^a$ & 1.7  &   2.7     &  0.0       &           &            & only sub-km nearby 3:1 \\
 410 & Chloris (CM)        & 8:3     & 1.70 & 1.19     &  399     & 1.7  &   0.9     &  2.7       &           &            & \\
 490 & Veritas (CM)        & 2:1     & 0.40 & $^a$     & $^a$     & 1.7  &  30.6     &  0.0       &           &            & \\
 668 & Dora (CM)           & 5:2     & 0.68 & 0.36     &  322     & 1.7  &   6.2     &  6.9       &           &            & \\
 778 & Theobalda (CM)      & 2:1     & 0.40 & $^a$     & $^a$     & 1.7  &  15.9     &  0.0       &           &            & \\
1128 & Astrid (CM)         & 5:2     & 0.68 & 0.23     &  786     & 1.7  &   8.1     &  2.4       &           &            & \\
3815 & K\"onig (CM)        & 3:1     & 1.83 & $^a$     & $^a$     & 1.7  &   3.9     &  0.0       &           &            & \\
\\                                                                                     
  10 & Hygiea (CI)         & 2:1     & 0.40 & 0.27$^u$ & 1899$^u$ & 2.1  &  20.0$^e$ &  2.8       &           &            & all CI are CI/IDP \\
  24 & Themis (CI)         & 2:1     & 0.40 & 0.62     & 1292     & 1.3  &  20.0$^e$ &  9.6       &           &            & \\
  31 & Euphrosyne (CI)     & 2:1     & 0.40 & 0.56     &  399     & 1.6  &  28.0$^e$ & 39.3       &           &            & high-$i$ C-type NEOs \\
 128 & Nemesis (CI)        & 8:3     & 1.70 & $^a$     & $^a$     & 3.5  &   6.3     &  0.0       &           &            & \\ 
 142 & Polana (CI)         & 3:1     & 1.83 & 5.12!    & 1261     & 1.7  &  14.2     & 57.7       &           &            & low-$i$ C-type NEOs \\
 302 & Clarissa (CI)       & 3:1     & 1.83 & $^a$     & $^a$     & 1.7  &   0.8     &  0.0       &           &            & \\
 375 & Ursula (CI)         & 2:1     & 0.40 & 0.30$^u$ & 2187$^u$ & 1.7  &  10.4$^e$ &  1.4       &           &            & \\
 569 & Misa (CI)           & 8:3     & 1.70 & 0.63$^u$ & 2002$^u$ & 1.7  &   2.8     &  0.9       &           &            & \\
 618 & Elfriede (CI)       & 2:1     & 0.40 & $^a$     & $^a$     & 1.7  &  10.6     &  0.0       &           &            & \\
 656 & Beagle (CI)         & 2:1     & 0.40 & $^a$     & $^a$     & 1.3  &   1.3     &  0.0       &           &            & \\
 702 & Alauda (CI)         & 2:1     & 0.40 & 0.30     & 1041     & 1.7  &  12.1     &  3.5       &           &            & \\
 845 & Naema (CI)          & 5:2     & 0.68 & 0.24     &  469     & 1.7  &   4.1     &  2.1       &           &            & \\
1726 & Hoffmeister (CI)    & 2:1     & 0.40 & 0.16     &  432     & 1.7  &  19.5     &  7.2       &           &            & \\
3556 & Lixiaohua (CI)      & 2:1     & 0.40 & $^a$     & $^a$     & 1.7  &  14.1$^e$ &  0.0       &           &            & \\
 283 & Emma (IDP)          & 9:4     & 0.40 & 0.22     & 1689     & 1.7  &   3.3$^e$ &  0.4       &           &            & \\
 363 & Padua (IDP)         & 8:3     & 1.70 & 0.13$^u$ & 2522$^u$ & 1.7  &   2.1     &  0.1       &           &            & \\
   2 & Pallas (B)          & 5:2     & 0.68 & 9.54!    &  824     & 2.9  &   0.3     &  3.5       &           &            & high-$i$ B-type NEOs \\
\\                                                                                                                         
 221 & Eos (CO/CV/CK)      & 7:3     & 0.20 & 0.28     & 1816     & 3.1  &  44.0     &  6.5       &           &            & \\
 729 & Watsonia (CO/CV/CK) & 5:2     & 0.68 & 1.24     & 1391     & 3.1  &   0.3     &  0.3       &           &            & CAI-like \\
\\                                                                                                                         
 298 & Baptistina (?)      & $\nu_6$ & 6.98 & 8.31     & 2833     & 3.1  &   3.0     &  8.8       &           &            & low-$i$ 1.00-$\mu$ NEOs \\
 396 & Aeolia (?)          & 5:2     & 0.68 & $^a$     & $^a$     & 1.7  &   1.9     &  0.0       &           &            & \\
4203 & Brucato (?)         & 8:3     & 1.70 & $^a$     & $^a$     & 1.7  &   2.1     &  0.0       &           &            & \\
\\
  22 & Kalliope (M)        & 5:2     & 0.68 & 0.31     & 1635     & 4.1  &   0.6     &  0.1       &           &            & \\
 293 & Brasilia (M)        & 5:2     & 0.68 & 0.83     &  285     & 3.5  &   4.4     & 12.8       &           &            & CH? \\
 369 & A\"eria (M)         & 8:3     & 1.70 & $^a$     & $^a$     & 3.5  &   0.8     &  0.0       &           &            & \\
 606 & Brang\"ane (M)      & 3:1     & 1.83 & $^a$     & $^a$     & 3.1  &   0.7     &  0.0       &           &            & CB? \\
1222 & Tina (M)            & 5:2     & 0.68 & 1.31     &  539     & 3.5  &   1.7     &  3.6       &           &            & \\
  44'& Nysa (E)            & 3:1     & 1.83 & 4.04     &  789     & 2.5  &   2.9     & 14.6       &           &            & faint Xn-type NEOs \\ 
 434 & Hungaria (E)        & 5:1     & 36.0 & 14.70!   & 1752     & 3.1  &   0.6     &  5.0       &           &            & high-$i$ E-type NEOs \\
\\
  87 & Sylvia (P)          & 2:1     & 0.40 & 0.33     & 1207     & 1.3  &   7.0$^e$ &  1.9       &           &            & \\
2732 & Witt (?)            & 5:2     & 0.68 & 0.15     & 2070     & 2.5  &   2.8     &  0.2       &           &            & \\
3200 & Phaethon (CY?)      & --      & --   & --       & --       & 1.7  &   --      &  3.0       &           &            & \\
4652 & Iannini (Aca/Lod)   & 11:4    & --   & $^a$     & $^a$     & 2.5  &   0.3     &  0.0       &           &            & \\
\hline
\end{tabular}
\end{table*}
\addtocounter{table}{-1}
\begin{table*}
\caption{continued.}
\centering
\tiny
\renewcommand{\arraystretch}{0.7}
\def\s#1{\color{gray}#1\color{black}}
\begin{tabular}{rllrrrrrrrrl}
\hline
\vrule width 0pt height 9pt                                                                   
     & HED                 &         &      &          &          &      &  11.4     &  29.2      & \s{ 16}   & \s{ 1.8\%} & \cite{Marsset_2022AJ....163..165M} \\
     & H                   &         &      &          &          &      &  23.6     &  29.8      & \s{--}    & \s{--}     & \cite{Broz_2023} \\
     & L                   &         &      &          &          &      &  10.6     &  25.1      & \s{--}    & \s{--}     & \cite{Broz_2023} \\
     & LL                  &         &      &          &          &      &  17.1     &  144.3     & \s{--}    & \s{--}     & \cite{Broz_2023} \\
\\
     & CM                  &         &      &          &          &      &  79.9     &  34.4      & \s{--}    & \s{--}     & \\
     & CI                  &         &      &          &          &      & 164.2     & 124.5      & \s{--}    & \s{--}     & \\
     & IDP                 &         &      &          &          &      &   5.4     &   0.5      & \s{--}    & \s{--}     & \\
     & \s{B}               &         &      &          &          &      &   0.3     &   3.5!     & \s{ 70}   & \s{ 7.6\%} & outliers of CI-like? \\
     & CO/CV/CK            &         &      &          &          &      &  44.3     &   6.8      & \s{  8}   & \s{ 0.9\%} & reclassified K- and L-type \\
     & \s{CR}              &         &      &          &          &      & --        & --         & \s{--}    & \s{--}     & CM-type? \\
     & \s{CH}              &         &      &          &          &      & --        & --         & \s{--}    & \s{--}     & M-type? \\
     & \s{CB}              &         &      &          &          &      & --        & --         & \s{--}    & \s{--}     & M-type? \\
     & \s{M}               &         &      &          &          &      &   8.2     &  16.5      & \s{ 29}   & \s{ 3.1\%} & \\
     & \s{E}               &         &      &          &          &      &   3.5     &  19.6      & \s{  5}   & \s{ 0.5\%} & \\
     & \s{P}               &         &      &          &          &      &   7.0     &   1.9!     & \s{ 26}   & \s{ 2.8\%} & outliers of CI-like? \\
     & \s{D}               &         &      &          &          &      & --        & --         & \s{ 45}   & \s{ 4.9\%} & Jupiter-family comets? \\
     & \s{Aca/Lod}         &         &      &          &          &      &   0.3     &   0.0      & \s{--}    & \s{--}     & \\
     & \s{?}               &         &      &          &          &      &   4.9     &   8.8      & \s{ 24}   & \s{ 2.6\%} & reclassified K- and L-type \\
     & \s{?}               &         &      &          &          &      &   2.1     &   0.0      & \s{--}    & \s{--}     & \\
\\
     & \multicolumn{4}{l}{CM+CI+IDP+B+CO/CV/CK+CR+CH+CB} &        &      & 304.1     & 178.5      & \s{303}   & \s{32.8\%} & \\
     & H+L+LL              &         &      &          &          &      &  51.3!    & 199.2      & \s{287}   & \s{31.1\%} & \\
\\                                                                                                    
     & all bodies          &         &      &          &          &      & 1360$^H$  &            & {\bf 925}$^H$ &      \\
\hline
\end{tabular}

\tablefoot{
For all families, we report
the neighbouring resonances,
the NEO life time~$\tau_{\rm g18}$ from \cite{Granvik_2018Icar..312..181G},
the NEO life times~$\tau_{\rm neo}$ from this work,
computed for 1-km bodies,
the main belt life times~$\tau_{\rm mb}$,
the volumetric density of simulated bodies,
the observed cumulative number~$N_{\rm mb}({>}\,1\,{\rm km})$ of main belt bodies,
the computed cumulative number~$N_{\rm neo}$ of NEOs and meteoroids.
Additional notes:
$^H$~\cite{Harris_2021Icar..36514452H},
$^N$~\cite{Nesvorny_2023AJ....166...55N},
$^a$~after ${\approx}\,100\,{\rm My}$ or more,
$^e$~extrapolated using a~collisional model,
$^u$~undersampled,
?~uncertain value,
!~exceptional value.
}
\end{table*}

\begin{table*}
\setlength{\tabcolsep}{5pt}
\centering
\caption{
Same as Tab.~\ref{tab:1km} for 1-m meteoroids.
}
\label{tab:1m}

\tiny
\renewcommand{\arraystretch}{0.7}
\def\s#1{\color{gray}#1\color{black}}

\begin{tabular}{rllrrrrrrrll}
&
&
&
&
1-m &
1-m &
&
1-m &
1-m &
1-m &
1-m &
\\
num. &
family &
res. &
$\tau_{\rm g18}$ &
$\tau_{\rm neo}$ &
$\tau_{\rm mb}$ &
$\rho$ &
$N_{\rm mb}$ &
$N_{\rm neo}$ &
$\Phi$ &
obs. &
notes \\
\vrule width 0pt depth 4pt
-- &
-- &
-- &
My &
My &
My &
${\rm g}\,{\rm cm}^{-3}$ &
$10^{10}$ &
$10^8$ &
$10^{-9}\,{\rm km}^{-2}\,{\rm y}^{-1}$ &
\% &
\\
\hline
\vrule width 0pt height 9pt
   4 & Vesta (HED)            & $\nu_6$ & 6.98 & 2.50     & 115      & 2.5  & 2-7           & 4.3-15      & 18-62          &          & mesosiderites \citep{Libourel_2023PSJ.....4..123L} \\ 
\\
  25 & Phocaea (H)            & $\nu_6$ & 6.98 & 7.24     & 114      & 2.5  & 0.5-1         & 3.2-6.4     & 3.2-6.5        &          \\
 170 & Maria (H)              & 3:1     & 1.83 & 1.82     &  98      & 2.5  & 0.8-2         & 1.5-3.7     & 2.3-5.7        &          \\ 
 808 & Merxia (H)             & 5:2     & 0.68 & 0.43     &  81      & 2.5  & 0.3-0.9       & 0.2-0.5     & 0.3-0.8        &          \\
 847 & Agnia (H)              & 5:2     & 0.68 & 0.34     & 103      & 2.5  & 1-2           & 0.3-0.7     & 1.6-3.7        &          \\
 158 & Koronis (H)            & 5:2     & 0.68 & 0.36     & 176      & 2.5  & 2-4           & 0.4-0.8     & 2.5-5.0        &          \\
 158'& Koronis$_{(2)}$ (H)    & 5:2     & 0.68 & 0.33     & 138      & 2.5  & $\sim$100-200 & 24-48       & 137-274        &          & $2.11^\circ$ dust band \\
 832 & Karin (H)              & 5:2     & 0.68 & 0.33     & 138      & 2.5  & 30-60         & 7.2-14      & 41-82          &          & {\em ditto\/} \\
\\                                                                                                        
  20 & Massalia (L)           & 3:1     & 1.83 & 3.83     & 139      & 2.5  & 0.4-1         & 1.1-2.8     & 8.0-20         &          \\ 
  20'& Massalia$_{(2)}$ (L)   & 3:1     & 1.83 & 3.83     & 139      & 2.5  & $\sim$10-20   & 28-55       & 200-400        &          & $1.43^\circ$ dust band \\
1272 & Gefion (L)             & 5:2     & 0.68 & 0.32     &  75      & 2.5  & 0.5-1.5       & 0.2-0.6     & 0.3-0.9        &          \\ 
   3 & Juno (L)               & 8:3     & 1.70 & 1.38     & 204      & 2.5  & 0.5-1.5       & 0.3-1.0     & 0.6-1.9        &          \\
\\                                                                             
   8 & Flora (LL)             & $\nu_6$ & 6.98 & 3.45     & 110      & 2.5  & 2-4           & 6.3-13      & 24-47          &          \\ 
  15 & Eunomia (LL)           & 3:1     & 1.83 & 1.56     & 199      & 2.5  & 1-6           & 0.8-4.7     & 1.3-7.4        &          \\ 
  44 & Nysa (LL)              & 3:1     & 1.83 & 1.79     & 114      & 2.5  & 0.5-0.8       & 0.8-1.3     & 6.3-10         &          \\
\\                                                                                                                                    
 144 & Vibilia (CM)           & 8:3     & 1.70 & 1.24     &  62      & 1.7  & 0.3-0.7       & 0.6-1.4     & 3.3-7.7        &          & \\ 
 145 & Adeona (CM)            & 8:3     & 1.70 & 1.55     & 157      & 2.3  & 2-5           & 2.0-4.9     & 4.2-13.9       &          & $11.7^\circ$ \\
 163 & Erigone (CM)           & 3:1     & 1.83 & 2.33!    &  77      & 1.7  & 0.3-1         & 0.9-3.0     & 5.0-12.2       &          & $4.8^\circ$ \\
 410 & Chloris (CM)           & 8:3     & 1.70 & 0.52     &  53      & 2.3  & 0.2-0.6       & 0.2-0.6     & 0.5-1.4        &          & \\
 490 & {\bf Veritas (CM)}     & 2:1     & 0.40 & 0.39     &  58$^1$  & 2.3  & 320-370!      & 215-249!    & 188-217        &          & $9.35^\circ$ dust band \\  
 668 & Dora (CM)              & 5:2     & 0.68 & 0.42     &  54      & 2.3  & 1.4-5         & 1.1-3.9     & 2.4-8.4        &          & $7.9^\circ$ \\
 778 & Theobalda (CM)         & 2:1     & 0.40 & 0.38     &  51      & 1.7  & 20-60!        & 15-45!      & 13-40          &          & $14.4^\circ$ \\
1128 & Astrid (CM)            & 5:2     & 0.68 & 0.60     & 134      & 1.7  & 1-3           & 0.4-1.3     & 2.2-7.1        &          & $0.6^\circ$ \\
3815 & {\bf K\"onig (CM)}     & 3:1     & 1.83 & 1.64     &  63      & 1.7  & 4-15          & 10-39!      & 22-84          &          & $9.0^\circ$! \\
\\                                                                                                                                     
  10 & Hygiea (CI)            & 2:1     & 0.40 & 0.36     & 161      & 1.6  & 6-14          & 1.3-3.1     & 0.9-2.2        &          & $5.1^\circ$, all CI are CI/IDP \\
  24 & Themis (CI)            & 2:1     & 0.40 & 0.43     & 145      & 1.6  & 10-30         & 3.1-9.4     & 6.6-20         &          & $1.1^\circ$ \\
  31 & {\bf Euphrosyne (CI)}  & 2:1     & 0.40 & 0.93     &  51      & 1.6  & 5-15          & 9.1-27!     & 7.0-21         &          & $26.5^\circ$ \\
 128 & Nemesis (CI)           & 8:3     & 1.70 & 0.49     & 155      & 3.8  & 0.6-2         & 0.2-0.6     & 0.5-1.5        &          & \\ 
 142 & {\bf Polana (CI)}      & 3:1     & 1.83 & 1.89     &  66      & 1.6  & 3-8           & 8.6-23      & 42-110!        &          & $3.3^\circ$ \\
 302 & Clarissa (CI)          & 3:1     & 1.83 & 2.44!    &  73      & 1.7  & 0.5-4         & 1.7-13      & 7.0-55         &          & $3.3^\circ$ \\
 375 & Ursula (CI)            & 2:1     & 0.40 & 0.71     & 185      & 1.6  & 3-10          & 1.2-3.8     & 1.0-3.3        &          & $16.4^\circ$ \\
 569 & Misa (CI)              & 8:3     & 1.70 & 1.08     &  80      & 1.7  & 1-6           & 1.4-8.1     & 7.3-42!        &          & $2.3^\circ$! \\
 618 & Elfriede (CI)          & 2:1     & 0.40 & 0.53     & 319      & 1.7  & 5-35          & 0.8-5.8     & 0.6-4.2        &          & $15.9^\circ$ \\
 656 & Beagle (CI)            & 2:1     & 0.40 & 0.20!    & 158      & 1.6  & 2-3           & 0.2-0.4     & 0.2-0.4        &          & cf. \cite{Marsset_2023} \\
 702 & Alauda (CI)            & 2:1     & 0.40 & 0.66     &  63      & 1.7  & 2-11          & 2.1-12      & 1.6-8.9        &          & $21.6^\circ$ \\
 845 & Naema (CI)             & 5:2     & 0.68 & 0.41     & 114      & 1.7  & 0.4-0.9       & 0.1-0.3     & 0.1-0.3        &          & \\
1726 & Hoffmeister (CI)       & 2:1     & 0.40 & 0.53     &  93      & 1.6  & 6-14          & 3.4-8.0     & 13-30          &          & $4.4^\circ$ \\
3556 & Lixiaohua (CI)         & 2:1     & 0.40 & 0.35     &  68      & 1.6  & 1.5-4.5       & 0.8-2.3     & 1.2-3.6        &          & $10.1^\circ$ \\
 283 & Emma (IDP)             & 9:4     & 0.40 & 0.28     & 174      & 1.7  & 0.8-3.5       & 0.1-0.6     & 0.1-0.7        &          & \\
 363 & Padua (IDP)            & 8:3     & 1.70 & 0.90     & 106      & 1.7  & 0.7-1.2       & 0.6-1.0     & 1.5-2.5        &          & \\
   2 & Pallas (B)             & 5:2     & 0.68 & 3.62     & 134      & 2.9  & 0.1-1         & 0.3-2.7     & 0.2-1.9        &          & $33.2^\circ$ \\
\\                                                                                                                                    
 221 & Eos (CO/CV/CK)         & 7:3     & 0.20 & 0.41     & 313!     & 3.1  & 10-30         & 1.3-3.9     & 2.1-6.3        &          & $9.9^\circ$ \\
 729 & Watsonia (CO/CV/CK)    & 8:3     & 1.70 & 1.88     &  53      & 3.1  & 0.1-0.3       & 0.4-1.1     & 0.5-1.3        &          & $17.4^\circ$! \\
\\
 298 & {\bf Baptistina (?)}   & $\nu_6$ & 6.98 & 5.11     &  85      & 3.1  & 0.6-1         & 3.6-6.0     & 11-19          &          & $6.0^\circ$! \\
 396 & {\bf Aeolia (?)}       & 5:2     & 0.68 & 0.56     &  58      & 1.7  & 3-10          & 2.9-9.7     & 9.7-33         &          & $3.5^\circ$ \\
4203 & Brucato (?)            & 8:3     & 1.70 & 4.48!    &  51      & 1.7  & 0.3-0.7       & 2.6-6.1     & 2.3-5.4        &          & $28.8^\circ$ \\
\\
  22 & Kalliope (M)           & 5:2     & 0.68 & 0.48     & 123      & 3.5  & 2-5$^S$       & 0.8-2.0     & 1.3-3.2        &          & \\
 293 & Brasilia (M)           & 5:2     & 0.68 & 0.68     &  86      & 3.8  & 1.5-3.5       & 1.2-2.8     & 1.7-3.9        &          & $15.0^\circ$ CH? \\
 369 & A\"eria (M)            & 8:3     & 1.70 & 1.11     & 346      & 3.8  & 1.5-3$^S$     & 0.5-1.0     & 0.9-1.8        &          & \\
 606 & {\bf Brang\"ane (M)}   & 3:1     & 1.83 & 1.55     & 102      & 3.1  & 0.5-5         & 0.8-7.6     & 2.0-19         &          & $9.6^\circ$! CB? \\
1222 & Tina (M)               & 5:2     & 0.68 & 1.72     &  21      & 3.5  & 0.1-0.4       & 0.8-3.3     & 0.9-3.8        &          & \\
  44'& Nysa (E)               & 3:1     & 1.83 & 1.79     & 114      & 2.5  & 0.5-0.8       & 0.8-1.3     & 6.3-10         &          & $3.1^\circ$ EH/EL? \\
 434 & Hungaria (E)           & 5:1     & 36.0 & 35.18!   & 300?     & 3.1  & 0.5-0.8       & 5.9-9.4     & 7.6-12         &          & $20.9^\circ$ aubrites \citep{Cuk_2014Icar..239..154C} \\
\\
  87 & Sylvia (P)             & 2:1     & 0.40 & 0.46     & 392      & 1.3  & 2-6           & 0.2-0.7     & 0.2-0.6        &          & $9.9^\circ$ \\
2732 & Witt (?)               & 5:2     & 0.68 & 0.91     & 118      & 2.5  & 1.5-4         & 1.2-3.1     & 3.5-9.0        &          & $5.8^\circ$ \\
3200 & Phaethon (CY?)         & --      & --   & --       & --       & 1.7  & --            & 2-15        & 6.3-48!        &          & Geminids, \citet{Maclennan_2024NatAs...8...60M} \\
4652 & Iannini (Aca/Lod)      & 11:4    & --   & 1.65     & 283      & 2.5  & $\sim$1-5     & 0.6-2.9     & 0.8-3.9        &          & $12.1^\circ$ dust band, pyroxene-rich \\
\hline
\end{tabular}
\end{table*}
\addtocounter{table}{-1}
\begin{table*}
\caption{continued.}
\centering
\tiny
\renewcommand{\arraystretch}{0.7}
\def\s#1{\color{gray}#1\color{black}}
\begin{tabular}{rllrrrrrrrrl}
\hline                                                                                                                             
\vrule width 0pt height 9pt                                                                                                        
     & HED                    &         &      &          &          &      & 2-7           & 4.3-15      & 18-62          & \s{6.0\%}   & \\
     & H                      &         &      &          &          &      & 135-270       & 37-74       & 188-377        & \s{34\%}    & \cite{Broz_2023} \\
     & L                      &         &      &          &          &      & 11-24         & 29-60       & 209-423        & \s{38\%}    & \cite{Broz_2023} \\
     & LL                     &         &      &          &          &      & 3.5-11        & 7.9-19      & 31-65          & \s{8.3\%}   & \cite{Broz_2023} \\
\\
     & CM                     &         &      &          &          &      & 349-460       & 246-348     & 241-391        & \s{1.8\%}   & cf. fragmentation! \\
     & CI                     &         &      &          &          &      & 46-157        & 34-117      & 88-304         & \s{0.4\%}   & cf. fragmentation! \\
     & IDP                    &         &      &          &          &      & 1.5-4.7       & 0.7-1.6     & 1.6-3.2        & \s{0.5\%}   & \\
     & \s{B}                  &         &      &          &          &      & 0.1-1         & 0.3-2.7     & 0.2-1.9        & \s{0.6\%}   & \\
     & CO/CV/CK               &         &      &          &          &      & 10-30         & 1.7-5.0     & 2.6-7.6        & \s{1.3\%}   & \\
     & \s{CR}                 &         &      &          &          &      & --            & --          & --             & \s{0.2\%}   & CM-type? \\
     & \s{CH}                 &         &      &          &          &      & --            & --          & --             & \s{0.05\%?} & M-type? \\
     & \s{CB}                 &         &      &          &          &      & --            & --          & --             & \s{0.05\%?} & M-type? \\
     & \s{M}                  &         &      &          &          &      & 5.6-15        & 4.1-15      & 6.8-31         & \s{4.1\%}   & irons \\
     & \s{E}                  &         &      &          &          &      & 1.5-2.4       & 6.7-11      & 14-22!         & \s{2.4\%}   & enstatite, reduced \\
     & \s{Aub}                &         &      &          &          &      & --            & --          & --             & \s{0.9\%}   & E-type \\
     & \s{EH}                 &         &      &          &          &      & --            & --          & --             & \s{0.8\%}   & E-type? \\
     & \s{EL}                 &         &      &          &          &      & --            & --          & --             & \s{0.7\%}   & E-type? \\
     & \s{P}                  &         &      &          &          &      & 2-6           & 0.2-0.7     & 0.2-0.6        & \s{--}      & \\
     & \s{D}                  &         &      &          &          &      & --            & --          & --             & \s{--}      & \\
     & \s{Aca/Lod}            &         &      &          &          &      & 1-5           & 0.6-2.9     & 0.8-3.9        & \s{0.2\%}   & acaplucoites, lodranites \\
     & \s{?}                  &         &      &          &          &      & 3.6-11        & 6.5-16      & 21-51          & \s{--}      & \\
     & \s{?}                  &         &      &          &          &      & 0.3-0.7       & 2.6-6.1     & 2.3-5.4        & \s{--}      & \\
\\
     & \multicolumn{4}{l}{CM+CI+IDP+B+CO/CV/CK+CR+CH+CB+?}&          &      & 411-665       & 281-461     & 355-760        & \s{4.4\%}   & \\
     & H+L+LL                 &         &      &          &          &      & 150-305       & 74-152      & 428-865        & \s{80\%}    & \\
\\                                                                                                                                     
     & all bodies             &         &      &          &          &      & 500-1400      & 200-300$^H$ & {\bf 740}$^N$  &             & \\
     & all bodies             &         &      &          &          &      &               &             & {\bf 890}$^C$  &             & \\
\hline
\end{tabular}

\tablefoot{
$^1$ Extended synthetic family of 1-m bodies, corresponding to a \hbox{$1/\!\sqrt{D}$} dependence;
$^S$ assuming $10$ times higher strength;
$^H$ \cite{Harris_2021Icar..36514452H};
$^N$ \cite{Nesvorny_2023AJ....166...55N}.
}
\end{table*}

\clearpage


\section{Discussion}

Our model ('METEOMOD'%
\footnote{\url{https://sirrah.troja.mff.cuni.cz/~mira/meteomod/}}%
) for populations of meteoroids and NEOs,
which has been constrained by the observed SFDs,
can be compared to various types of other observations.
We organize the discussion around three major science questions to which we attempt to provide elements of answer.

\def\q#1{\vskip\baselineskip\noindent{\it Question: #1\/}}
\q{Are our proposed source regions of CC meteorites compatible with laboratory and remote measurements?}

\subsection{IRAS dust bands}\label{iras}

\begin{figure}
\centering
\includegraphics[width=9cm]{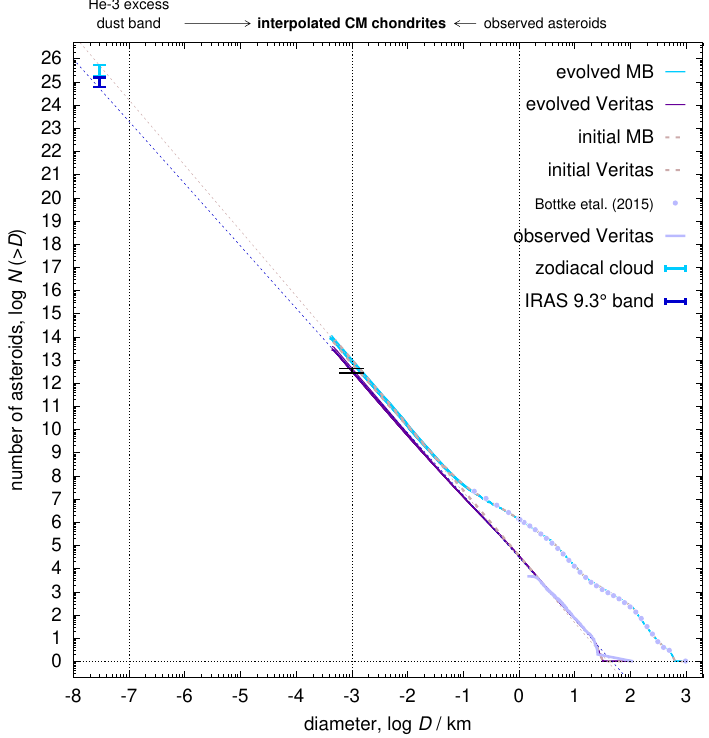}
\caption{
Observed SFD of the Veritas family (\textcolor{gray}{gray}, solid).
Its extrapolation (dotted)
is supported by the observed $9.3^\circ$ IRAS dust band
\citep{Nesvorny_2006Icar..181..107N},
at approximately $30\,\mu{\rm m}$ (error bar).
Moreover, the initial SFD (\textcolor{pantone}{Pantone 7633C}, dashed)
is supported by the excess of extraterrestrial He-3
\citep{Farley_2006Natur.439..295F},
which 8.3\,My ago 'overshot' the background level by a~factor of~${\sim}3$.
The interpolation to metre sizes (error bar)
implies the abundance of CM chondrite meteoroids.
}
\label{Veritas_DUST}
\end{figure}

The prominent $9.3^\circ$ IRAS dust band
is known to be associated with the CM-like Veritas family
(\citealt{Nesvorny_2006Icar..181..107N}).
If the observed SFD is extrapolated with a $-2.7$ slope from kilometre sizes to $100\,\mu{\rm m}$,
it 'hits' the observed dust band abundance
(see Fig.~\ref{Veritas_DUST}).
If the SFD is {\em interpolated\/} to 1\,m,
it corresponds to the number of CM meteoroids
in our model.

The inclinations of
K\"onig ($9.0^\circ$) and
Brang\"ane ($9.6^\circ$)
are so similar to that of Veritas ($9.3^\circ$)
that their dust bands might be 'hidden' within that of Veritas.
Since the band is more than several degrees wide,
the model of \cite{Nesvorny_2006Icar..181..107N} can surely be adapted
to encompass all three families.
Perhaps, there is another hidden band related to the Misa family ($2.3^\circ$),
contributing to the $2.1^\circ$ band,
even though its age indicates that the band must be 'extinct'.

Fainter bands have imprecise inclinations
(Fig.~\ref{aei7_sykes}), the only exception being the $12.1^\circ$ J/K band,
associated to the family of (4562) Iannini
(also denoted (1547) Nele; \citealt{Carruba_2018MNRAS.477.1308C}).

The ${\approx}\,15^\circ$ M/N dust band
may be linked to the family of (293) Brasilia 
(also named after (1521) Seinajoki; \citealt{Nesvorny_2003ApJ...591..486N})
as its SFD is steep down to the observational limit.
However, the effect of Ceres resonance indicates an age of 100\,My,
implying that the band should be more-or-less extinct.
Nevertheless, having both prominent and extinct bands is expected.



\subsection{CRE ages}\label{cre}

\begin{figure}
\centering
\includegraphics[width=8.5cm]{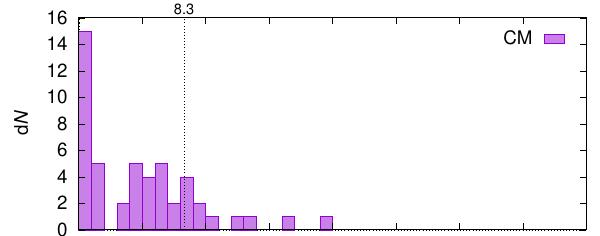}
\includegraphics[width=8.5cm]{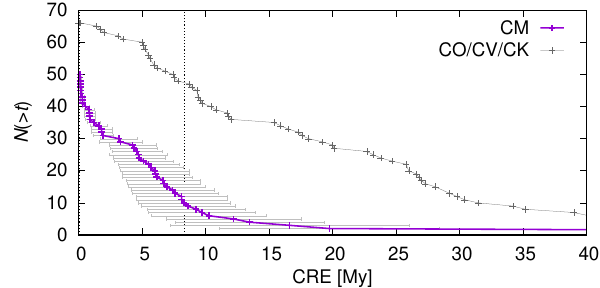}
\includegraphics[width=8.5cm]{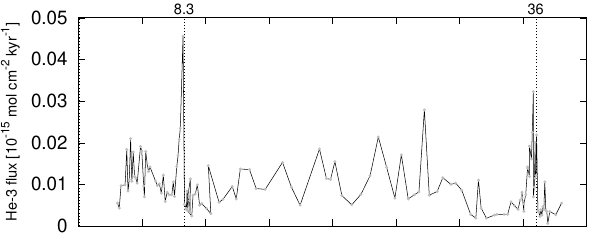}
\includegraphics[width=8.5cm]{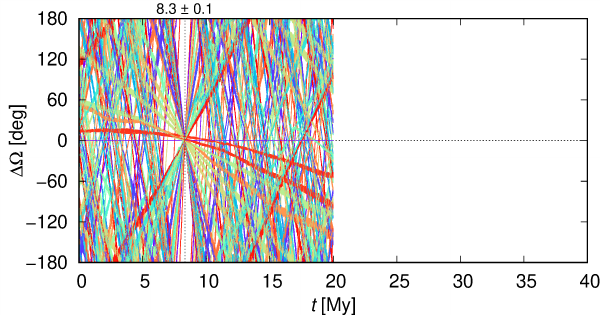}
\caption{
CRE ages of CM chondrites (top)
\citep{Krietsch_2021GeCoA.310..240K}
exhibit an increase on a cumulative histogram,
i.e., a peak on a differential histogram.
On the contrary, CV, CO, and CK chondrites
\citep{Scherer_2000M&PS...35..145S}
exhibit a continuous distribution over the last $40\,{\rm My}$.
This increase/peak corresponds to the late miocene peak
of extraterrestrial helium-3 in terrestrial sediments
\citep{Farley_2006Natur.439..295F} (middle)
and to the age of the CM-like Veritas family,
$(8.3\pm0.1)\,{\rm My}$,
as confirmed by convergence of orbits (bottom).
\color{gray}
A plot with a sum of probability distributions presented in
\citet{Krietsch_2021GeCoA.310..240K}, fig.~9
is misleading; uncertainties of old CRE ages are large,
the probability is low, it seems like there are 'none'
meteorites having old CRE ages.
On the contrary, our cumulative histogram demonstrates
that there are numerous meteorites with old CRE ages.
\color{black}
}
\label{cre_CM_hist2}
\end{figure}

CRE ages of CM chondrites
(\citealt{Krietsch_2021GeCoA.310..240K}; Fig.~\ref{cre_CM_hist2})
indicate an onset at 8 to 9\,My
corresponding to the age of the Veritas family (8.3\,My), in agreement with our model.
The ages seem to indicate another onset at about 2\,My.
This may correspond to a secondary collision, which occur regularly in our collisional model.

CY (i.e., reclassified CI) chondrites
\citep{King_2019ChEG...79l5531K}
exhibit very short CRE ages (${<}\,1.3\,{\rm My}$).
We did not find any CI-like family which could be so young.
Instead, it is likely that CI-like meteoroids
transported to the NEO space
were thermally altered in a similar way as (3200) Phaethon
\citep{Maclennan_2024NatAs...8...60M}
-- due to its extreme eccentric orbit,
high irradiation flux,
leading to additional fragmentation (M. Granvik, pers. comm.),
hence shortening of CRE ages.

CO, CV and CK chondrites
\citep{Scherer_2000M&PS...35..145S}
exhibit a continuous distribution over the last $40\,{\rm My}$,
especially if all the three types are considered together.
This corresponds to a typical collisional time scale,
expected for old families like Eos.
A similarity to the distribution of L or LL chondrites
confirms this conclusion,
as well as a similar strength to ordinary chondrites.

CB chondrites are rare. Nevertheless,
Gujba has an age $(26\pm 7)\,{\rm My}$,
Bencubbin ${\sim}\,27\,{\rm My}$,
Isheyevo ${\sim}\,34\,{\rm My}$
\citep{Rubin_2003GeCoA..67.3283R,Ivanova_2008M&PS...43..915I},
all of which might be compatible with the Brang\"ane family;
even though we classified it rather as M-type.
CH chondrites are likely related to CB,
having a similar formation age and environment
\citep{Ivanova_2008M&PS...43..915I,Wolfer_2023arXiv231012749W},
but their CRE ages are relatively short, 1 to 12\,My
\citep{Eugster_2006mess.book..829E},
indicating a distinct parent body.

\subsection{Pre-atmospheric orbits of CM chondrites}\label{preatmospheric}

\begin{figure}
\centering
\includegraphics[width=9cm]{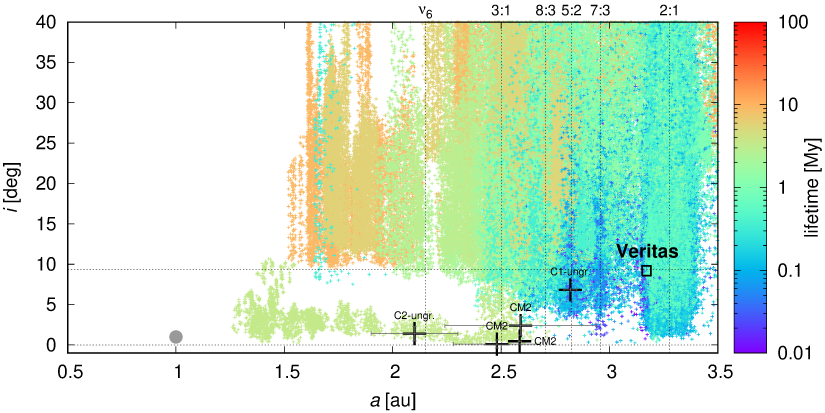}
\includegraphics[width=9cm]{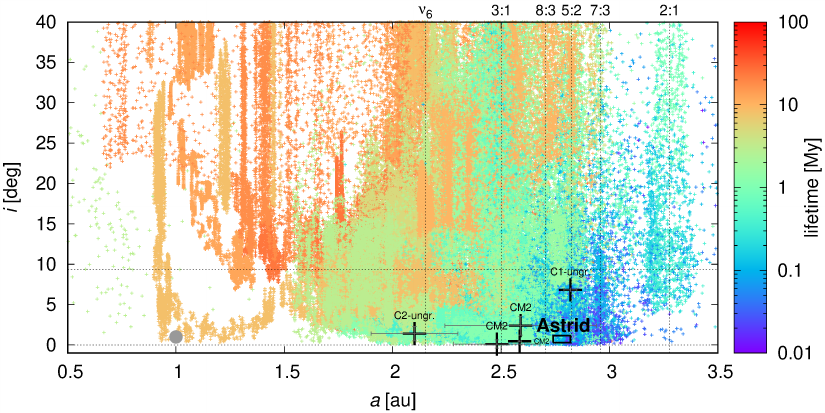}
\caption{
Pre-atmospheric orbits of CM, C1- and C2-ungrouped meteorites
(\citealt{Meier_2023}; black crosses with error bars).
Their maximum semimajor axis is well beyond the 3:1 mean-motion resonance
($a \lesssim 2.82\,{\rm au}$)
and their inclination is low
($i \lesssim 2^\circ$).
For comparison,
synthetic orbits of 1-m meteoroids from Veritas (top) and Astrid (bottom) are plotted,
whenever an orbit is in the NEO region
($a(1+e) > 0.983\,{\rm au} \land a(1-e) < 1.3\,{\rm au}$).
Colours correspond to the lifetimes in the NEO region.
It is difficult to scatter orbits from high $\rightarrow$ low inclinations,
and {\em vice versa\/}.
Low-inclination orbits also have much larger probability~$p_{\rm i}$
of collision with the Earth.
Nevertheless, the flux $\Phi = p_{\rm i} N_{\rm neo}$
from Veritas is larger than from Astrid,
according to our model.
Consequently, we favor Veritas as the main source of CM chondrites.
}
\label{ai_CM}
\end{figure}

Pre-atmospheric orbits of CM2 chondrites
(\citealt{Meier_2023}; Fig.~\ref{ai_CM})
are rare, but even with a limited sample of 3 orbits,
it probably shows the most probable orbits
from an underlying distribution.
Orbits have low inclinations $i \lesssim 2^\circ$,
semimajor axes up to $a \simeq 2.6\,{\rm au}$.
These are important indicators,
because {\em some\/} meteorites should be found close to their source.

According to Tab.~\ref{tab:1m},
one of the promising sources could be the Astrid family,
with $i \simeq 0.6^\circ$ and $a \simeq 2.787\,{\rm au}$.
It readily produces low-inclination orbits of meteoroids
and these have the largest collisional probability with the Earth.
However, the respective SFD of Astrid is too shallow.

On the contrary, the SFD of Veritas is so steep
that it should produce a much larger flux anyway
(cf. $\Phi$ in Tab.~\ref{tab:1m}).
Some meteoroids from Veritas are indeed able to reach low-inclination orbits
(see Fig.~\ref{ai_CM}).
This is even more common if the ejection velocity field from Veritas
was more extended for metre-sized bodies than for kilometre ones.
However, the size dependence of $v(D)$ can hardly be $1/D$,
because particles at $100\,\mu{\rm m}$ would be totally dispersed
and we would see no dust bands.
Instead, it could be proportional to ${\approx}1/\!\sqrt{D}$.

\subsection{Entry speeds from FRIPON}\label{fripon}

\begin{figure}
\centering
\includegraphics[width=8cm]{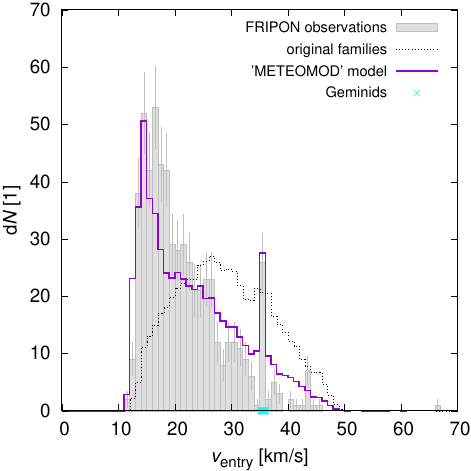}
\caption{
Histogram of entry speeds observed by FRIPON (\textcolor{gray}{gray})
and computed from our model ('METEOMOD'; \textcolor{violet}{violet}),
with appropriate weighting by collisional probabilities.
For comparison, underlying (unweighted) distribution from the original families
is plotted (dotted).
The uncertainties were estimated as poissonian.
The total number of events was 602.
}
\label{impvel_nooutliers}
\end{figure}

\begin{figure}
\centering
\includegraphics[width=9cm]{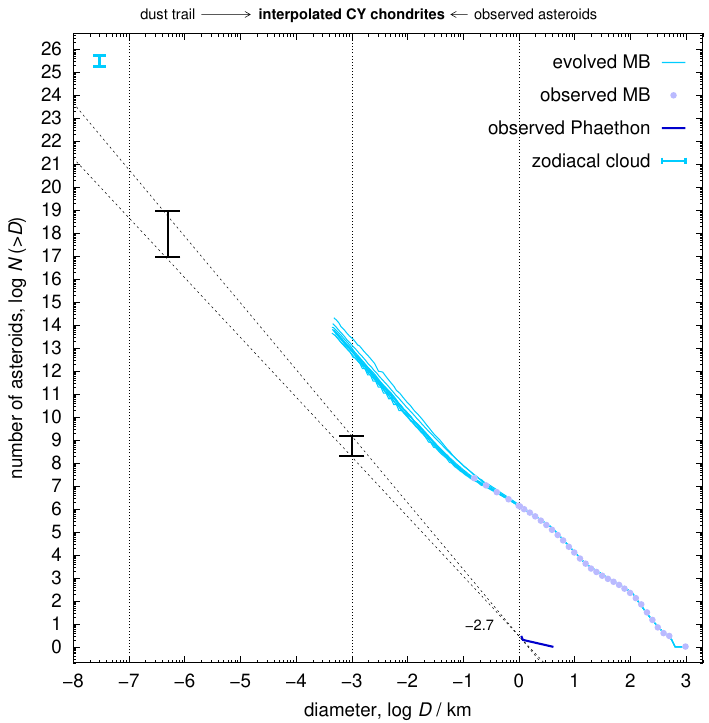}
\caption{
Same as Fig.~\ref{Veritas_DUST} for (3200) Phaethon.
Two additional NEOs were included in this SFD,
(155140) and (225416);
\citep{Maclennan_2021Icar..36614535M}.
Its extrapolation (dotted)
is supported by the observed dust trail \citep{Battams_2022ApJ...936...81B}.
Its interpolation (error bar)
explains the abundance of Geminids.
Cf.~Fig.~\ref{impvel_nooutliers}.
}
\label{Phaethon_DUST}
\end{figure}

Since 2016, the worldwide FRIPON camera network \citep{Colas_2020A&A...644A..53C} observes on a nightly basis
incoming bolides, allowing the determination of their orbits and entry velocity, among other things. 
We considered detections of bolides recorded since 2016 by at least 3 stations.
According to an ablation model,
these events correspond to bolides with initial mass $m_{\rm i} \ge 10\,{\rm g}$.
We removed outliers, with
uncertain values of mass, $\sigma_{m_i}/m_{\rm i} > 3$, or
unrealistic values of the ablation parameter,
$B < 10^{-11}\,{\rm m}^2\,{\rm J}^{-1}$.
From these 602 events,
a histogram of entry speeds was constructed
(Fig.~\ref{impvel_nooutliers}).
It is skewed towards low speeds
(the minimum is the escape speed from Earth, $11.2\,{\rm km}\,{\rm s}^{-1}$).
At high speeds, there is no 'cometary source';
all massive meteoroids seem to be asteroidal.
The only prominent shower is Geminids,
related to asteroid (3200) Phaethon
(\citealt{Whipple_1983IAUC.3881....1W,Hanus_2018A&A...620L...8H,Battams_2022ApJ...936...81B};
see Fig.~\ref{Phaethon_DUST}).
This NEO source was also added in Tab.~\ref{tab:1m}.

A comparison with our model is not straightforward.
Meteoroids from asteroid families have a range of speeds relative to Earth,
typically between $20$ to $30\,{\rm km}\,{\rm s}^{-1}$
(cf.~Figs.~\ref{impvel_nooutliers}, \ref{impvel}).
Nevertheless, entry speeds must be properly weighted,
since we compare it to observations of bodies entering the Earth's atmosphere.
Thus for each family~$j$,
we computed a histogram $\d N_j(v,v+\d v)$ of $v_i$ values
weighted by individual collisional probabilities $p_i$
(not the mean $p$, from Tab.~\ref{tabc2}).
Then we computed a sum of histograms,
weighted by the corresponding meteoroid fluxes (from Tab.~\ref{tab:1m}):
\begin{equation}
\d N(v, v+\d v) = \sum_j\Phi_j\,\d N_j(v, v+\d v)\,.
\end{equation}
The resulting histogram is surprisingly similar to the observations
(Fig.~\ref{impvel_nooutliers}).
There are only minor differences,
e.g., the synthetic histogram exhibits a bit narrower peak
and a bit higher tail.
Partly, this might be attributed to an observational bias,
e.g., due to an instability of the ablation model
(cf. outliers).
Moreover, it is known that the mass determination from fireball data
is especially difficult for higher speeds, or lower decelerations.
Nevertheless, we consider our model ('METEOMOD') to be
in agreement with bright bolides.

\begin{figure}
\centering
\includegraphics[width=9cm]{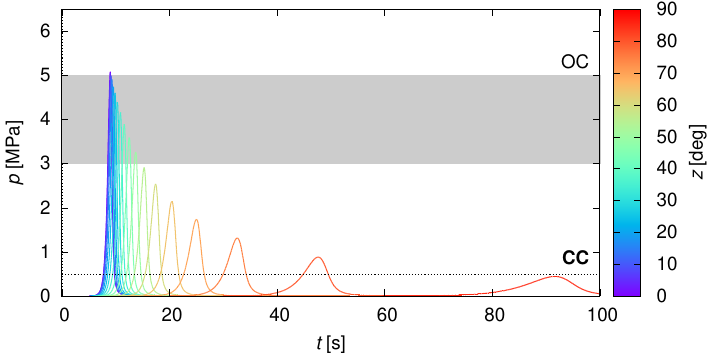}
\caption{
Dynamical pressure~$p$ on a meteoroid during its atmospheric entry,
depending on the zenith angle~$z$ (in degrees).
A basic ablation model was used, with the following parameters:
initial mass $m = 1000\,{\rm kg}$,
entry speed $v = 20\,{\rm km}\,{\rm s}^{-1}$,
density $\rho = 1700\,{\rm kg}\,{\rm m}^{-3}$,
ablation efficiency $\eta = 0.02$,
specific heat $l = 10^6\,{\rm J}\,{\rm kg}^{-1}$,
shape factor $A = 1.22$ (i.e., sphere),
friction coefficient $C = 0.5$.
The resulting
terminal mass $m' = 0.3\,{\rm kg}$.
The strength of
OC meteoroids (3 to 5\,MPa; \citealt{Borovicka_2019M&PS...54.1024B}; gray area)
is comparable to $p$.
CC meteoroids (${\approx}\,0.5\,{\rm MPa}$; dotted line)
are substantially weaker than $p$.
If fragmentation occurs, the ablation process becomes much more efficient;
CC meteorites only survive for trajectories with $z \gtrsim 85^\circ$.
}
\label{bolid3_zenith}
\end{figure}


\q{Are CC meteorite fall statistics driven by atmospheric, thermal or collisional fragmentation?}

\subsection{Atmospheric fragmentation}\label{atmospheric}

We find that CI-like and CM-like meteoroids should be nearly
as abundant as ordinary chondrite (OC) like meteoroids at the top of the atmosphere. 
Yet, OCs are 40 times more abundant than CI and CM chondrites among falls.
Hereafter, we explore the effects of various mechanisms that may explain
this discrepancy between our model and observations.

Atmospheric entry has long been proposed as the main bias to explain the paucity of CI and CM falls
with respect to OC falls, for instance, a consequence of their difference in terms of strength.
Fragmentation of CC meteoroids occurs at a low dynamical pressure
($10^{-2}\,{\rm MPa}$; \citealt{Borovicka_2019M&PS...54.1024B}).
While this material ('original matrix') cannot survive,
fragments logically have a smaller volume,
a lower number of flaws,
hence a higher strength,
comparable to a typical peak dynamical pressure during an atmospheric entry
($3$ to $5\,{\rm MPa}$; \citealt{Borovicka_2019M&PS...54.1024B}).

Of course, the strength of meteorites is generally different
\citep{Pohl_2020M&PS...55..962P}.
If considering tensile strength,
which is typically 10 times lower than the compressive strength,
the mean values of meteorites classes are as follows:
CI      2,
CV      3.5,
CM      7,
LL     10,
H      15,
L      30, and
irons 400\,MPa.
Fragments surviving the atmospheric entry do have strengths
higher than the peak dynamical pressure.
The key question, however,
is not "at what pressure?",
but "into how many pieces?", and
eventually "how to find them?".
This determines whether a bolide will result in a fall.

Using a basic ablation model
\citep{Pecina_1983BAICz..34..102P,Broz_2013}
one can verify that the peak dynamical pressure~$p$
depends on the zenith angle~$z$%
\footnote{i.e., not the orbital inclination~$i$;
even high-$i$ orbits enter the atmosphere at low-$z$,
and {\em vice versa\/}}
(Fig.~\ref{bolid3_zenith}).
If CC meteoroids have approximately 10 times lower strength than OC
(as indicated above),
they should only survive for trajectories with $z \gtrsim 85^\circ$.
This amounts to a factor of 10 in the adjacent solid angle,
$\omega = 2\pi(1-\cos z)$.

An additional bias relates to entry speeds,
which differ substantially for individual sources (Fig.~\ref{impvel}).
Bodies with excessive speeds (${\gtrsim}\,30\,{\rm km}\,{\rm s}^{-1}$) are eliminated by ablation.
On the other hand, a low-speed part ${\lesssim}\,20\,{\rm km}\,{\rm s}^{-1}$
of the distributions is strongly preferred.
Surviving CM chondrites are thus likely related to low-speed Veritas meteoroids.
Relatively,
Polana (CI) is preferred over Euphrosyne,
etc.
This certainly contributes to the fact that OCs prevail over CCs,
especially when Koronis (H) and Massalia (L) produce a lot of meteoroids
with low speeds.


\subsection{Thermal cracking}\label{cracking}

Other mechanisms such as
thermal cracking \citep{Granvik_2016Natur.530..303G}, or
water desorption (Granvik, pers. comm.)
might be at play.
In-situ observations of the asteroid Bennu
\citep{Delbo_2022NatGe..15..453D}
revealed angular boulders and oriented cracks,
induced by the diurnal cycle.
This is a direct evidence of thermal cracking on metre scales.

On metre-sized bodies, a thermal stress of the order of MPa develops
\citep{Capek_2010A&A...519A..75C},
especially if the heliocentric distance is below 0.3\,au,
or the rotation frequency is below 1\,Hz,
or the obliquity $\gamma \sim 90^\circ$,
since then one hemisphere points towards the Sun for a long time.
Evolution of the spin due to the YORP effect
\citep{Rubincam_2000Icar..148....2R,Capek_2004Icar..172..526C}
might temporarily lead to conditions favouring disruptions.
If carbonaceous bodies have strengths less than a~MPa
\citep{Borovicka_2019M&PS...54.1024B},
this should be commonplace
and only strong bodies should enter the atmosphere.
However, meteoroids like Geminids still exhibit low strengths
\citep{Henych_2024A&A...683A.229H},
which is in contradiction with dispersal of weak bodies.
Subsequently, these bodies might be disrupted by tides
\citep{Walsh_2008Icar..193..553W,Granvik_2024ApJ...960L...9G},
if they encounter the Earth.

Regardless of CI and CM meteoroids being disrupted prior to or during atmospheric entry,
they contribute to the abundance of interplanetary dust particles (IDPs),
in particular, chondritic-smooth (CS), hydrous, asteroidal IDPs,
which comprise about 50\% of all IDPs at ${<}50\,\mu{\rm m}$
\citep{Bradley_2007}.

Micrometeorites recovered in Antarctica
\citep{Genge_2020P&SS..18704900G}
also indicate
a comparable amount of C- vs. S-type asteroidal materials (60 vs. 40\%)
at sizes ${>}300\,\mu{\rm m}$.
Chondritic-porous (CP), anhydrous, cometary IDPs
are already negligible at these sizes.
Given that the micrometeoroid flux is directly proportional to the metre-size flux,
since the SFDs are in a collisional equilibrium,
our model (Tab.~\ref{tab:1m}) appears fully in agreement with this observation.


\subsection{Collisional fragmentation}\label{fragmentation}

Alternatively, if C-types are less strong than S-types,
their collisional evolution and their current SFD might be systematically different.
If sub-km C-type bodies are more easily disrupted than S-type bodies,
the local slope of the SFD should be more shallow.
Consequently, the population of metre-sized bodies should decrease,
as well as that of $100\,\mu{\rm m}$-sized dust grains.

However, this is in contradiction with the observations of dust bands,
the $9.3^\circ$ band being the most prominent one (see above). 
It follows that a collisional cascade cannot have destroyed all sub-km C-type bodies;
otherwise we would not observe this prominent dust band.

In fact, C-type asteroids represent the vast majority of sub-km bodies in the asteroid belt
\citep{Marsset_2022AJ....163..165M}.
It does not make sense to decrease the strength of the Veritas family,
without decreasing the strength of 'everything' else.
According to our tests, assuming a 10 times lower strength of all sub-km bodies
implies similar SFDs (relatively to each other; Veritas w.r.t. main belt) at sub-km sizes.
Consequently, collisional fragmentation on its own cannot explain
the paucity of CC meteorites.


\begin{figure*}
\centering
\begin{tabular}{c@{\kern-.3cm}c}
\includegraphics[height=8.0cm]{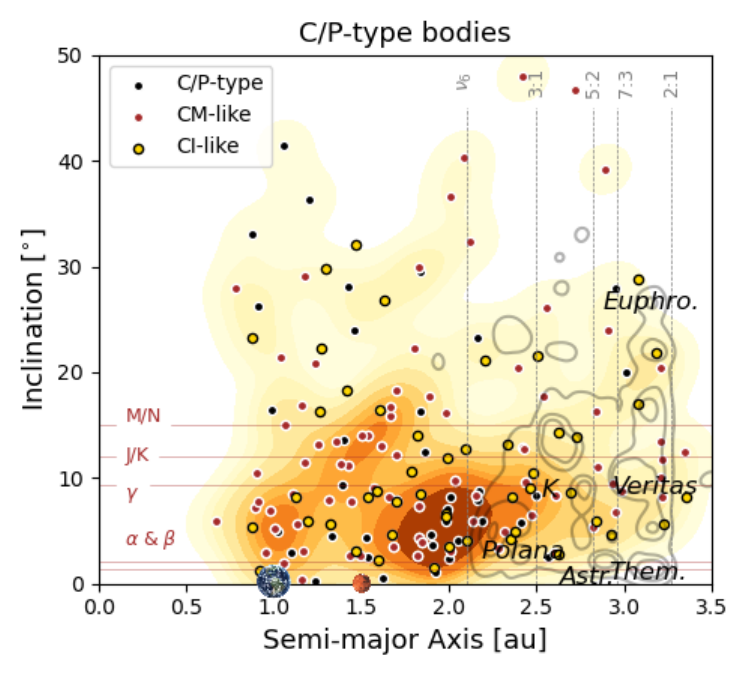} &
\includegraphics[height=8.0cm]{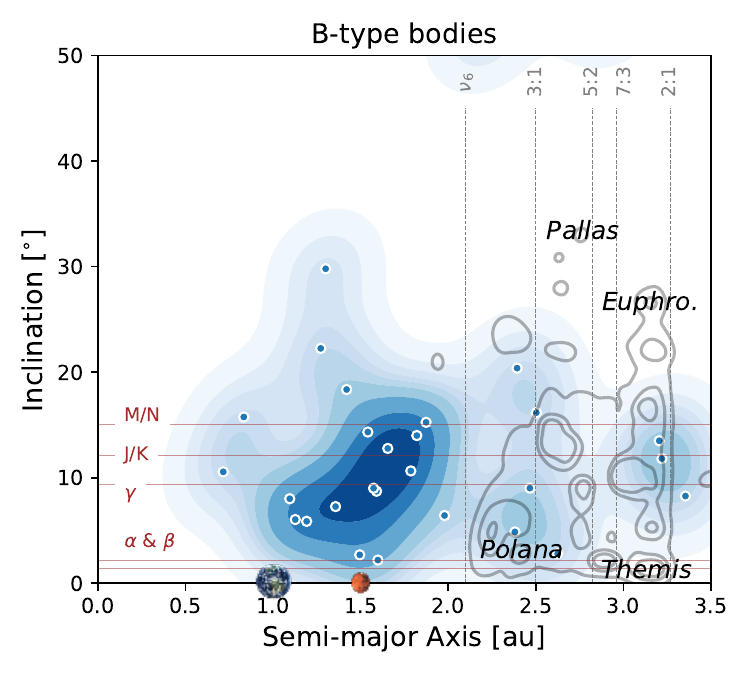} \\[-0.3cm]
\includegraphics[height=8.3cm]{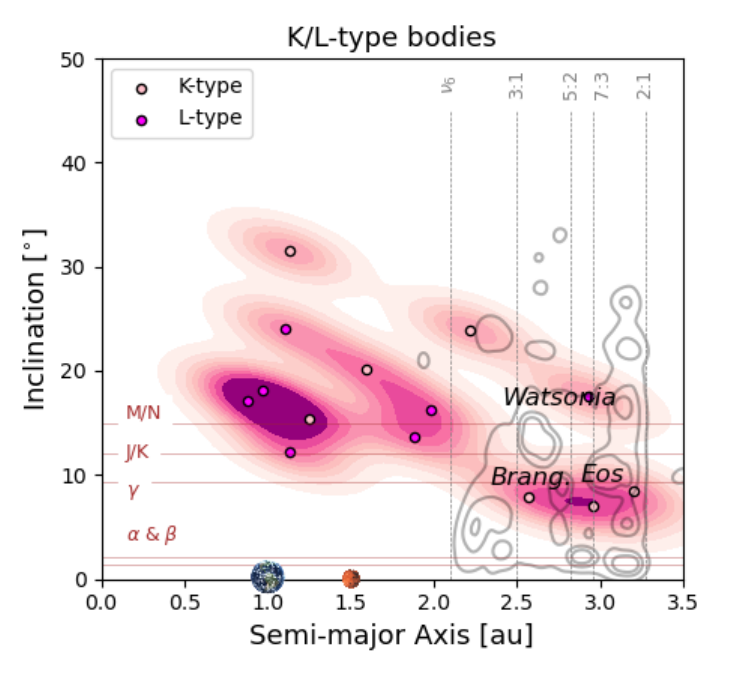} &
\includegraphics[height=8.3cm]{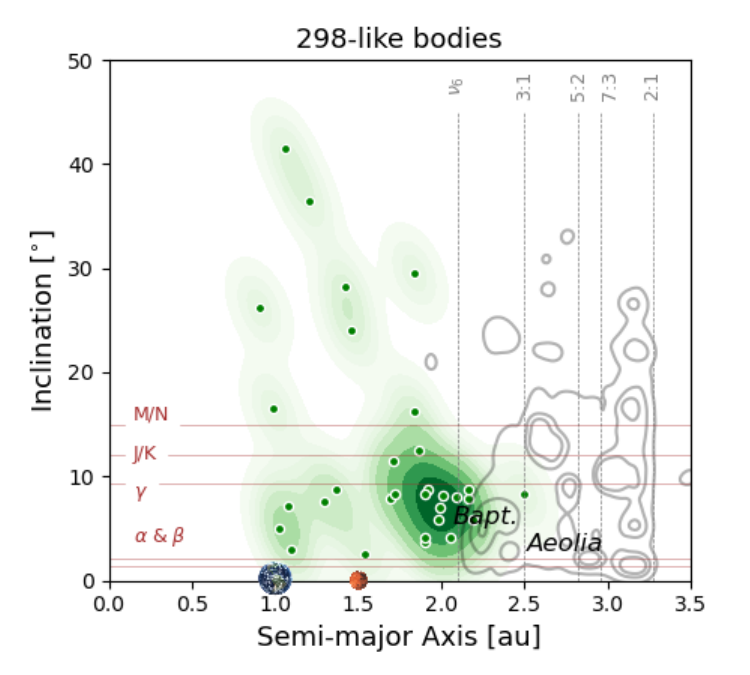} \\[-0.3cm]
\includegraphics[height=8.0cm]{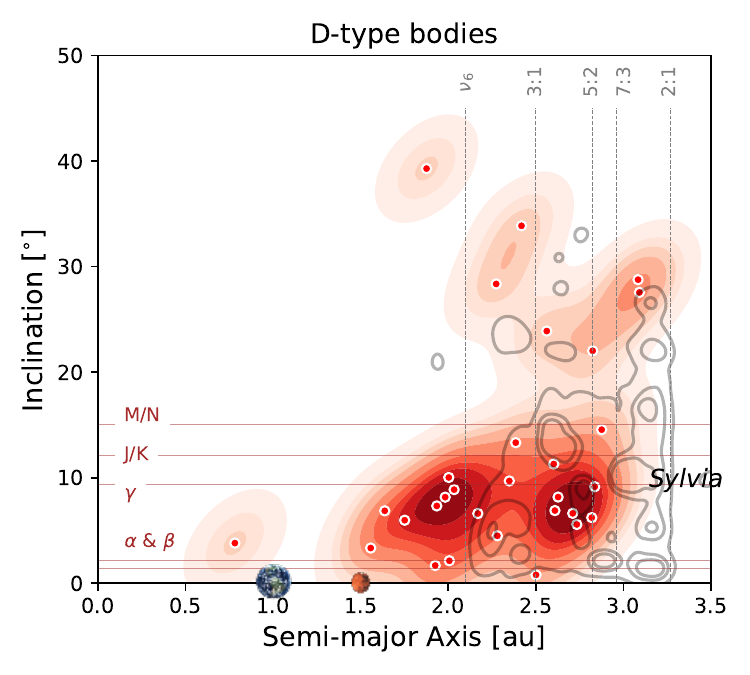} &
\includegraphics[height=8.0cm]{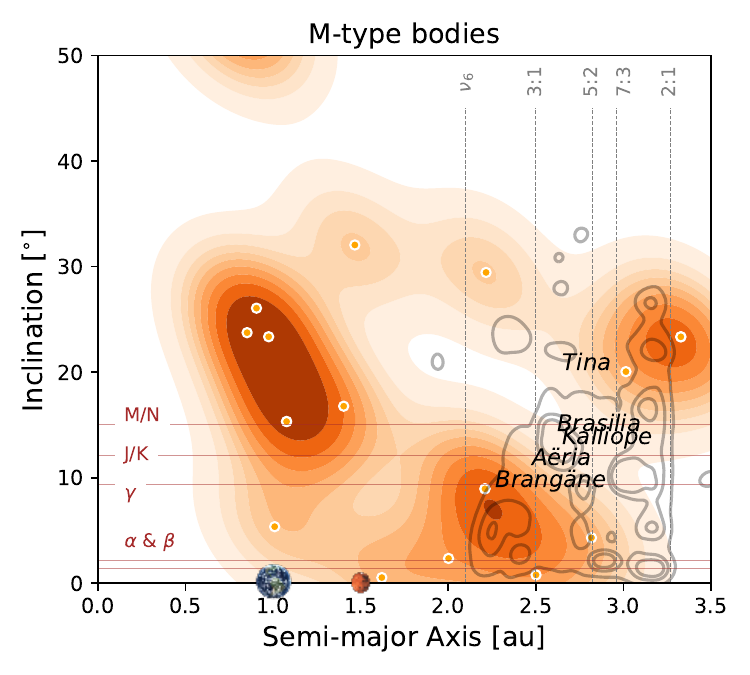} \\[-0.3cm]
\end{tabular}
\caption{
Observed orbital distributions of C/P-, B-, K/L-, Baptistina-like, D- and M-type NEOs.
The semimajor axis $a$ versus inclination~$i$ plot
is a good indicator of the source,
because NEO orbits are initially close to their source,
before they disperse due to planetary perturbations.
For reference, the locations of selected asteroid families are indicated
(`K' corresponds to K\"onig).
}
\label{NEO_orbits_compo_C}
\end{figure*}

\q{Is our synthetic population of kilometre-sized bodies compatible with spectroscopic observations?}

\subsection{Taxonomy of NEO orbits}\label{taxonomy}

A comparison of our synthetic population of kilometre-sized CI- and CM-like NEOs
with the observed orbital distribution is important,
as it is a direct test of our transport model.

The orbits of CI- and CM-like NEOs (Fig.~\ref{NEO_orbits_compo_C})
exhibit an 'extreme' spread of inclinations up to $40^\circ$,
and semimajor axes up to $3.2\,{\rm au}$.
The distribution seems bimodal,
with a low-inclination component ($i \lesssim 5^\circ$);
and a more scattered high-inclination component ($i \gtrsim 20^\circ$).
According to our model (Tab.~\ref{tab:1km}),
most CI- and CM-like NEOs should originate from the Polana ($3.3^\circ$)
and Euphrosyne ($26.5^\circ$) families;
CM-like NEOs being significantly less abundant than CI-like ones. 
It follows that our model is in perfect agreement with observations. 

Regarding B-type NEOs,
a few of them possess a high albedo ($p_V\geq 0.1$) and do have high inclinations;
these bodies 'certainly' originate from Pallas.
However, the orbits of the low albedo B-type population are similar to those of CI- and CM-like NEOs,
albeit without the low-inclination component from Polana ($3.3^\circ$).
These bodies must originate from other CI-like families,
{\em ergo\/}, B-type being compositionally similar to C-types
(e.g., \citealt{Vernazza_2015ApJ...806..204V,Marsset_2016A&A...586A..15M}).

CO/CV/CK-like NEOs (K-types), have medium- to high-inclination orbits.
The only sizable family at the correct inclination is Eos ($9.9^\circ$).

Bodies having a characteristic shallow 1.00-micron band,
which are relatively rare,
preferentially have low-inclination orbits,
which cluster close to 2\,au (Fig.~\ref{NEO_orbits_compo_C}).
According to our model, they must originate from the Baptistina family ($6.0^\circ$).

CH- and CB-like bodies are metal rich
and they might appear as M-types.
About 10 of them should be observable from Brasilia ($15^\circ$).
This seems to be compatible with the observed percentage of M-type NEOs
\citep{Marsset_2022AJ....163..165M}.



\subsection{Common origin of Ryugu and Bennu}\label{ryugu}

The JAXA Hayabusa~2 and NASA OSIRIS-Rex sample return missions have targeted two C-type NEOs,
namely Ryugu and Bennu, respectively \citep{Watanabe_2019Sci...364..268W,Lauretta_2019Natur.568...55L}.
Ongoing analysis of the object' samples in terrestrial laboratories have revealed that
CI chondrites are their closest analogs \citep{Yokoyama_2023Sci...379.7850Y}.

\begin{figure}
\centering
\includegraphics[width=9cm]{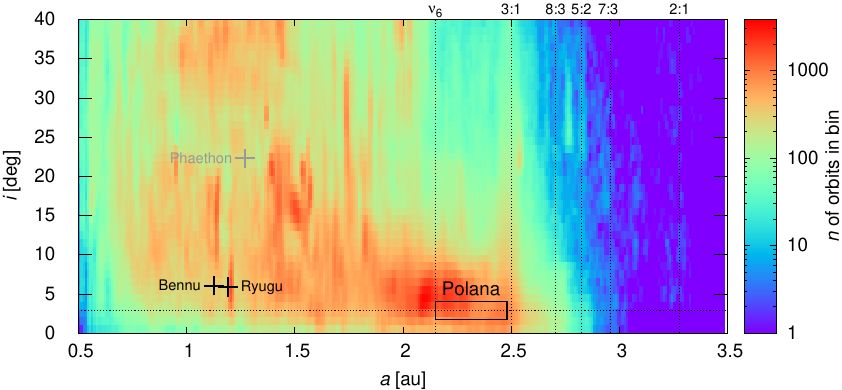}
\includegraphics[width=9cm]{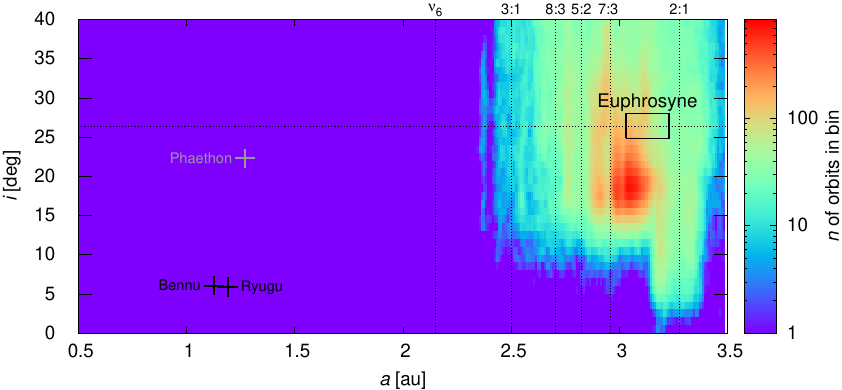}
\caption{
Same as Fig.~\ref{ai_CM} for the Polana (top) and Euphrosyne (bottom) families.
Colours correspond to the numbers of orbits in bins.
The orbits of both Ryugu and Bennu (crosses) are compatible with Polana.
Even (3200) Phaethon on a high-eccentricity orbit
($e = 0.89$)
is compatible with Polana.
}
\label{polana-2_map}
\end{figure}

Applying our model to Ryugu and Bennu,
it appears that there is a ${\gtrsim}\,90\%$ probability that both objects
originate from the CI-like Polana family
(Fig.~\ref{polana-2_map}).
Even without a-priori knowledge of CI vs. CM classification,
this is the only option,
since Euphrosyne (CI) is too distant
and Adeona (CM) is too inclined.
It follows that these NEOs sample the same parent body.
Our results are fully consistent with previous findings of
\cite{Bottke_2015Icar..247..191B,Bottke_2020AJ....160...14B}.

In case a future sample return mission aims to probe the diversity among CI-like asteroids,
we suggest to target high-inclination CI-like NEOs originating from the Euphrosyne family.
This will of course have a cost, due to a higher $\Delta v$.

\section{Conclusions}

In this work, we constructed a complex model linking
NEOs and meteorites with their sources in the asteroid belt.
Our approach is very different from previous works
(e.g., \citealt{Granvik_2018Icar..311..271G,Granvik_2018Icar..312..181G,Nesvorny_2023AJ....166...55N})
because we use a-priori knowledge of families, their taxonomy and SFDs.
Our sources correspond to individual families
(not to resonances),
because otherwise the 'critical' information
about the original location ($a$, $i$) is lost.
The taxonomy is important, because the orbital distributions
for various classes do differ
(cf. Figs.~\ref{ai_CM}, \ref{NEO_orbits_compo_C}).
The SFD is equally important, because some sources dominate the others
and they must serve as appropriate weights.

For kilometre-sized NEOs, our model indicates that most of them
originate from just two families:
Polana and Euphrosyne
(Fig.~\ref{pie}).
Their reflectance spectra are compatible with CI chondrites.
Other primitive classes P and D do not have sufficient sources in the main belt,
but they rather originate as comets \citep{DeMeo_2008Icar..194..436D}.

On the other hand, the flux of metre-sized carbonaceous chondrites
is dominated by just one family:
Veritas
(cf. Fig.~\ref{pie}).
Its reflectance spectrum is compatible with CM chondrites.
This certainly contributes to the prevalence of CM over CI chondrites.
Even though the observed ratio ${\rm CM}/({\rm CI{+}\hbox{C-ungr.}}) \simeq 2$ is higher than our prediction 1.6,
it is still within uncertainties
on the side of our model (i.e., a factor of 2) and
on the side of observations.

Regarding the total flux of meteorites,
estimated to $740\times 10^{-9}\,{\rm km}^{-2}\,{\rm y}^{-1}$
\citep{Nesvorny_2023AJ....166...55N}
about half of it comes from S-types
\citep{Broz_2023},
the other half from C-types
(this work).
No other sources (e.g., comets) are needed to explain it.

However, the overall meteorite falls statistics
\citep{Gattacceca_2022M&PS...57.2102G}
shows different proportions of OC and CC meteorites:
OC 80\%,
CC 4.4\%.
In particular, the observed ratio
${\rm OC}/({\rm CM{+}CI{+}\hbox{C-ungr.}}) \simeq 30$
is too high compared to our prediction 1.3.
This can only be explained by a difference in fragmentation
in the upper atmosphere
\citep{Borovicka_2019M&PS...54.1024B}
together with ablation at relatively high entry speeds
(${\gtrsim}\,30\,{\rm km}\,{\rm s}^{-1}$; cf.~Fig.~\ref{impvel}),
or by thermal fragmentation.
Furthermore, there are substantial differences within carbonaceous classes;
CO/CV/CK must be much stronger than CM, CI, C-ungr.
On the other hand, our model does not indicate dramatic differences
between CM, CI, C-ungr. classes in terms of fragmentation,
since their proportions is largely a natural outcome of
the sources, locations and transport.

\vskip\baselineskip

\begin{acknowledgements}
We thank an anonymous referee for a number of constructive comments.
We thank Chrysa Avdellidou for providing spectroscopic observations of K\"onig.
This work has been supported by the Czech Science Foundation through grant
21-11058S (M.~Bro\v z).
The Chimera HPC cluster at Charles University (MFF UK)
was used for computations.
FRIPON\footnote{\url{https://www.fripon.org}} is funded by the ANR grant N.13-BS05-0009-03, carried by
the Paris Observatory,
Mus\'eum National d'Histoire Naturelle,
Paris-Saclay University and
Pyth\'eas institute
(LAM,
CEREGE).
Vigie-Ciel is part of the
65 Millions d'Observateurs project, carried by
the Mus\'eum National d'Histoire Naturelle,
funded by the French Investissements d'Avenir program.
FRIPON data are hosted and processed at Institut Pyth\'eas IT department (SIP).
A mirror is also hosted at IMCCE
(Institut de M\'ecanique Céleste et de Calcul des \'Eph\'em\'erides).
\end{acknowledgements}


\bibliographystyle{aa}
\bibliography{references}


\clearpage
\appendix

\section{Enstatite chondrites and achondrites}

When our model is consistently applied to E-type families
(Nysa, Hungaria),
it produces NEO and meteoroid populations
which are compatible with the observed abundance of E-type NEOs
as well as enstatite chondrites and achondrites,
also known as aubrites
(see Tabs.~\ref{tab:1km}, \ref{tab:1m}).
It is important to know that the spread of albedos in the Nysa family is huge
(from 0.1 to 0.6).
Approximately half of bodies have high albedos,
$p_{\rm V} > 0.3$,
i.e., characteristic of aubrites.
At least a part of the other half
might appear as EH- or EL-chondrite-like.%
\footnote{Additionally, a sizable M-type within the family is (135) Hertha.}

\begin{figure}[h]
\centering
\includegraphics[width=9cm]{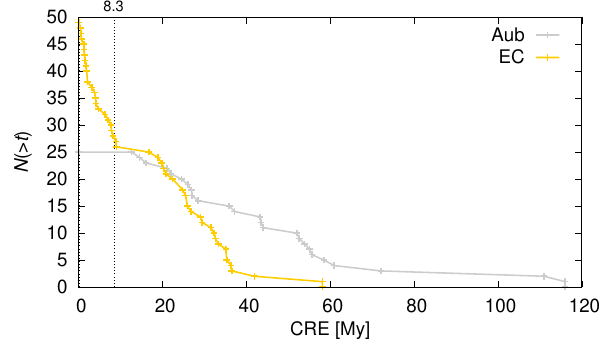}
\caption{
Same as Fig.~\ref{cre_CM_hist2} for enstatite chondrites (EC) and achondrites (Aub).
}
\label{cre_Aub_hist2}
\end{figure}

For EL, EH chondrites,
the CRE ages reach up to 40\,My
(Fig.~\ref{cre_Aub_hist2}; \citealt{Patzer_2001M&PS...36..947P}),
corresponding to a collisional cascade.
This is compatible with Nysa (age ${\gg}\,40\,{\rm My}$, presumably).
The second onset coincides with the age of Veritas.

For aubrites \citep{Keil_2010ChEG...70..295K},
the CRE ages reach up to ${\approx}\,110\,{\rm My}$,
but an onset is observed at 60\,My
(Fig.~\ref{cre_Aub_hist2}; \citealt{Lorenzetti_2003GeCoA..67..557L}).
This is fully compatible with the Hungaria family,
which is more-or-less isolated from the asteroid belt,
hence the collisional lifetime is long
\citep{Cuk_2014Icar..239..154C}.

\begin{figure}
\centering
\includegraphics[height=8.0cm]{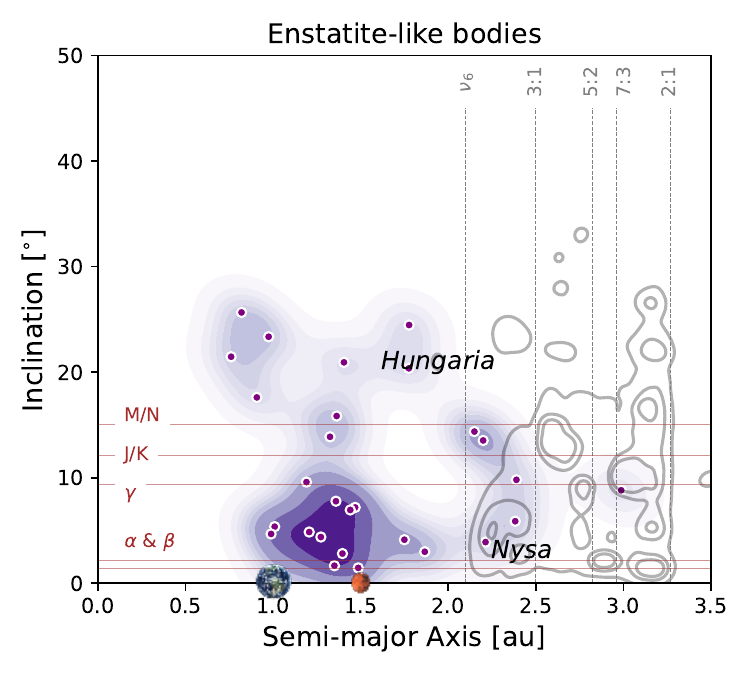}
\caption{
Same as Fig.~\ref{NEO_orbits_compo_C} for E-type NEOs.
}
\label{NEO_orbits_compo_Enstatite}
\end{figure}

E-type NEOs,
or more specifically Xe- or Xn-types,
often exhibit a distinct spectral feature,
which is characteristic of (44)~Nysa.
Those which are located at low inclinations (${\approx}\,5^\circ$)
'certainly' originate from the Nysa family
(see Fig.~\ref{NEO_orbits_compo_Enstatite}).
The remaining population at high inclinations
is easily explained by the Hungaria family.

\section{Supplementary figures}

Apart from figures referred to in the main text,
we show identified asteroid families in Fig.~\ref{aei2}
and histograms of entry speeds for individual families in Fig.~\ref{impvel}.

\begin{figure}
\centering
\includegraphics[width=8cm]{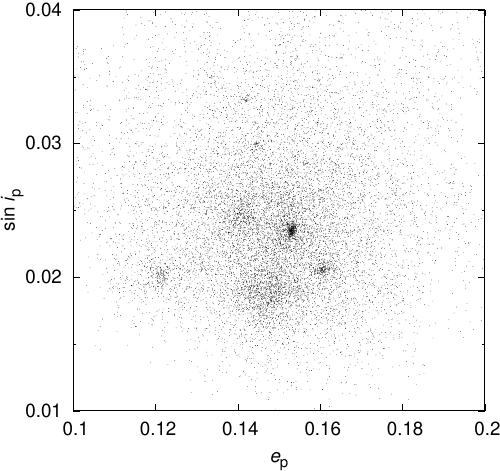}
\caption{
Structure of the Themis family
in the proper eccentricity~$e_{\rm p}$ versus proper inclination~$\sin i_{\rm p}$.
Every asteroid is represented by a dot.
Concentrations of dots indicate up to 7 sub-families.
The most prominent one is related to (656) Beagle.
The least prominent ones are very dispersed in the proper semimajor axis~$a_{\rm p}$,
due to the Yarkovsky effect.
Nevertheless, it is a confirmation of an ongoing collisional cascade.
}
\label{24_Themis_ei}
\end{figure}

\begin{figure}
\centering
\begin{tabular}{c}
\kern0.5cm Vesta (HED), 1100\,My \\
\includegraphics[width=8.5cm]{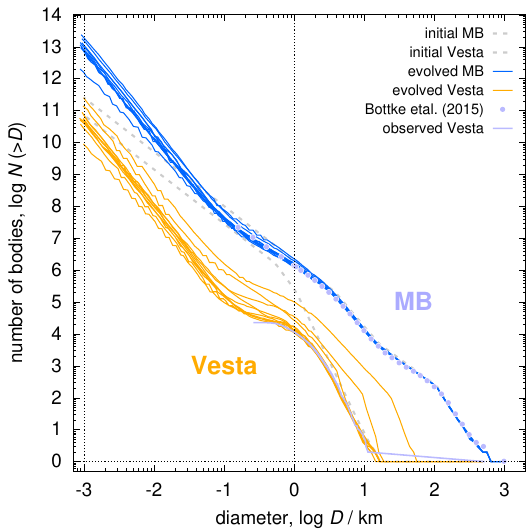} \\
\end{tabular}
\caption{
SFDs of the main belt (\textcolor{blue}{blue}) and the Vesta family (\textcolor{yellow}{yellow}),
used for calibration of our collisional model.
The observed SFDs (\citealt{Bottke_2015Icar..247..191B,Harris_2015aste.book..835H} and this work)
are in agreement with the synthetic SFDs,
where the observational bias is negligible ($D\gtrsim 1\,{\rm km}$).
At sub-km sizes, the SFD of the Vesta family is shallow,
similarly as the main belt, due to a collisional cascade.
Additionally, the model is constrained
by the observed SFD of NEOs, and
cratering record at (4) Vesta.
}
\label{sfd_1100}
\end{figure}

\begin{figure}
\centering
\begin{tabular}{c}
\kern0.5cm NEOs, steady-state \\
\includegraphics[width=8.5cm]{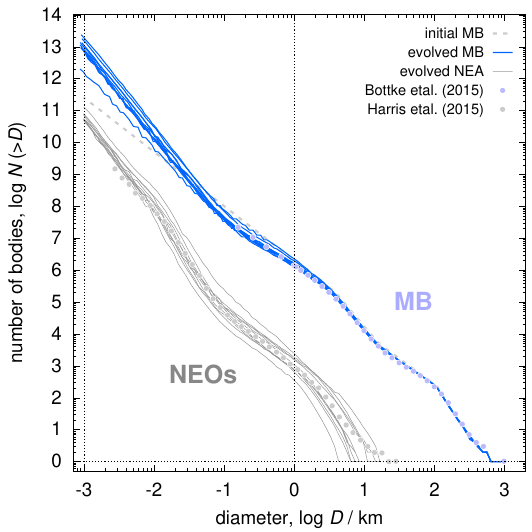} \\
\end{tabular}
\caption{
Same as Fig.~\ref{sfd_1100} for the NEOs.
}
\label{sfd_1100_NEA}
\end{figure}

\begin{figure}
\centering
\centering
\begin{tabular}{c}
\kern1cm 3.2\,My \\
\includegraphics[width=8.5cm]{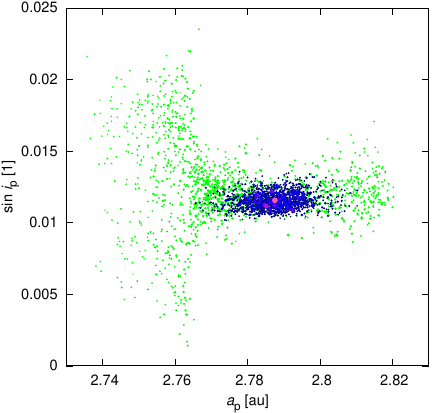} \\
\kern1cm 150\,My \\
\includegraphics[width=8.5cm]{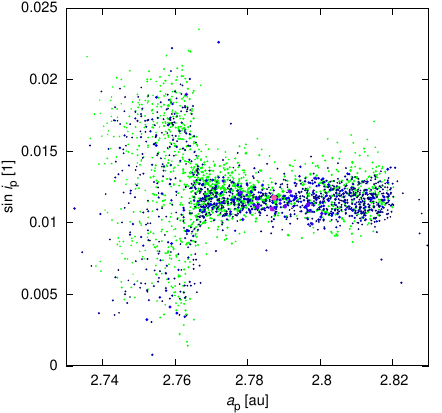} \\
\end{tabular}
\caption{
Astrid family is more than 150\,My old according to orbital evolution.
The observed family (\textcolor{green}{green}) has a substantial spread of inclinations~$\sin i_{\rm p}$
below the semimajor axis~$a_{\rm p} = 2.763\,{\rm au}$.
The synthetic family (\textcolor{blue}{blue}; top) is initially compact,
in accord with a low escape speed ($v_{\rm esc} \doteq 25\,{\rm m}\,{\rm s}^{-1}$).
In the course of evolution,
it is modified by the $g-g_{\rm C}$ secular resonance with Ceres
\citep{Novakovic_2016IAUS..318...46N}.
At the time ${\gtrsim}150\,{\rm My}$ (bottom),
it corresponds to the observed distribution.
Such an old age indicates that the SFD must already be shallow.
}
\label{astrid-1_CERES_0003.200}
\end{figure}

\begin{figure}
\centering
\centering
\begin{tabular}{c}
\kern1cm 3.2\,My \\
\includegraphics[width=8.5cm]{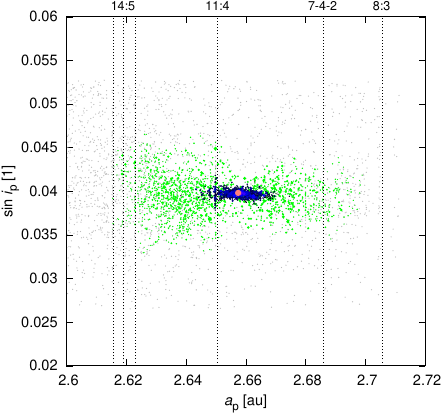} \\
\kern1cm 100\,My \\
\includegraphics[width=8.5cm]{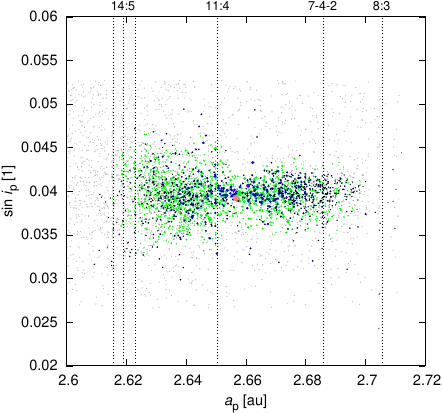} \\
\end{tabular}
\caption{
Same as Fig.~\ref{astrid-1_CERES_0003.200} for the Misa family.
Its orbital distribution indicates an age up to 100\,My,
due to interactions with the 11:4 mean-motion resonance with Jupiter.
}
\label{misa-2_ai_0003.200}
\end{figure}

\begin{figure*}
\centering
\centering
\begin{tabular}{c@{}c@{}c}
\kern1cm observed &
\kern1cm initial conditions &
\kern1cm 60\,My, regolith, YORP \\
\includegraphics[width=6.0cm]{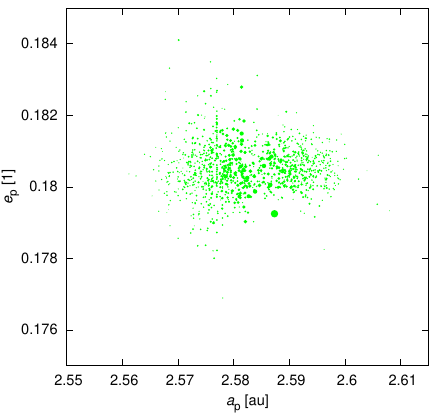} &
\includegraphics[width=6.0cm]{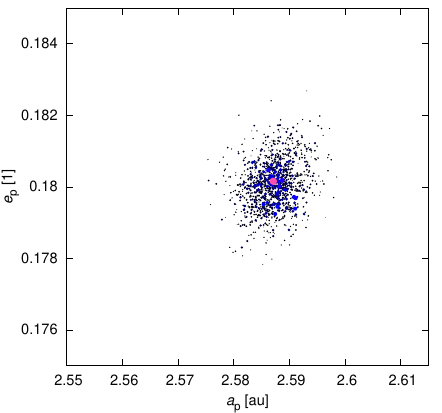} &
\includegraphics[width=6.0cm]{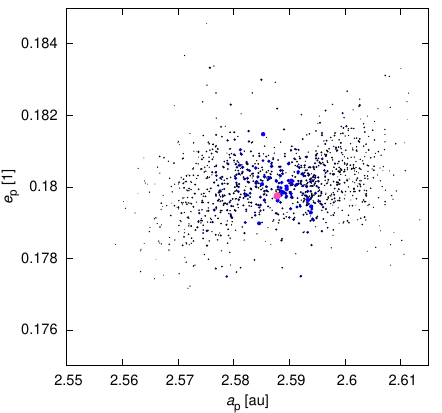} \\
\kern1cm 1000\,My, no regolith, YORP &
\kern1cm 200\,My, no regolith, no YORP &
\kern1cm close encounter \\
\includegraphics[width=6.0cm]{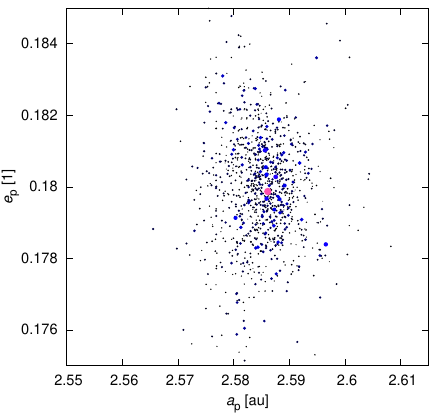} &
\includegraphics[width=6.0cm]{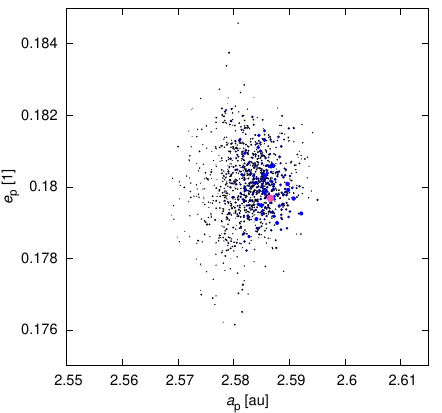} &
\includegraphics[width=6.0cm]{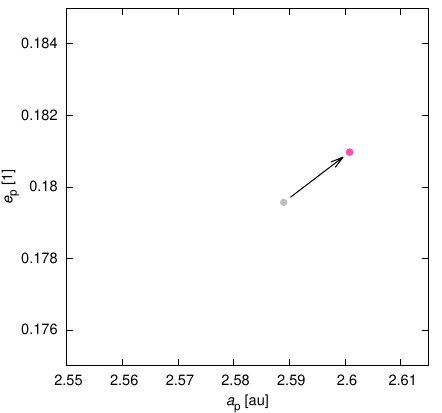} \\
\end{tabular}
\caption{
Brang\"ane family asteroids must have surfaces covered with regolith.
The observed family (\textcolor{green}{green})
in the proper semimajor axis~$a_{\rm p}$ versus the proper eccentricity~$e_{\rm p}$
exhibits a depletion of small bodies in the centre.
The synthetic family (\textcolor{blue}{blue}) was initially more compact.
After about 60\,My of orbital evolution,
it exhibits the same depletion due to the YORP effect
\citep{Vokrouhlicky_2006Icar..182..118V}.
The value of thermal conductivity was assumed low,
$K = 10^{-3}\,{\rm W}\,{\rm m}^{-1}\,{\rm K}^{-1}$,
corresponding to regolith.
An alternative model,
assuming iron composition and no regolith
($K = 40\,{\rm W}\,{\rm m}^{-1}\,{\rm K}^{-1}$),
was excluded,
because small bodies were not depleted, but spread due to chaotic diffusion in~$e$.
A suppression of the YORP effect
was also excluded,
because small bodies drifted towards small~$a$,
due to the seasonal variant of the Yarkovsky drift.
The offset of (606) Brang\"ane with respect to other asteroids
is most likely due to a random close encounter with (4) Vesta.
}
\label{brangane-1_VESTA_ae_obs}
\end{figure*}

\begin{figure*}
\centering
\begin{tabular}{c@{\kern0.1cm}c@{\kern0.1cm}c}
\kern1cm CM-chondrite &
\kern1cm CI-chondrite, steep &
\kern1cm CI-chondrite, shallow \\
\includegraphics[width=5.9cm]{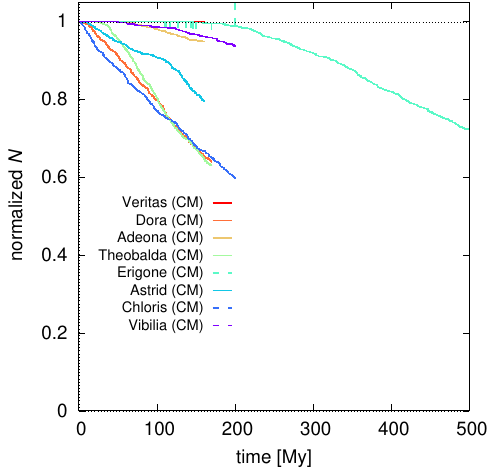} &
\includegraphics[width=5.9cm]{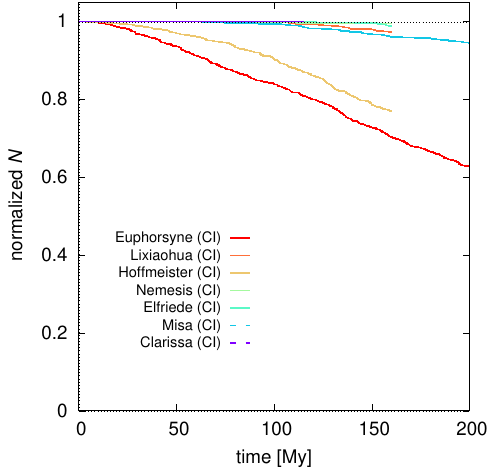} &
\includegraphics[width=5.9cm]{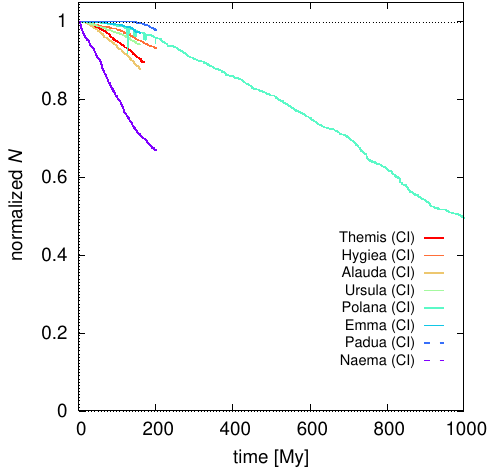} \\
\kern1cm CO/CV/CK-type &
\kern1cm M-type &
\kern1cm unknown \\
\includegraphics[width=5.9cm]{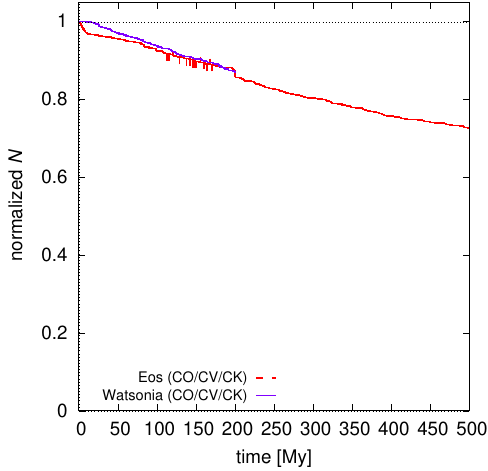} &
\includegraphics[width=5.9cm]{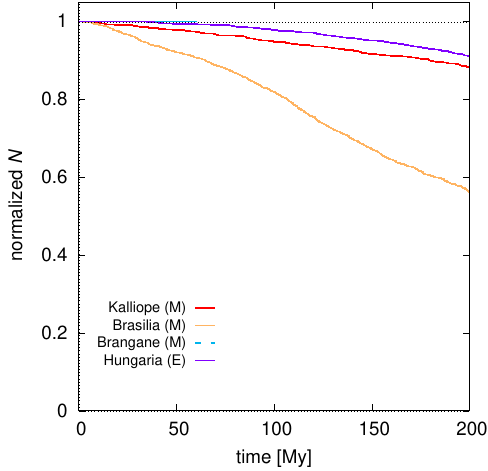} &
\includegraphics[width=5.9cm]{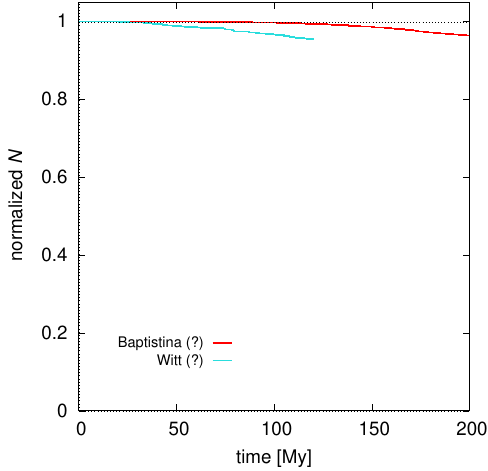} \\
\end{tabular}
\caption{
Dynamical decay of individual populations in the main belt,
computed for kilometre-sized bodies.
Our dynamical model included
perturbations by 11 massive bodies (Sun, planets, Ceres, Vesta),
mean-motion, secular, or three-body resonances,
the Yarkovsky effect,
the YORP effect,
or collisonal reorientations.
From the dependence of (normalized) number of bodies $N$ on time~$t$,
the decay time scales $\tau_{\rm mb}$ were estimated
(cf.~Tab.~\ref{tab:1km}).
Some of the dependencies do not decay at all,
because the source population is far from any resonances.
}
\label{nbody_observed_decay_CM}
\end{figure*}

\begin{figure*}
\centering
\begin{tabular}{c@{\kern0.1cm}c@{\kern0.1cm}c}
\kern1cm CM-chondrite &
\kern1cm CI-chondrite, steep &
\kern1cm CI-chondrite, shallow \\
\includegraphics[width=5.9cm]{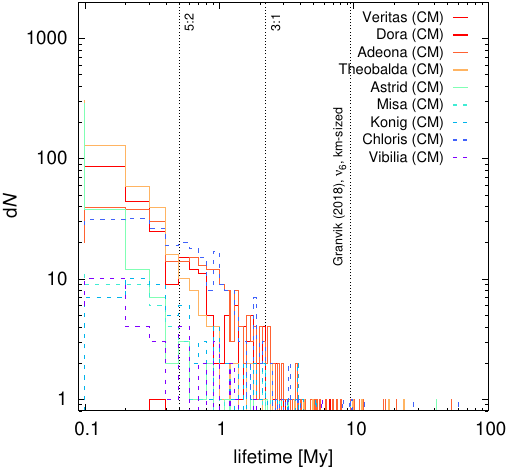} &
\includegraphics[width=5.9cm]{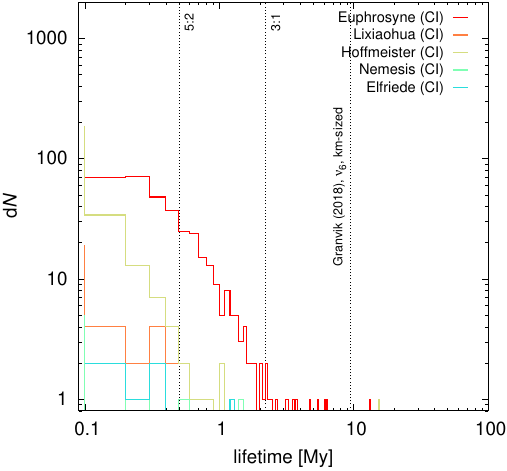} &
\includegraphics[width=5.9cm]{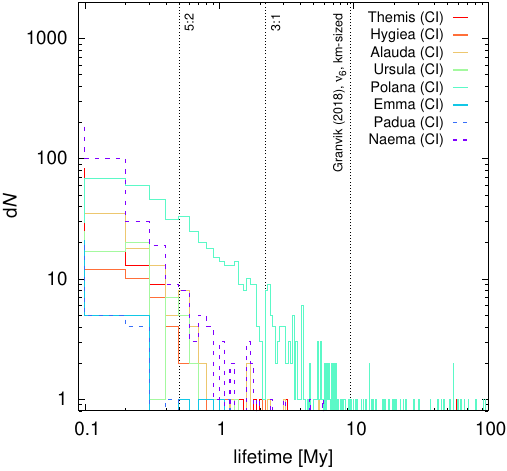} \\
\kern1cm CO/CV/CK-type &
\kern1cm M-type &
\kern1cm unknown \\
\includegraphics[width=5.9cm]{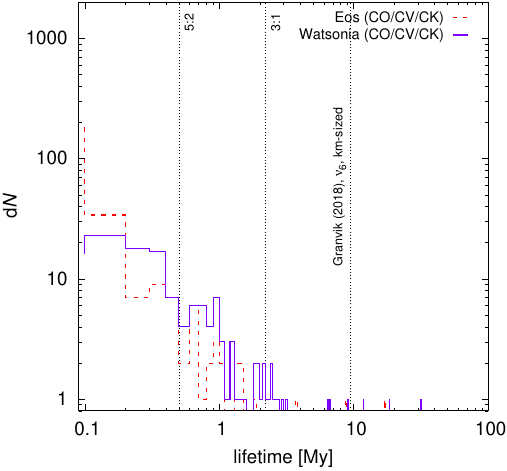} &
\includegraphics[width=5.9cm]{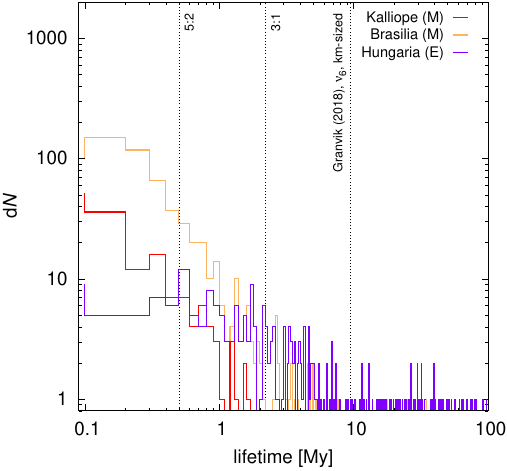} &
\includegraphics[width=5.9cm]{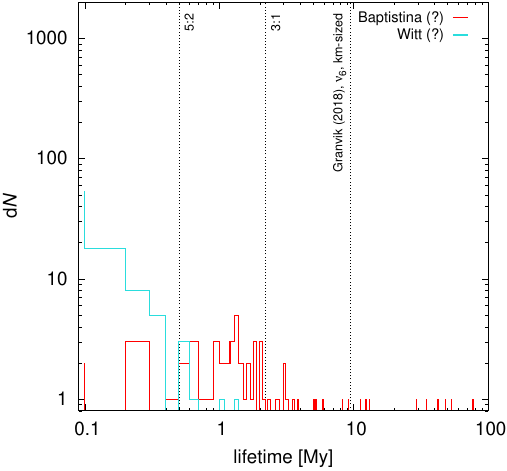} \\
\end{tabular}
\caption{
Lifetime histograms of individual NEO populations,
computed for kilometre-sized bodies.
Originally, these bodies were located in the main belt
(see Fig.~\ref{nbody_observed_decay_CM})
and subsequently transported to the NEO region
using our dynamical model.
From the histograms,
the {\em mean\/} lifetimes $\bar\tau_{\rm neo}$ were estimated
(cf.~Tab.~\ref{tab:1km}).
Some of the histograms contain a low number of bodies,
because it is indeed limited in the source population.
}
\label{nbody_observed_nea3_CM}
\end{figure*}

\begin{figure*}
\centering
\begin{tabular}{c@{\kern0.1cm}c@{\kern0.1cm}c}
\kern1cm CM-chondrite &
\kern1cm CI-chondrite,steep &
\kern1cm CI-chondrite, shallow \\
\includegraphics[width=5.9cm]{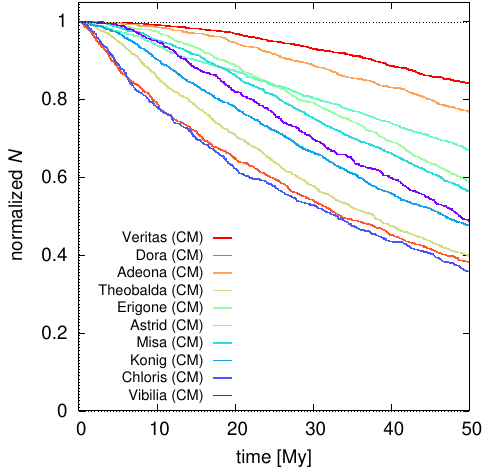} &
\includegraphics[width=5.9cm]{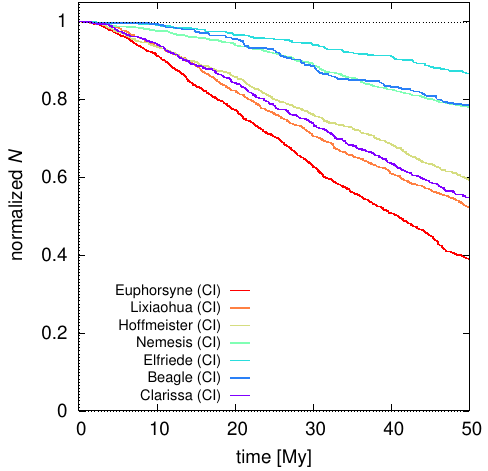} &
\includegraphics[width=5.9cm]{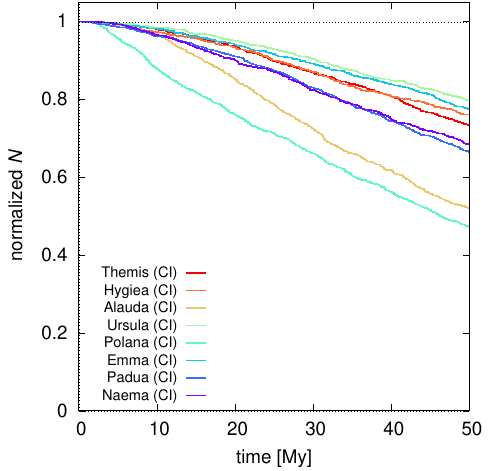} \\
\kern1cm CO/CV/CK-type &
\kern1cm M-type &
\kern1cm unknown \\
\includegraphics[width=5.9cm]{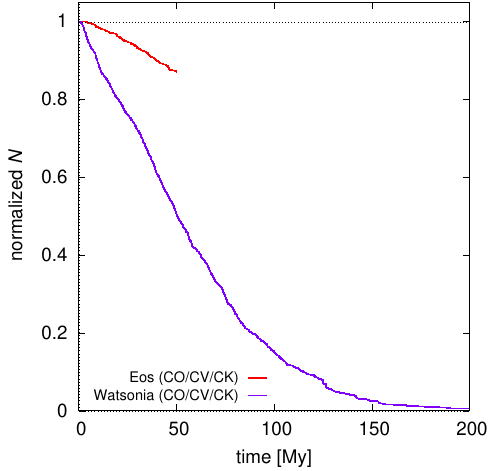} &
\includegraphics[width=5.9cm]{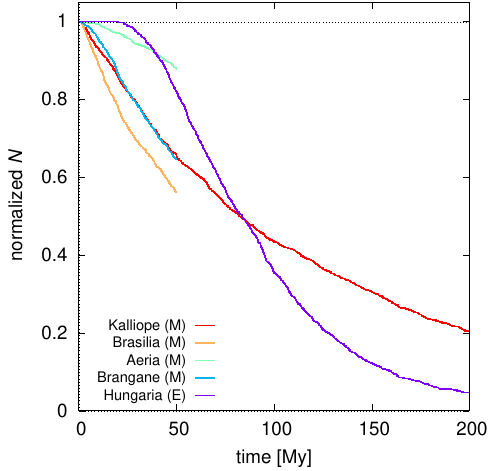} &
\includegraphics[width=5.9cm]{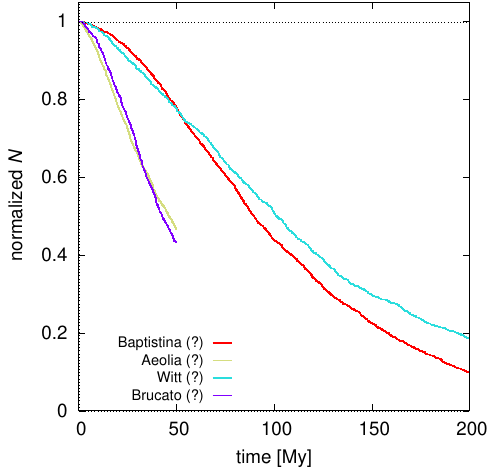} \\
\end{tabular}
\caption{
Same as Fig.~\ref{nbody_observed_decay_CM} for metre-sized bodies.
}
\label{nbody_metresized_decay_CM}
\end{figure*}

\begin{figure*}
\centering
\begin{tabular}{c@{\kern0.1cm}c@{\kern0.1cm}c}
\kern1cm CM-chondrite &
\kern1cm CI-chondrite, steep &
\kern1cm CI-chondrite, shallow \\
\includegraphics[width=5.9cm]{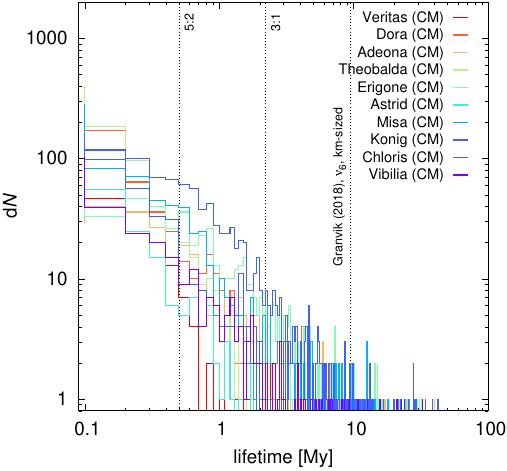} &
\includegraphics[width=5.9cm]{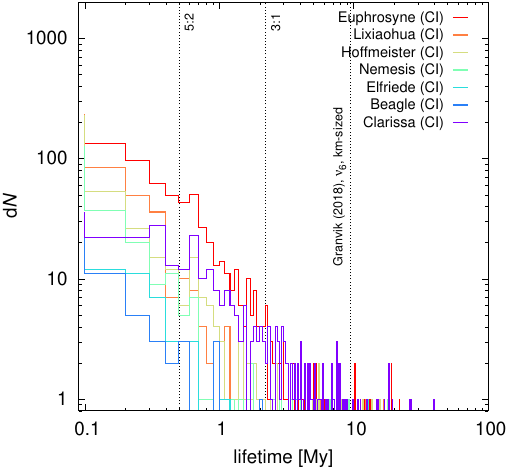} &
\includegraphics[width=5.9cm]{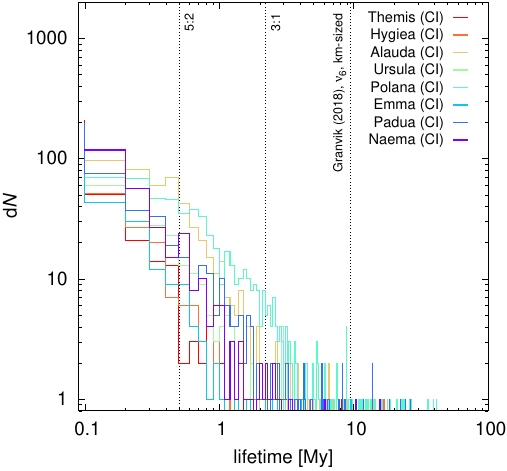} \\
\kern1cm CO/CV/CK-type &
\kern1cm M-type &
\kern1cm unknown \\
\includegraphics[width=5.9cm]{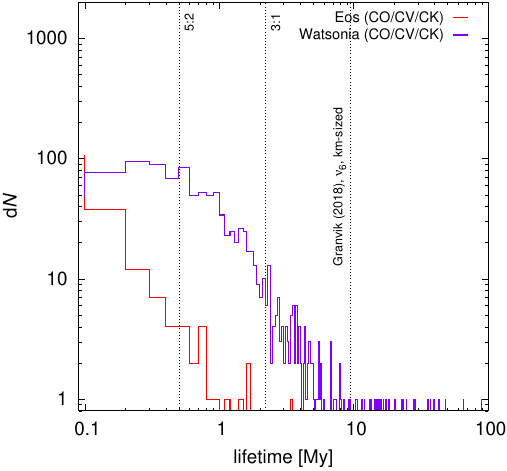} &
\includegraphics[width=5.9cm]{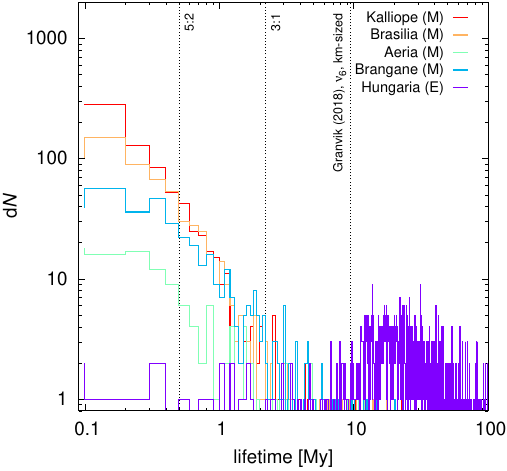} &
\includegraphics[width=5.9cm]{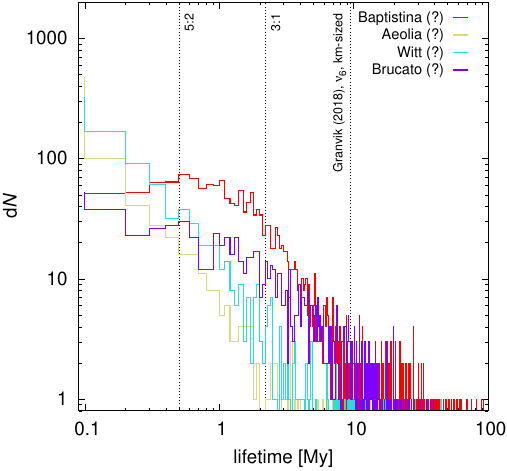} \\
\end{tabular}
\caption{
Same as Fig.~\ref{nbody_observed_nea3_CM} for metre-sized bodies.
}
\label{nbody_metresized_nea3_CM}
\end{figure*}

\begin{figure*}
\centering
\begin{tabular}{c@{\kern0.1cm}c@{\kern0.1cm}c@{}c}
\kern0.5cm Adeona (CM), 350\,My &
\kern0.5cm {\bf Aeolia (?)}, 25\,My &
\kern0.5cm A\"eria (M), 350\,My &
\\
\includegraphics[width=5.9cm]{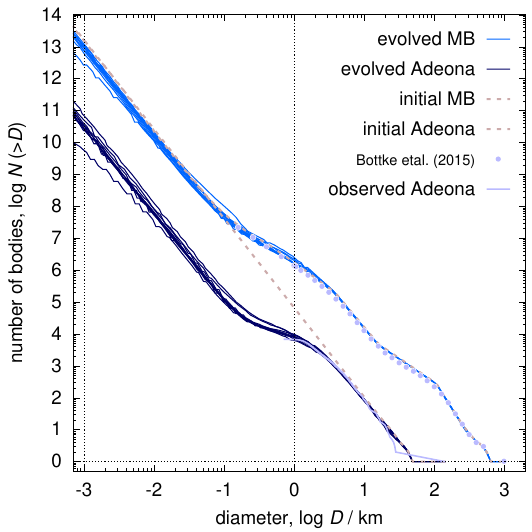} &
\includegraphics[width=5.9cm]{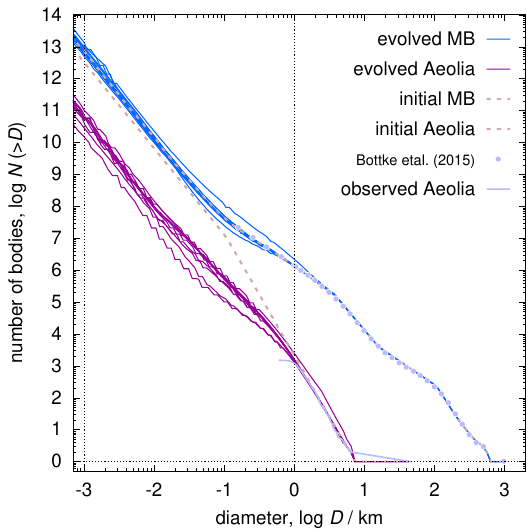} &
\includegraphics[width=5.9cm]{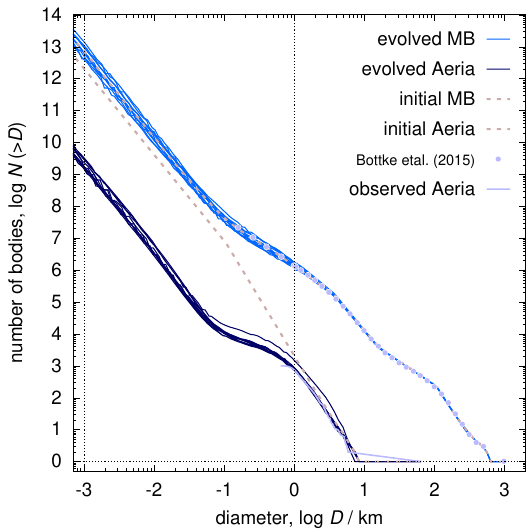} &
\\[0.1cm]
\kern0.5cm Alauda (CI), 2500\,My &
\kern0.5cm Astrid (CM), 150\,My &
\kern0.5cm {\bf Baptistina (?)}, 300\,My &
\\
\includegraphics[width=5.9cm]{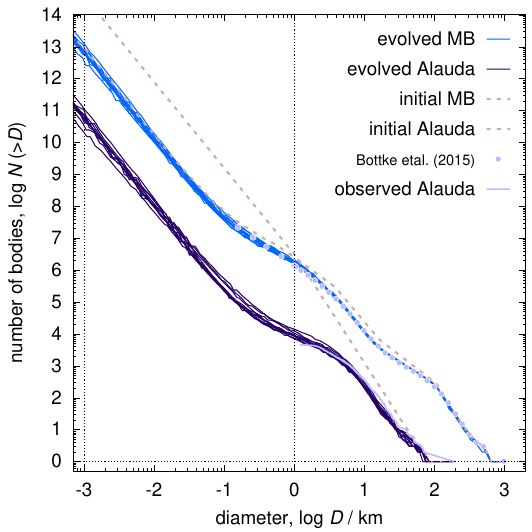} &
\includegraphics[width=5.9cm]{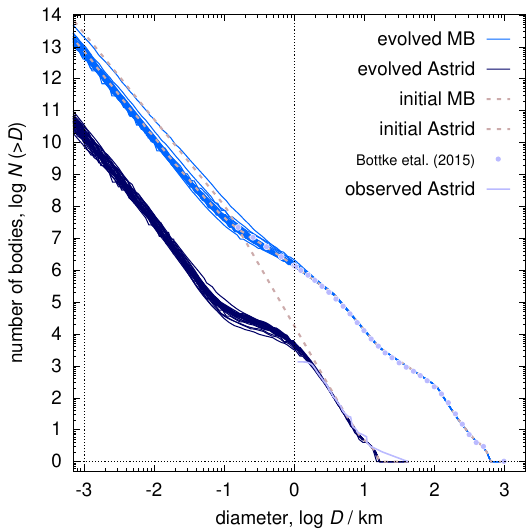} &
\includegraphics[width=5.9cm]{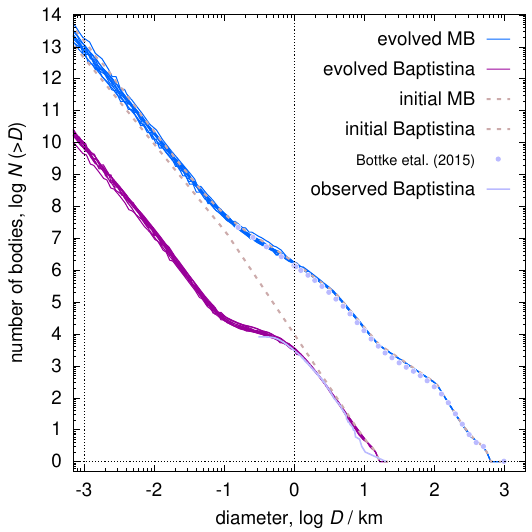} &
\\[0.1cm]
\kern0.5cm Beagle (CI), 30\,My &
\kern0.5cm {\bf Brang\"ane (M)}, 20-50\,My &
\kern0.5cm Brasilia (M), 100\,My &
\\
\includegraphics[width=5.9cm]{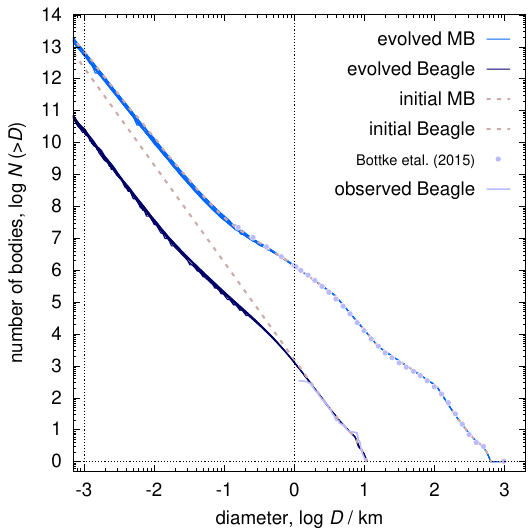} &
\includegraphics[width=5.9cm]{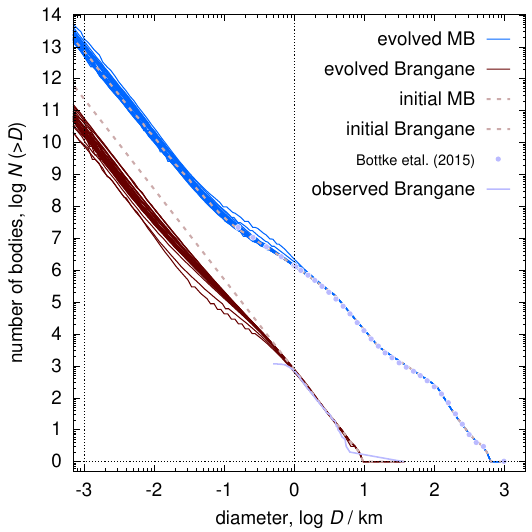} &
\includegraphics[width=5.9cm]{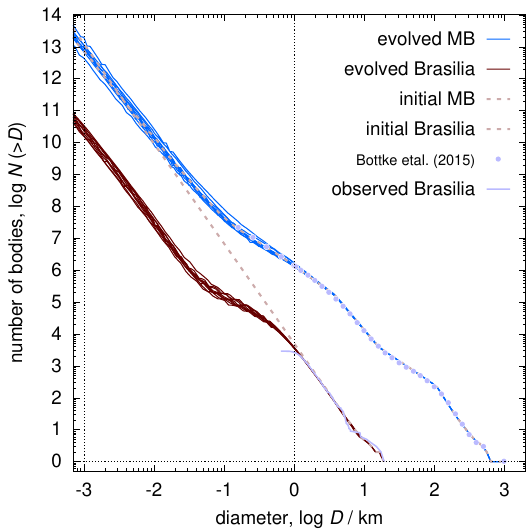} &
\\[0.1cm]
\end{tabular}
\caption{
Synthetic SFDs of the C-type asteroid families derived from our collisional model.
Each panel contains:
the initial main belt,
the initial family (\textcolor{pantone}{Pantone 7633C}, dotted),
evolved main belt (\textcolor{blue}{blue}),
evolved family (different colours),
observed main belt \citep{Bottke_2015aste.book..701B},
observed family (\textcolor{gray}{gray}, solid).
The SFDs between 1 and 10\,km were initially a smooth power-law.
They evolved due to collisions
and exhibit a characteristic slope change at about 5\,km,
which is observed (see Tab.~\ref{tab:q}).
Each model was run 10 times to account for stochasticity.
The best-fit age is reported on top (see Tab.~\ref{tab2}).
}
\label{sfds}
\end{figure*}

\addtocounter{figure}{-1}
\begin{figure*}
\centering
\begin{tabular}{c@{\kern0.1cm}c@{\kern0.1cm}c@{}c}
\kern0.5cm Brucato (?), 500\,My &
\kern0.5cm Chloris (CM), 1100\,My &
\kern0.5cm Clarissa (CI), 50\,My &
\\
\includegraphics[width=5.9cm]{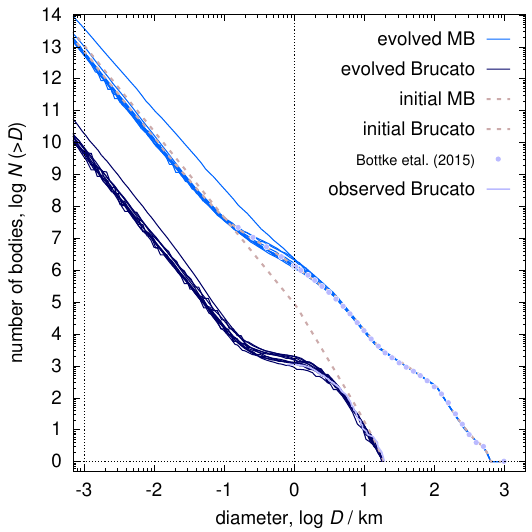} &
\includegraphics[width=5.9cm]{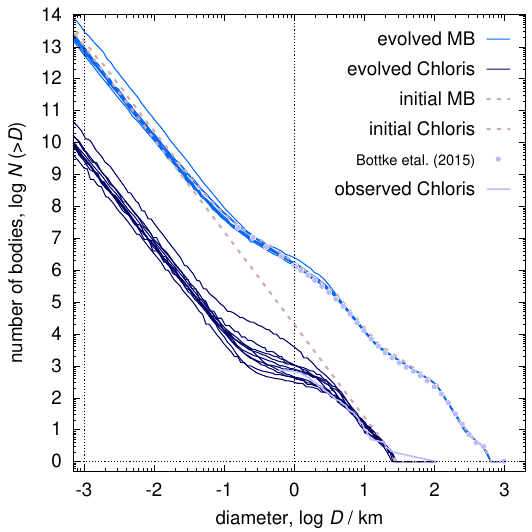} &
\includegraphics[width=5.9cm]{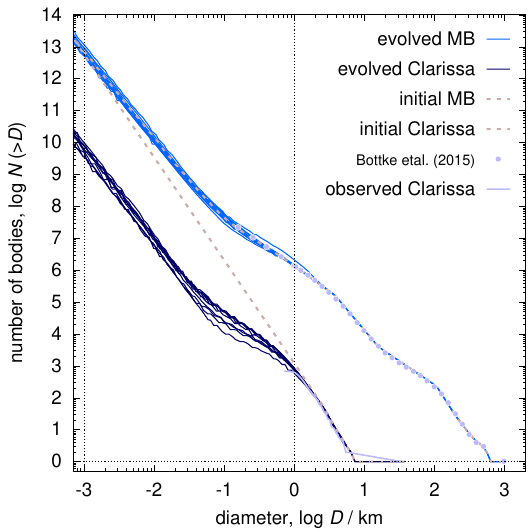} &
\\[0.1cm]
\kern0.5cm Dora (CM), 250\,My &
\kern0.5cm Elfriede (CI), 35-100\,My &
\kern0.5cm Emma (IDP), 1200\,My &
\\
\includegraphics[width=5.9cm]{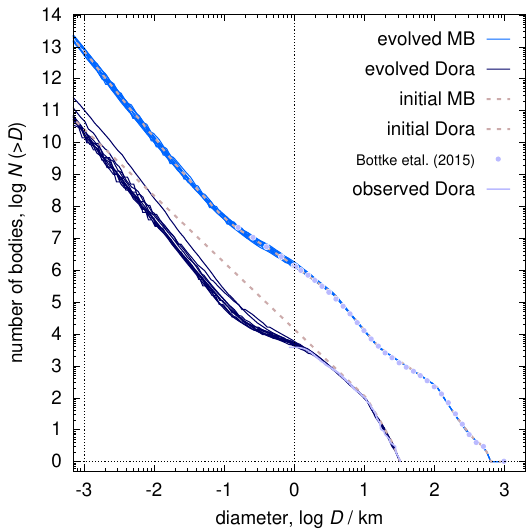} &
\includegraphics[width=5.9cm]{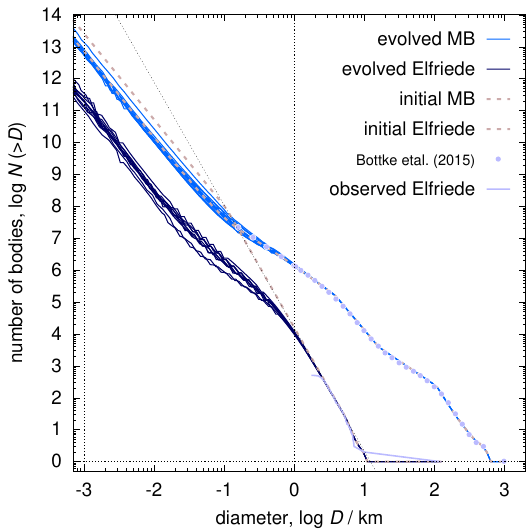} &
\includegraphics[width=5.9cm]{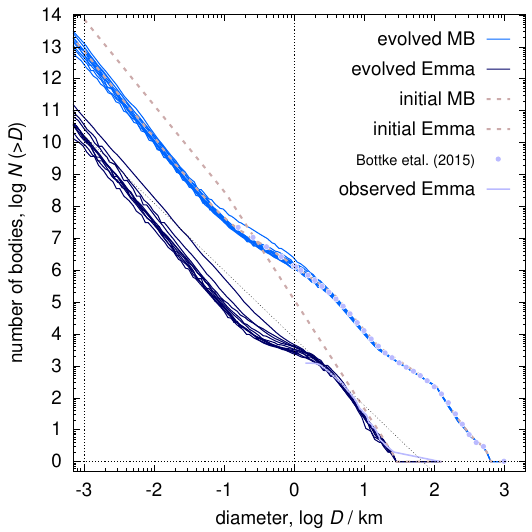} &
\\[0.1cm]
\kern0.5cm Eos (CO/CV/CK), 1800\,My &
\kern0.5cm Erigone (CM), 500\,My &
\kern0.5cm {\bf Euphrosyne (CI)}, 800\,My &
\\
\includegraphics[width=5.9cm]{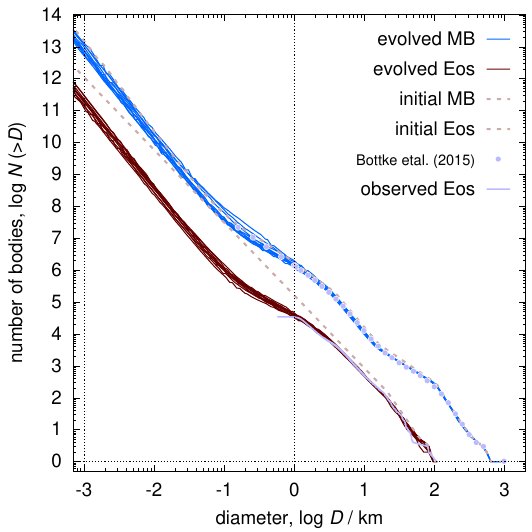} &
\includegraphics[width=5.9cm]{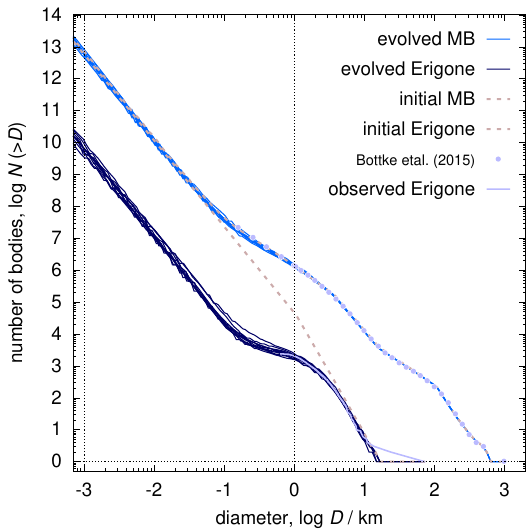} &
\includegraphics[width=5.9cm]{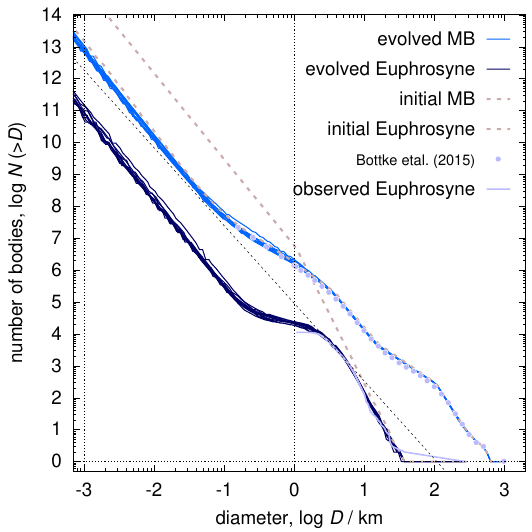} &
\\[0.1cm]
\end{tabular}
\caption{continued.}
\end{figure*}

\addtocounter{figure}{-1}
\begin{figure*}
\centering
\begin{tabular}{c@{\kern0.1cm}c@{\kern0.1cm}c@{}c}
\kern0.5cm Hoffmeister (CI), 100\,My &
\kern0.5cm Hungaria (E), 150\,My &
\kern0.5cm Hygiea (CI), 2500\,My &
\\
\includegraphics[width=5.9cm]{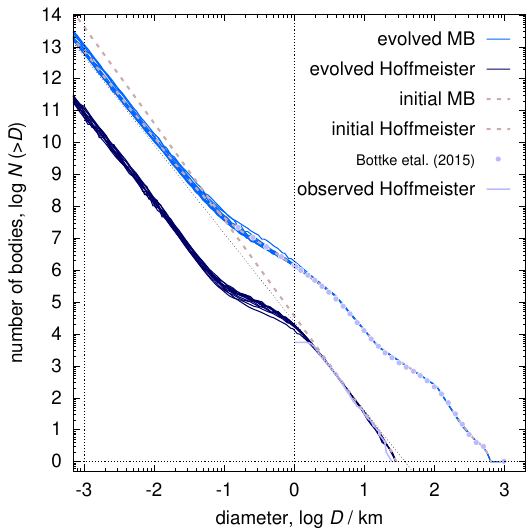} &
\includegraphics[width=5.9cm]{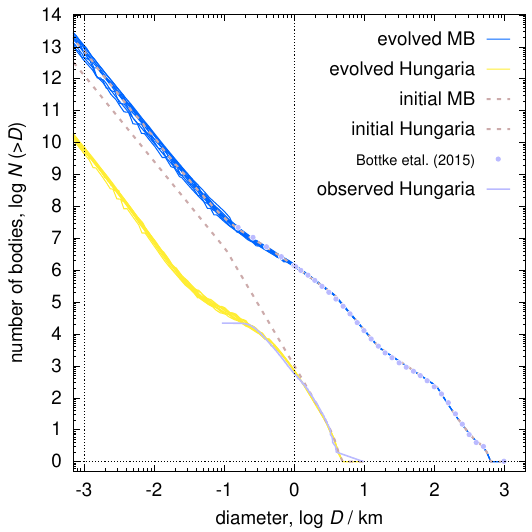} &
\includegraphics[width=5.9cm]{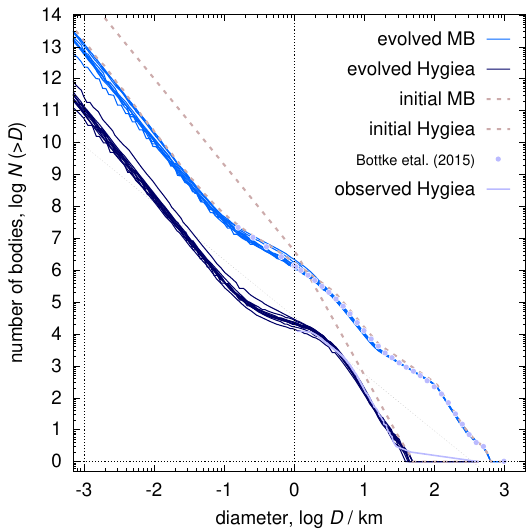} &
\\[0.1cm]
\kern0.5cm Iannini (Aca/Lod), 7.5\,My &
\kern0.5cm Kalliope (M), 900\,My &
\kern0.5cm {\bf K\"onig (CM)}, 20\,My &
\\
\includegraphics[width=5.9cm]{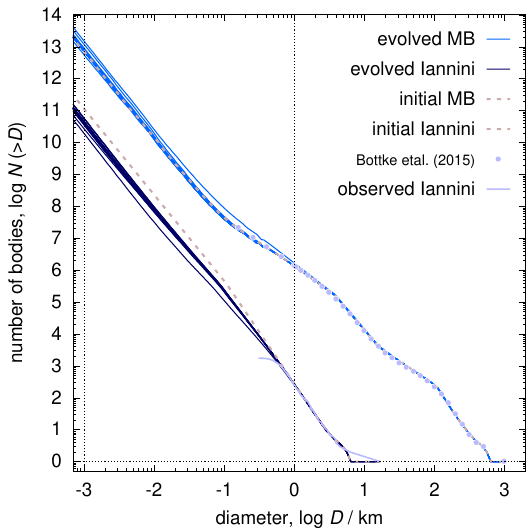} &
\includegraphics[width=5.9cm]{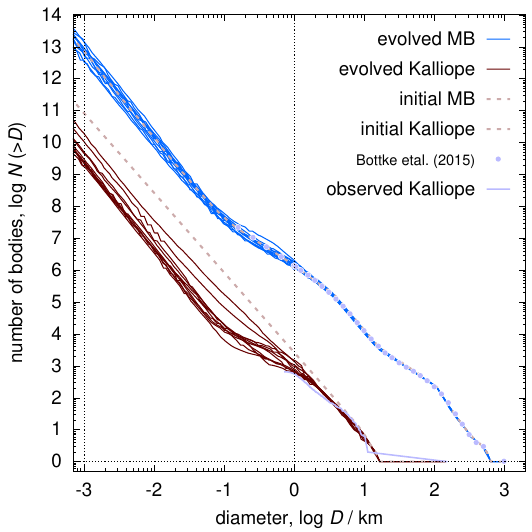} &
\includegraphics[width=5.9cm]{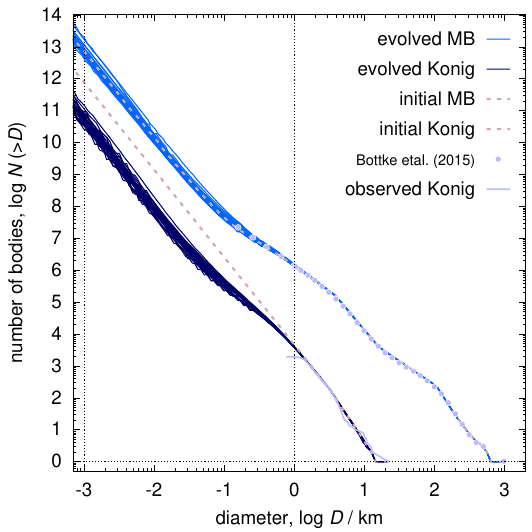} &
\\[0.1cm]
\kern0.5cm Lixiaohua (CI), 150\,My &
\kern0.5cm Misa (CI), 50\,My &
\kern0.5cm Naema (CI), 1500\,My &
\\
\includegraphics[width=5.9cm]{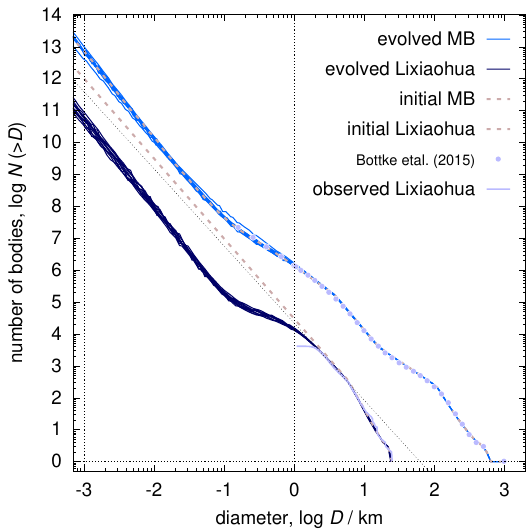} &
\includegraphics[width=5.9cm]{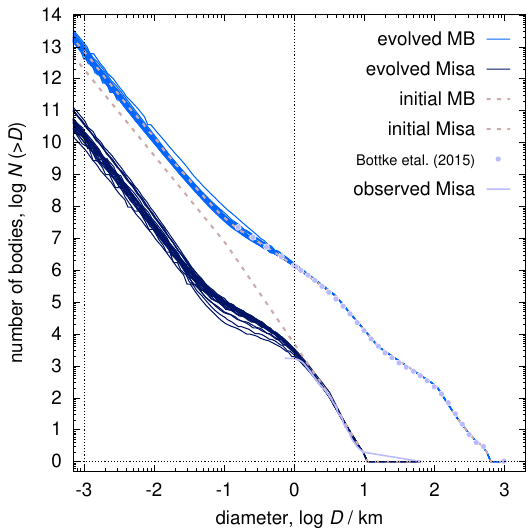} &
\includegraphics[width=5.9cm]{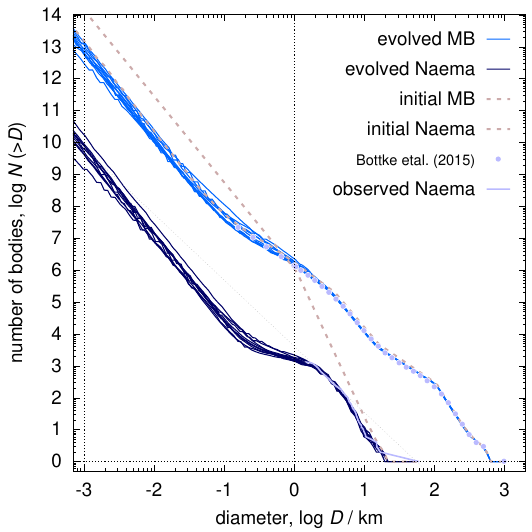} &
\\[0.1cm]
\end{tabular}
\caption{continued.}
\end{figure*}

\addtocounter{figure}{-1}
\begin{figure*}
\centering
\begin{tabular}{c@{\kern0.1cm}c@{\kern0.1cm}c@{}c}
\kern0.5cm Nemesis (CI), 1300\,My &
\kern0.5cm Padua (IDP), 900\,My &
\kern0.5cm Pallas (B), 2500\,My &
\\
\includegraphics[width=5.9cm]{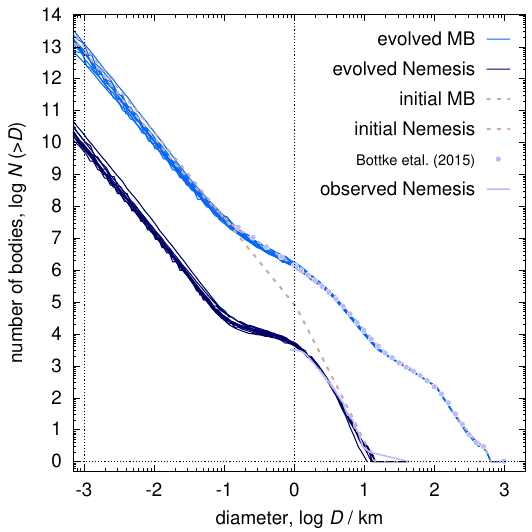} &
\includegraphics[width=5.9cm]{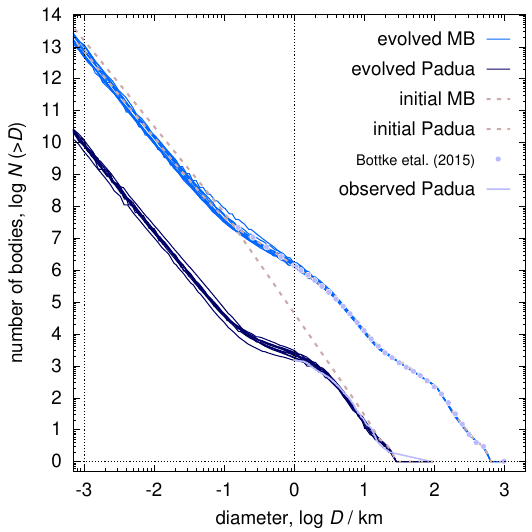} &
\includegraphics[width=5.9cm]{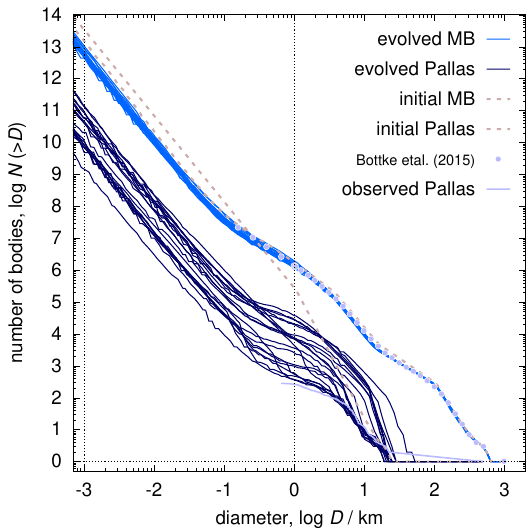} &
\\[0.1cm]
\kern0.5cm {\bf Polana (CI)}, 200\,My &
\kern0.5cm Themis (CI), 2000\,My &
\kern0.5cm Theobalda (CM), 20\,My &
\\
\includegraphics[width=5.9cm]{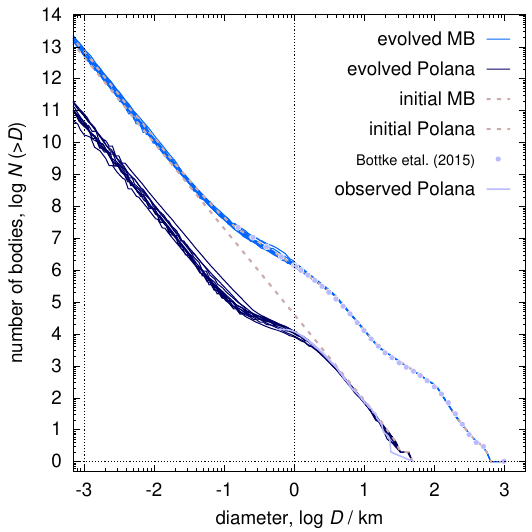} &
\includegraphics[width=5.9cm]{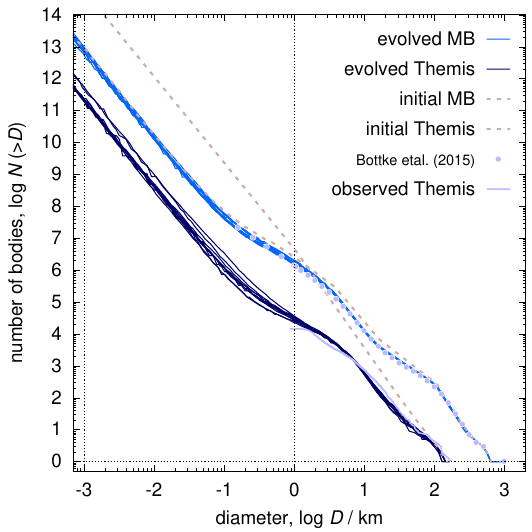} &
\includegraphics[width=5.9cm]{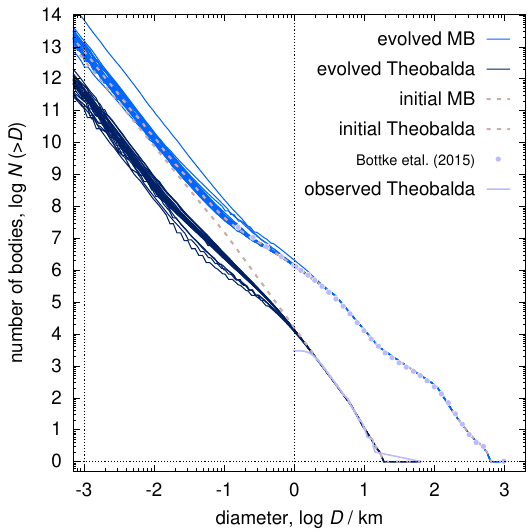} &
\\[0.1cm]
\kern0.5cm Tina (M), 800\,My &
\kern0.5cm Ursula (CI), 1800\,My &
\kern0.5cm {\bf Veritas (CM)}, 8.3\,My &
\\
\includegraphics[width=5.9cm]{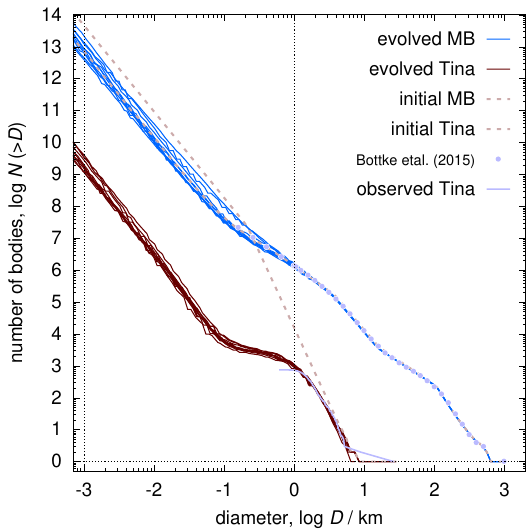} &
\includegraphics[width=5.9cm]{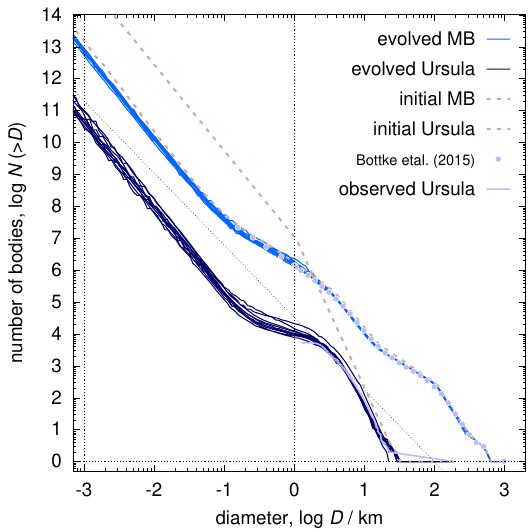} &
\includegraphics[width=5.9cm]{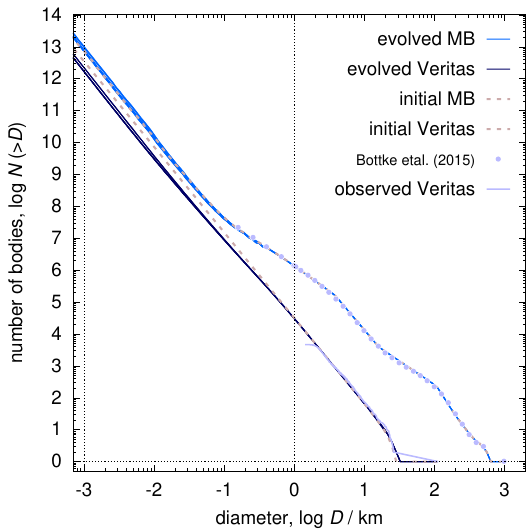} &
\\[0.1cm]
\end{tabular}
\caption{continued.}
\end{figure*}

\addtocounter{figure}{-1}
\begin{figure*}
\centering
\begin{tabular}{c@{\kern0.1cm}c@{\kern0.1cm}c@{}c}
\kern0.5cm Vibilia (CM), 700\,My &
\kern0.5cm Watsonia (CO/CV/CK), 2500\,My &
\kern0.5cm Witt (?), 70\,My &
\\
\includegraphics[width=5.9cm]{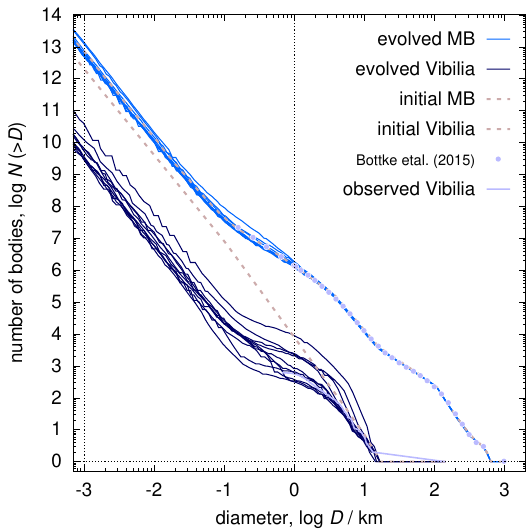} &
\includegraphics[width=5.9cm]{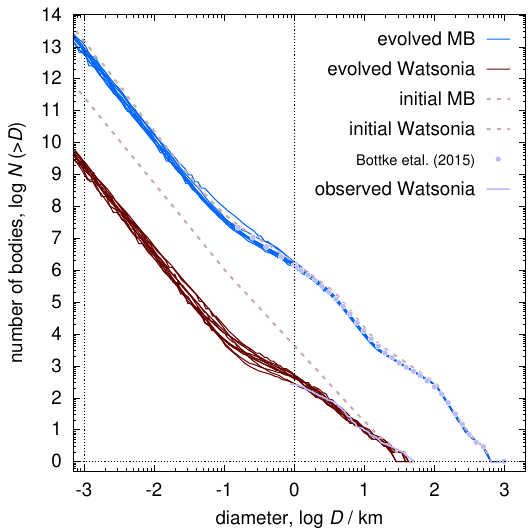} &
\includegraphics[width=5.9cm]{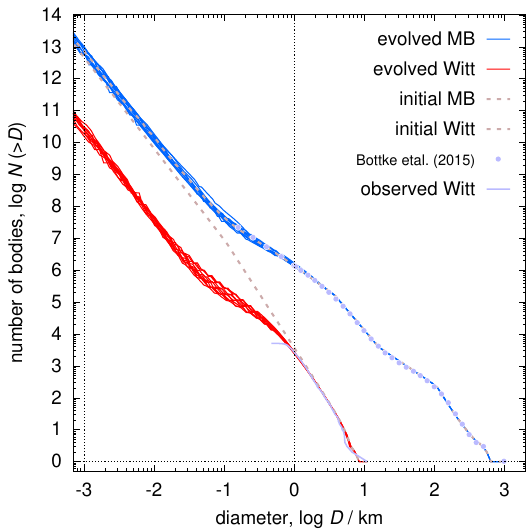} &
\\[0.1cm]
\end{tabular}
\caption{continued.}
\end{figure*}


\begin{figure*}[h]
\centering
\begin{tabular}{c@{\kern0.1cm}c@{\kern0.1cm}c@{}c}
\kern0.5cm Adeona (CM) &
\kern0.5cm {\bf Aeolia (?)} &
\kern0.5cm A\"eria (M) &
\\
\includegraphics[width=5.9cm]{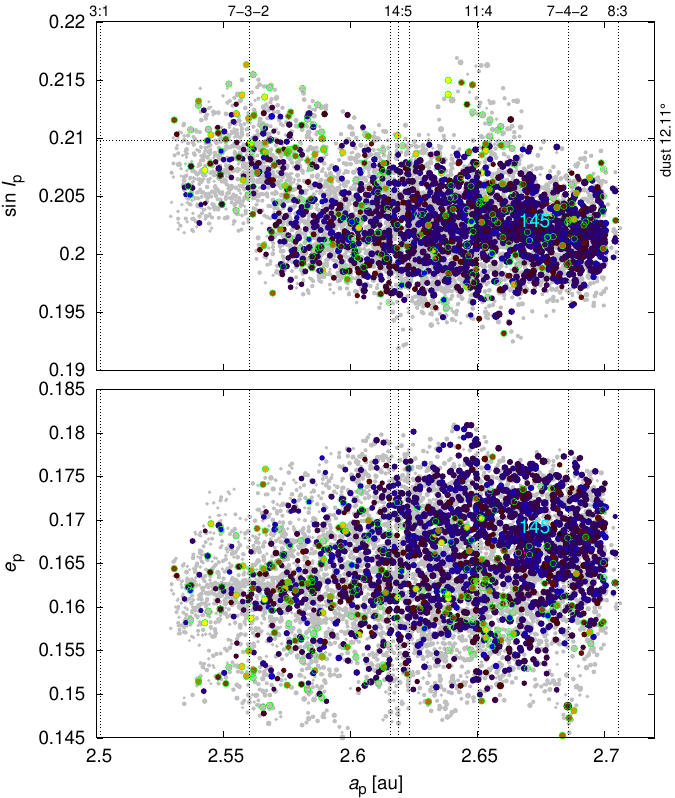} &
\includegraphics[width=5.9cm]{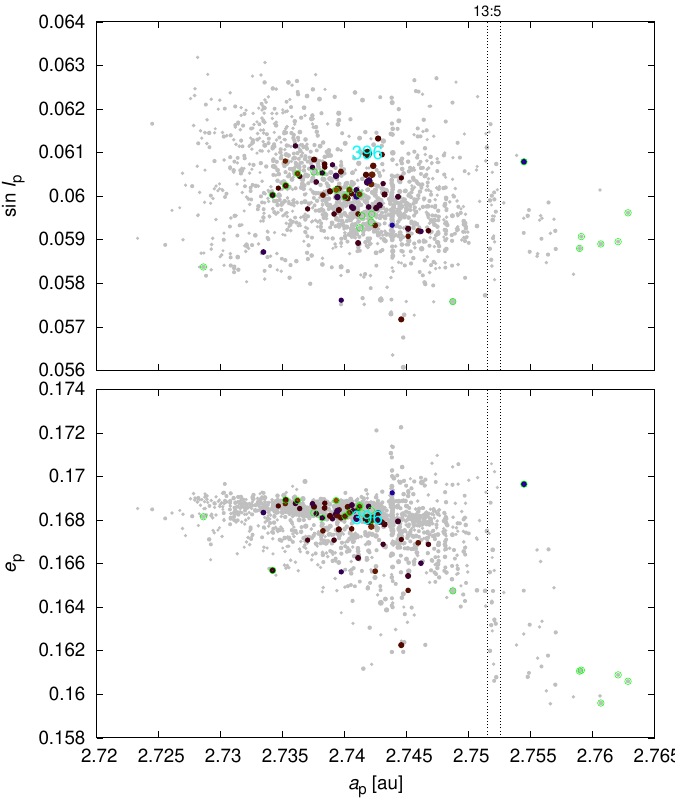} &
\includegraphics[width=5.9cm]{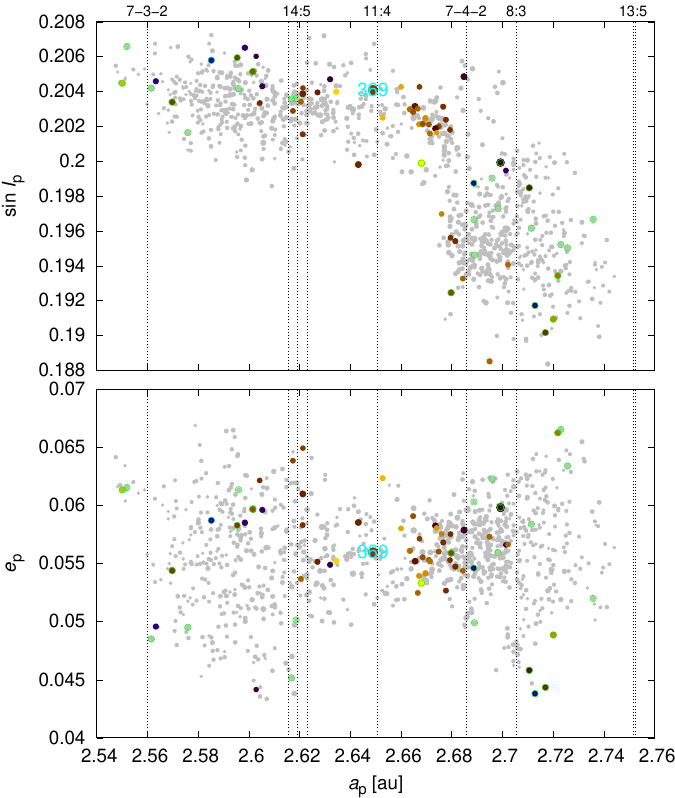} &
\\[0.1cm]
\kern0.5cm Alauda (CI) &
\kern0.5cm Astrid (CM) &
\kern0.5cm {\bf Baptistina (?)} &
\\
\includegraphics[width=5.9cm]{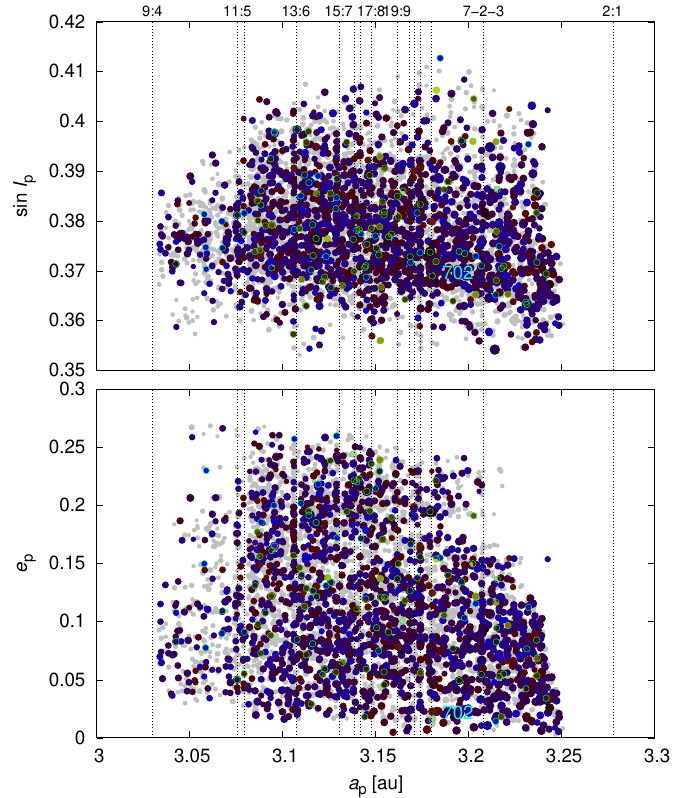} &
\includegraphics[width=5.9cm]{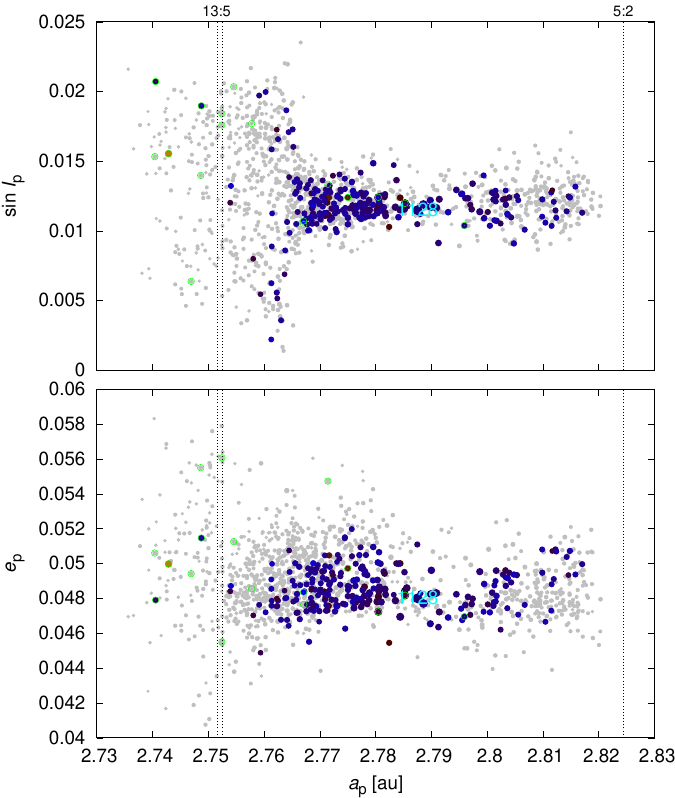} &
\includegraphics[width=5.9cm]{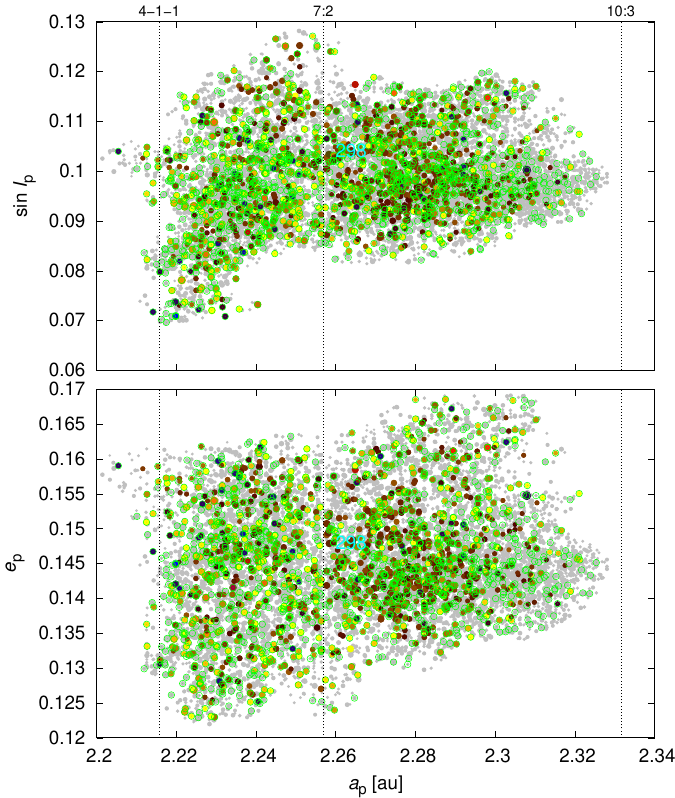} &
\\[0.1cm]
\kern0.5cm Beagle (CI) &
\kern0.5cm {\bf Brang\"ane (M)} &
\kern0.5cm Brasilia (M) &
\\
\includegraphics[width=5.9cm]{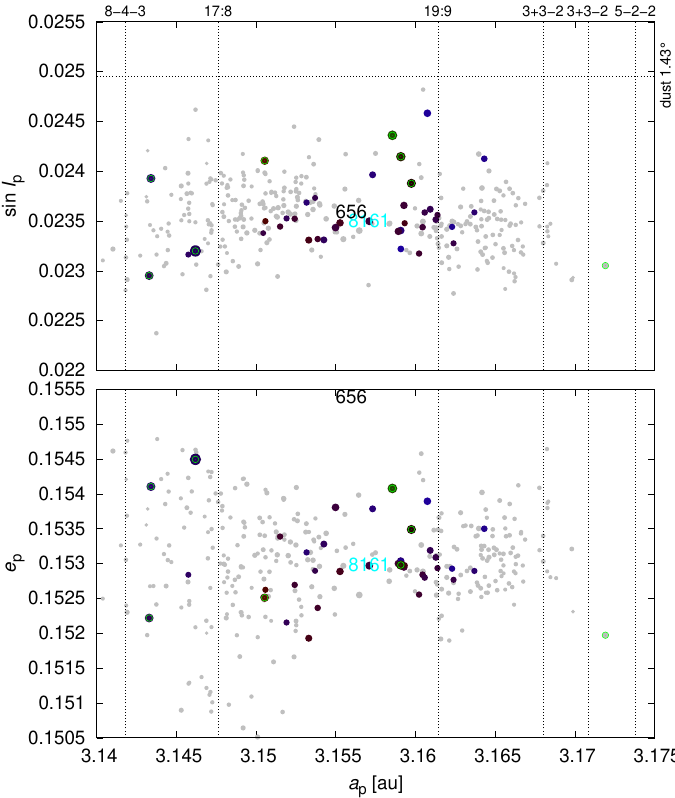} &
\includegraphics[width=5.9cm]{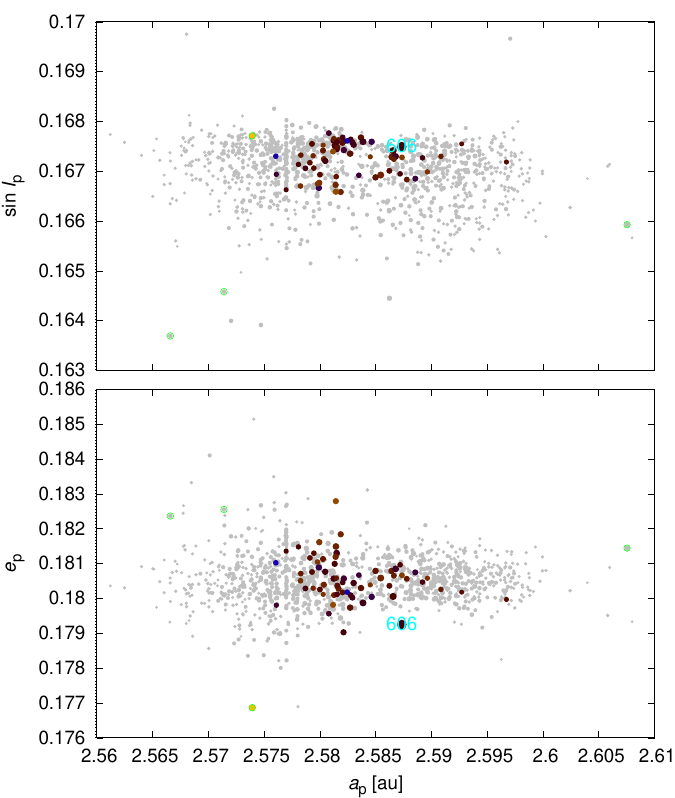} &
\includegraphics[width=5.9cm]{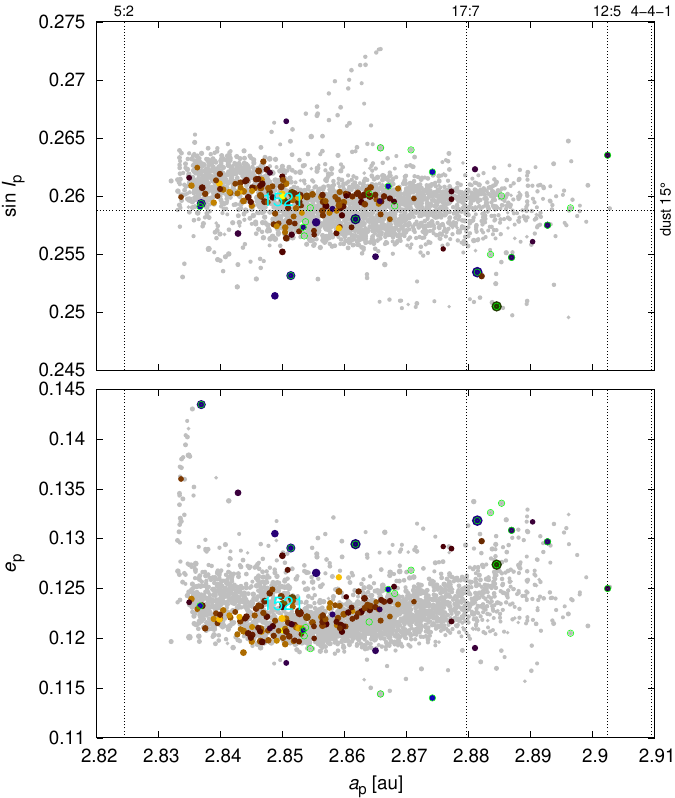} &
\\[0.1cm]
\end{tabular}
\caption{
C-type families as identified in this work.
The proper semimajor axis $a_{\rm p}$ vs. the proper eccentricity $e_{\rm p}$
and vs. the proper inclination $\sin I_{\rm p}$ are plotted.
Colours correspond to the geometric albedo~$p_{\rm V}$
(\color{blue}blue\color{black}$\,\rightarrow\,$\color{brown}brown\color{black}$\,\rightarrow\,$\color{yellow}yellow\color{black}).
Major mean-motion and three-body resonances (vertical dotted lines),
as well as identified interlopers (\color{green}green\color{black}\ circles) are indicated.
}
\label{aei2}
\end{figure*}

\addtocounter{figure}{-1}
\begin{figure*}
\centering
\begin{tabular}{c@{\kern0.1cm}c@{\kern0.1cm}c@{}c}
\kern0.5cm Brucato (?) &
\kern0.5cm Chloris (CM) &
\kern0.5cm Clarissa (CI) &
\\
\includegraphics[width=5.9cm]{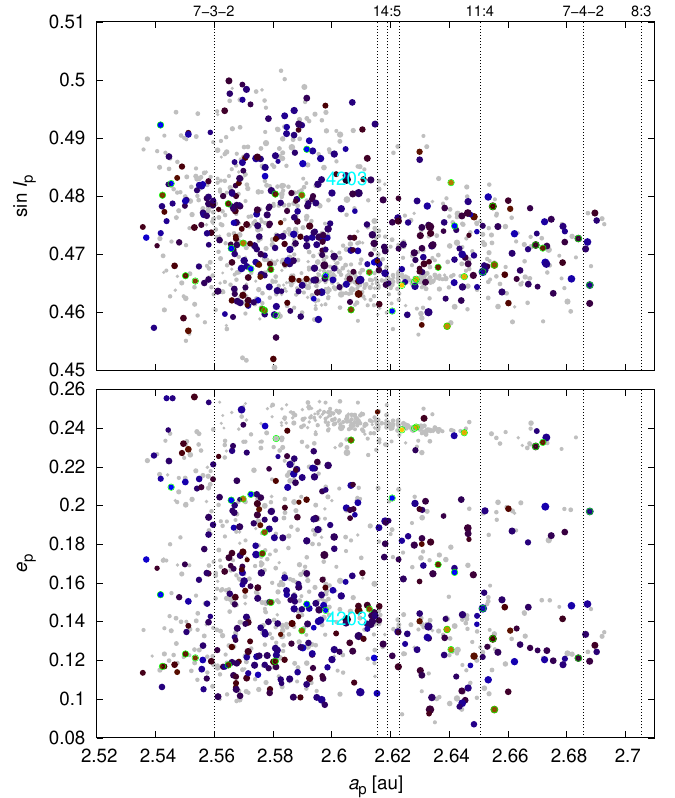} &
\includegraphics[width=5.9cm]{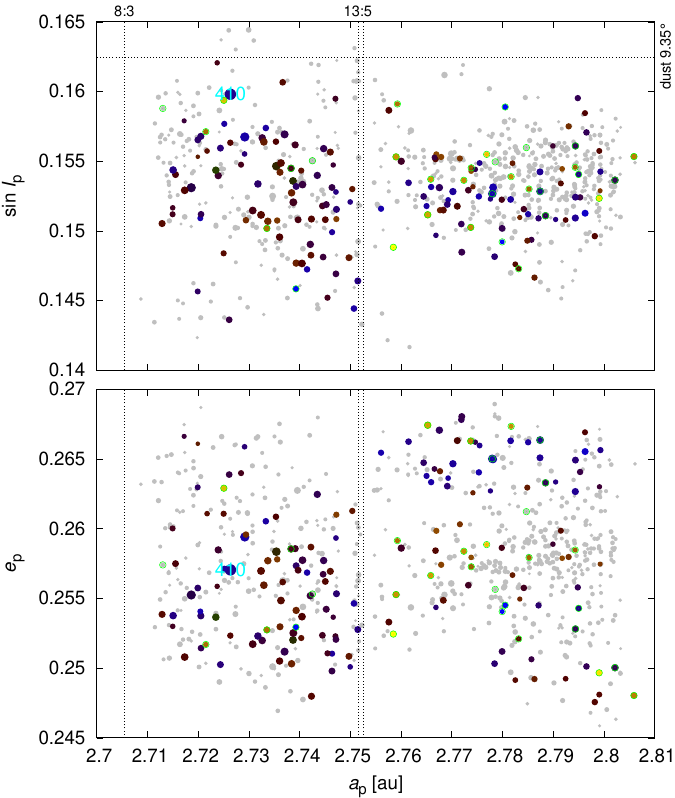} &
\includegraphics[width=5.9cm]{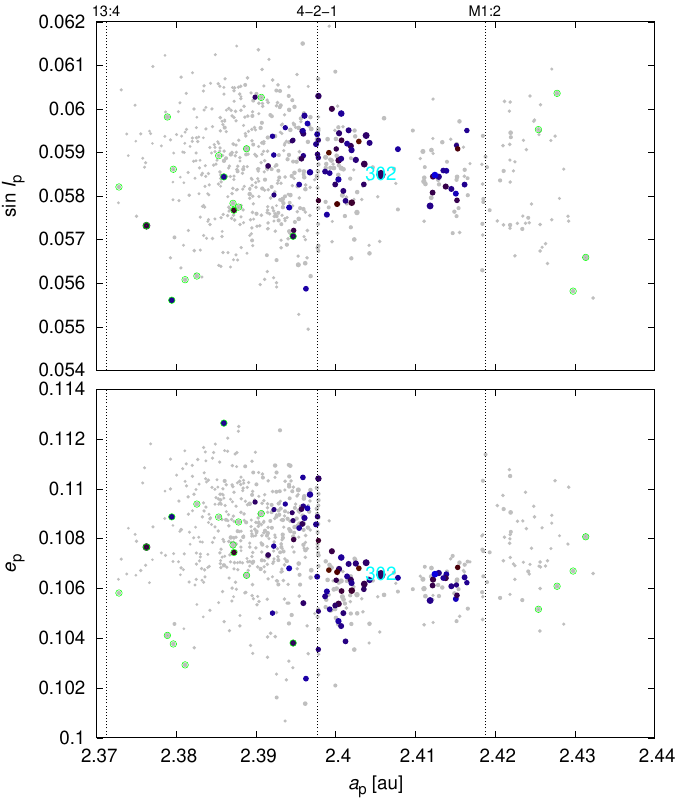} &
\\[0.1cm]
\kern0.5cm Dora (CM) &
\kern0.5cm Elfriede (CI) &
\kern0.5cm Emma (IDP) &
\\
\includegraphics[width=5.9cm]{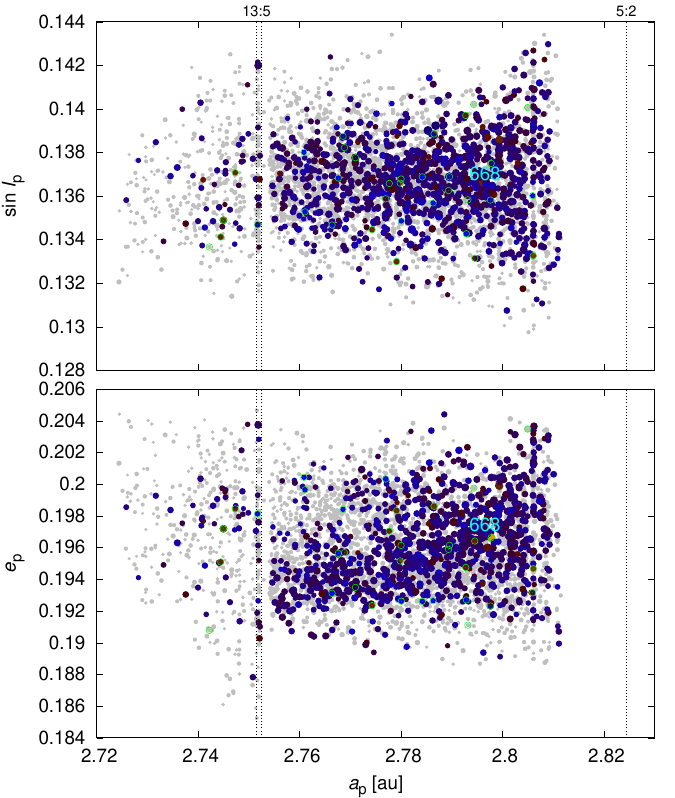} &
\includegraphics[width=5.9cm]{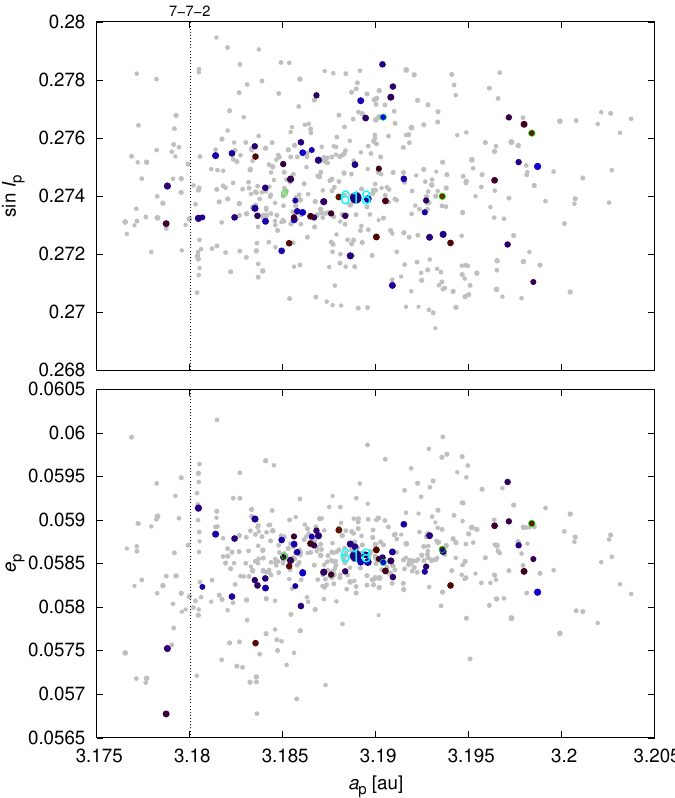} &
\includegraphics[width=5.9cm]{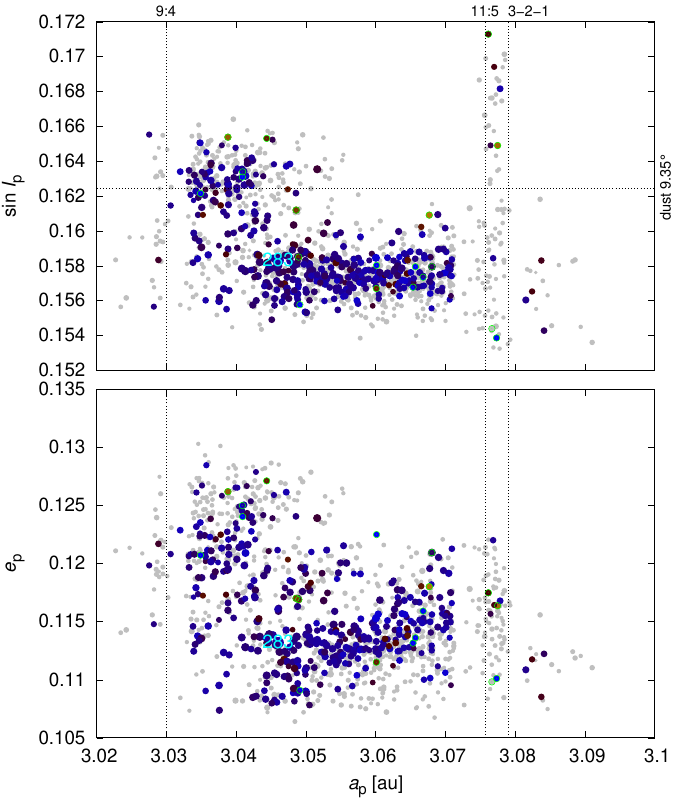} &
\\[0.1cm]
\kern0.5cm Eos (CO/CV/CK) &
\kern0.5cm Erigone (CM) &
\kern0.5cm {\bf Euphrosyne (CI)} &
\\
\includegraphics[width=5.9cm]{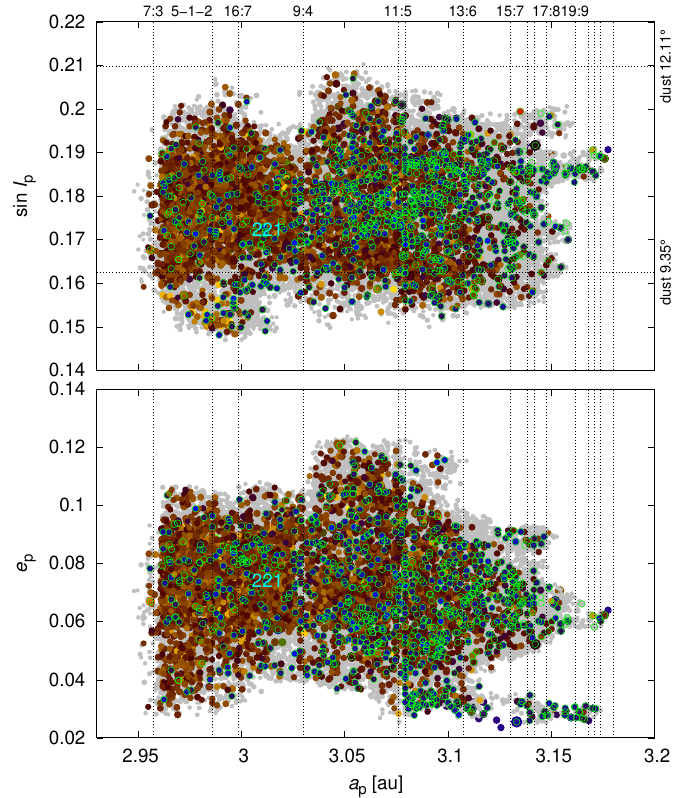} &
\includegraphics[width=5.9cm]{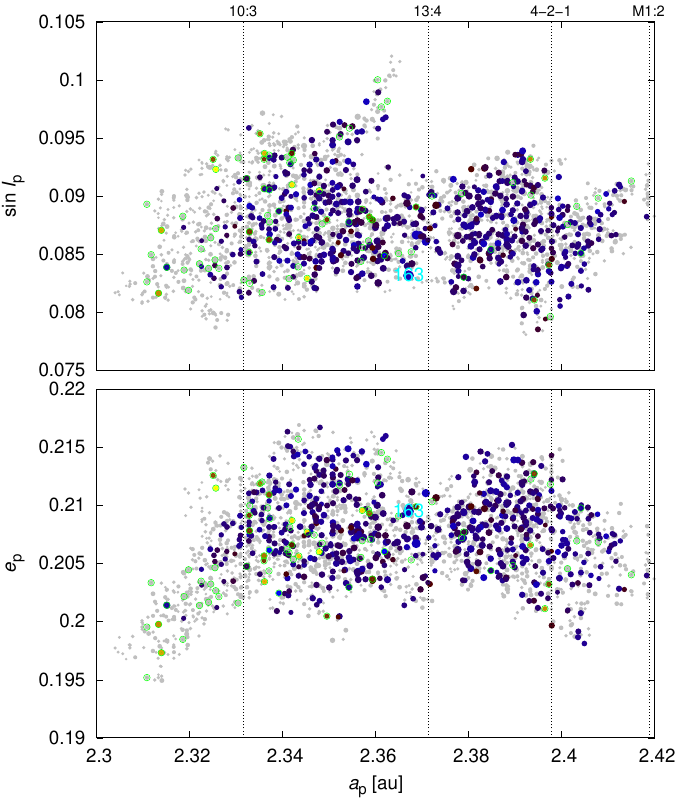} &
\includegraphics[width=5.9cm]{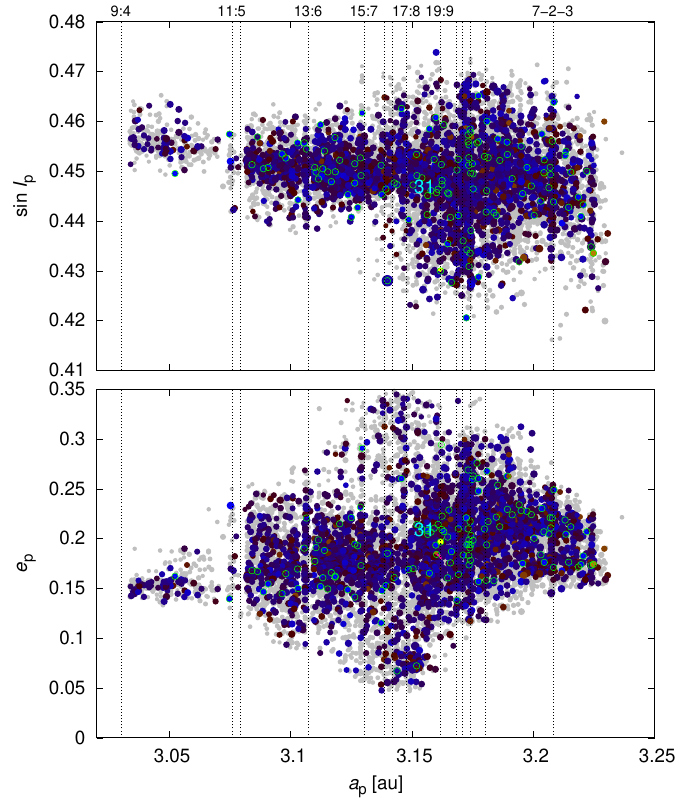} &
\\[0.1cm]
\end{tabular}
\caption{continued.}
\end{figure*}

\addtocounter{figure}{-1}
\begin{figure*}
\centering
\begin{tabular}{c@{\kern0.1cm}c@{\kern0.1cm}c@{}c}
\kern0.5cm Hoffmeister (CI) &
\kern0.5cm Hungaria (E) &
\kern0.5cm Hygiea (CI) &
\\
\includegraphics[width=5.9cm]{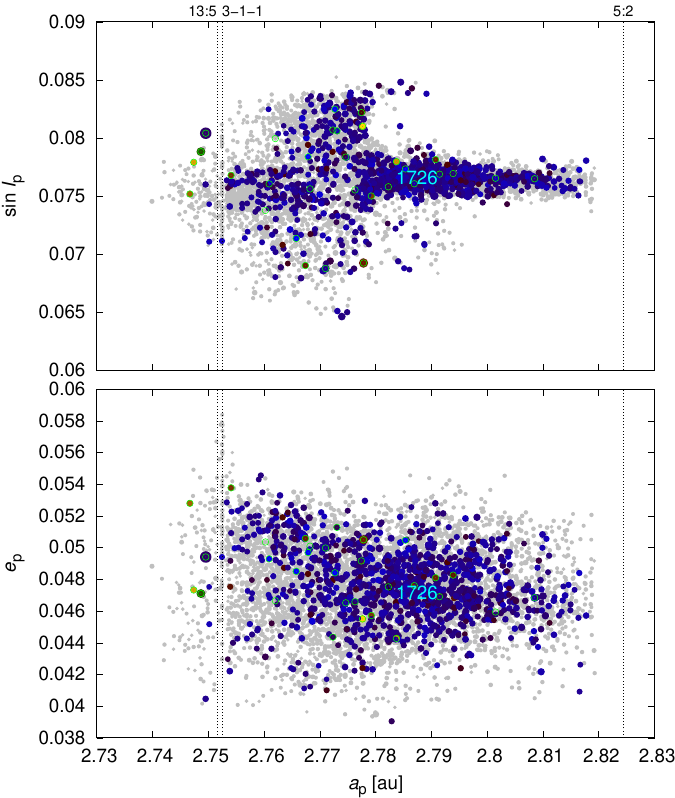} &
\includegraphics[width=5.9cm]{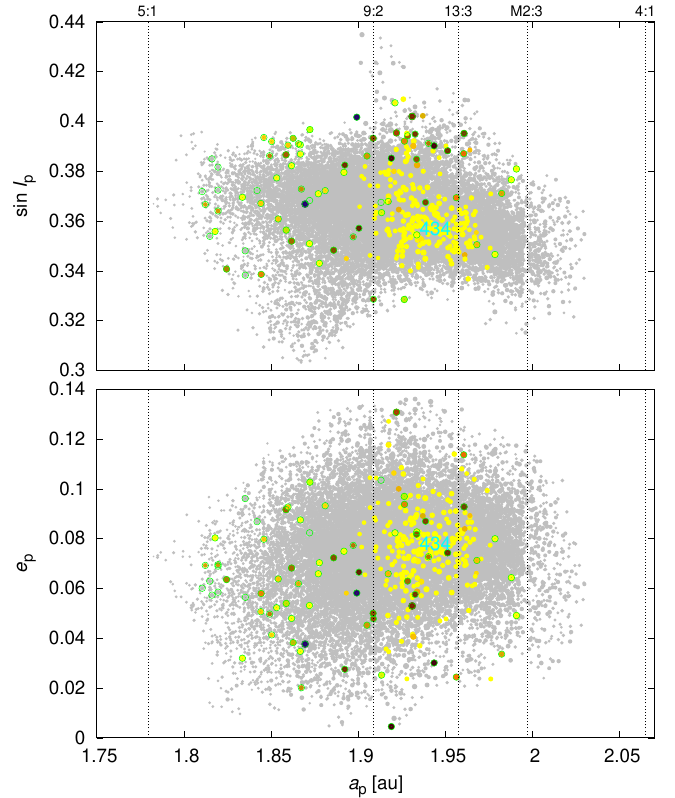} &
\includegraphics[width=5.9cm]{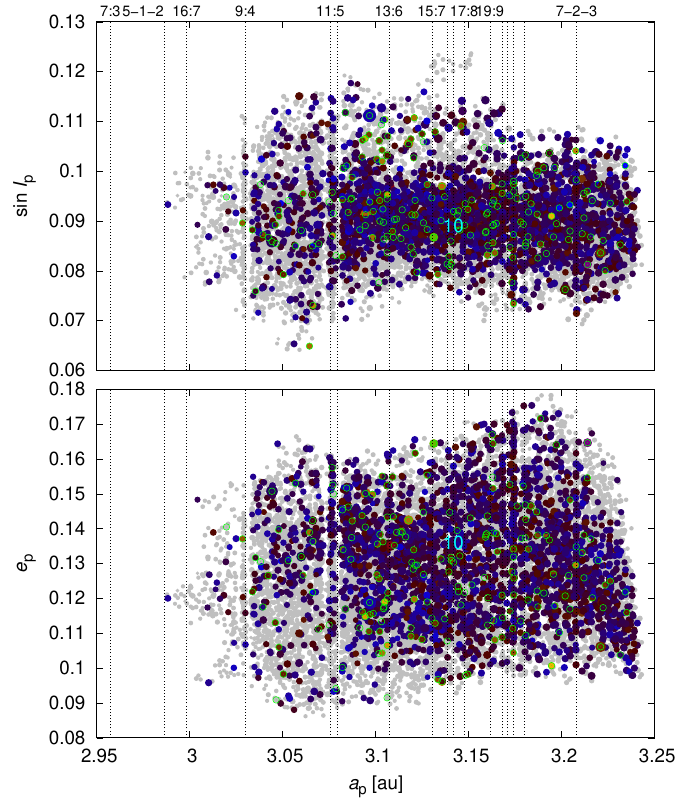} &
\\[0.1cm]
\kern0.5cm Iannini (Aca/Lod) &
\kern0.5cm {\bf K\"onig (CM)} &
\kern0.5cm Nemesis (CI) &
\\
\includegraphics[width=5.9cm]{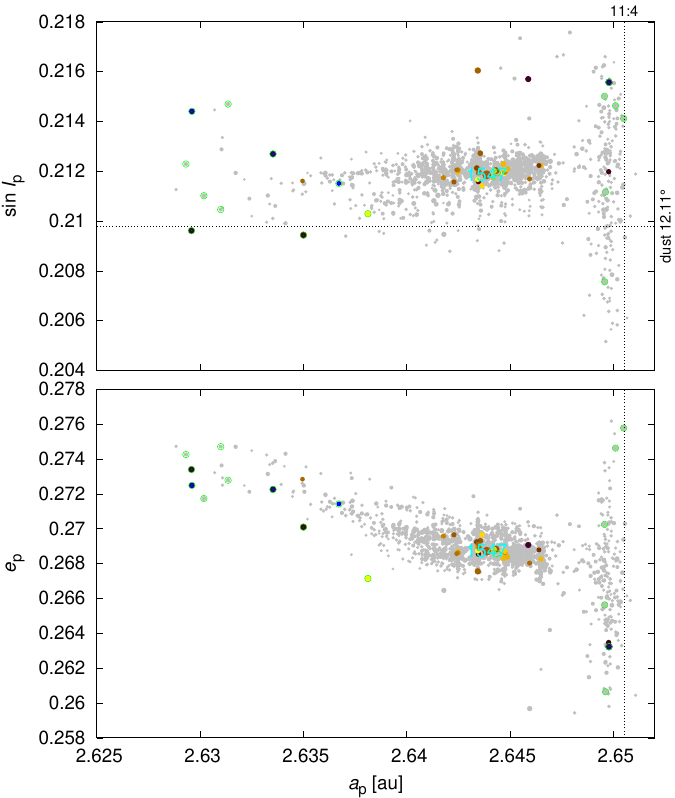} &
\includegraphics[width=5.9cm]{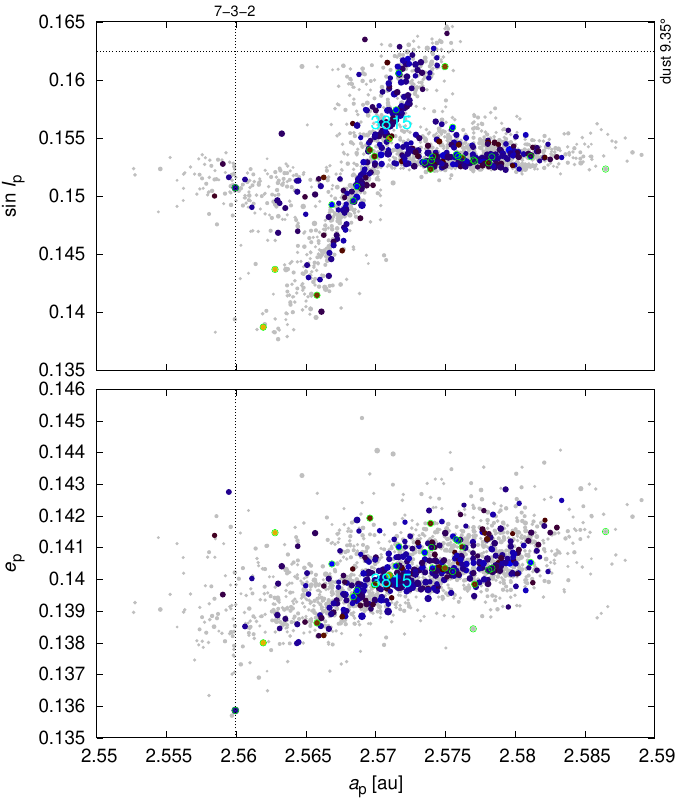} &
\includegraphics[width=5.9cm]{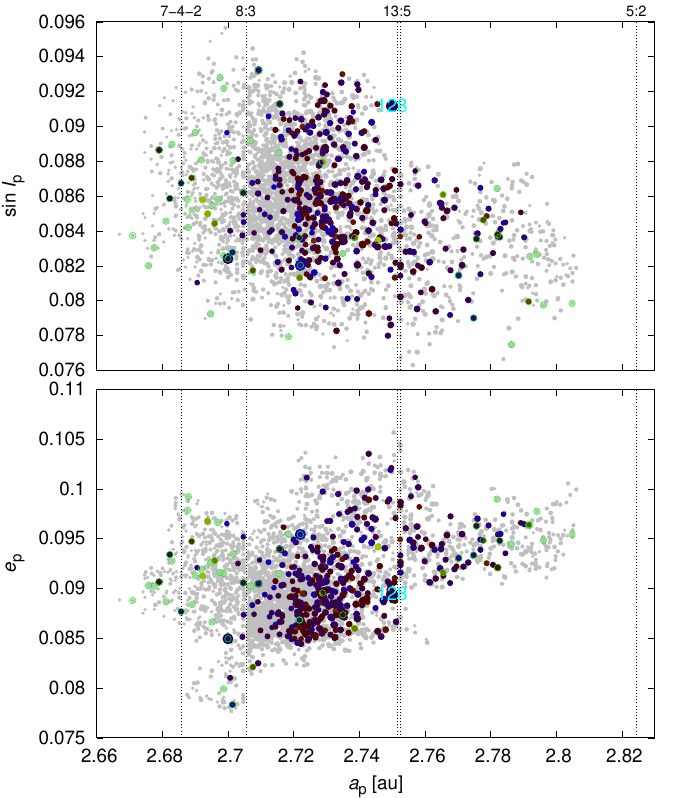} &
\\[0.1cm]
\kern0.5cm Kalliope (M) &
\kern0.5cm Lixiaohua (CI) &
\kern0.5cm Misa (CI) &
\\
\includegraphics[width=5.9cm]{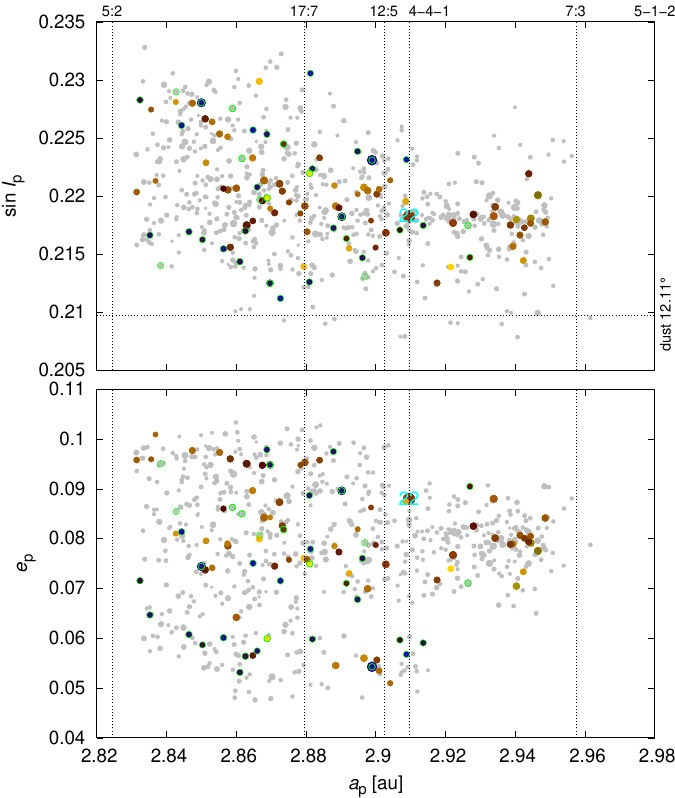} &
\includegraphics[width=5.9cm]{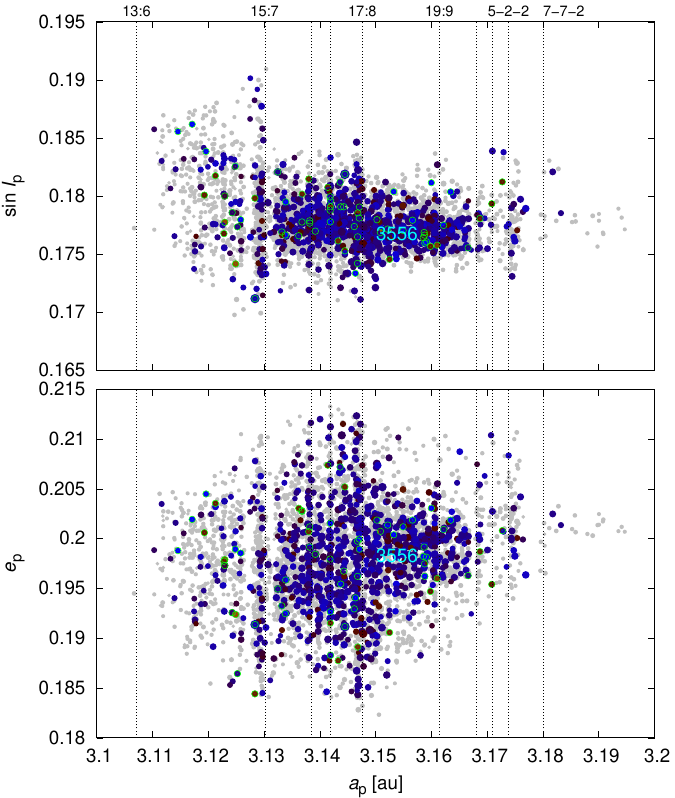} &
\includegraphics[width=5.9cm]{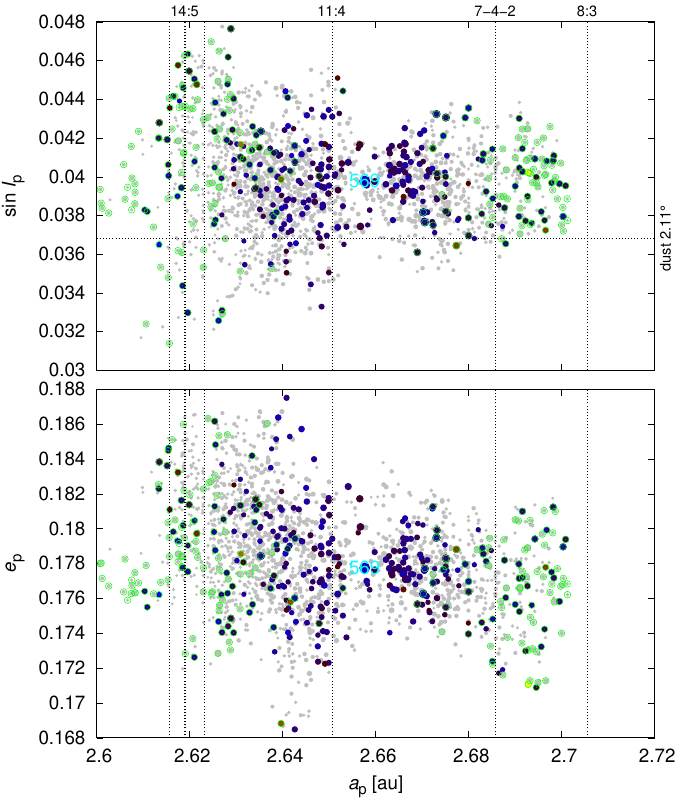} &
\\[0.1cm]
\end{tabular}
\caption{continued.}
\end{figure*}

\addtocounter{figure}{-1}
\begin{figure*}
\centering
\begin{tabular}{c@{\kern0.1cm}c@{\kern0.1cm}c@{}c}
\kern0.5cm Naema (CI) &
\kern0.5cm Padua (IDP) &
\kern0.5cm Pallas (B) &
\\
\includegraphics[width=5.9cm]{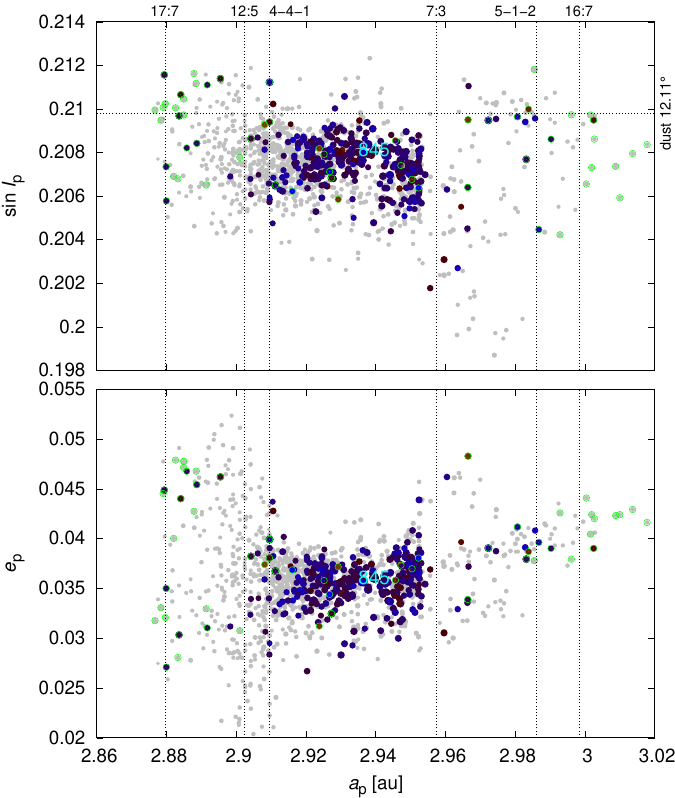} &
\includegraphics[width=5.9cm]{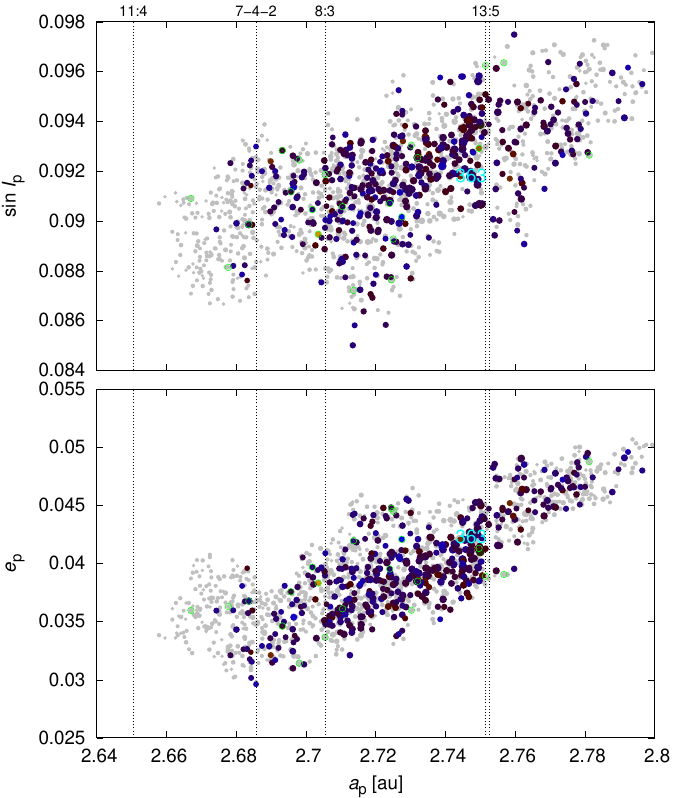} &
\includegraphics[width=5.9cm]{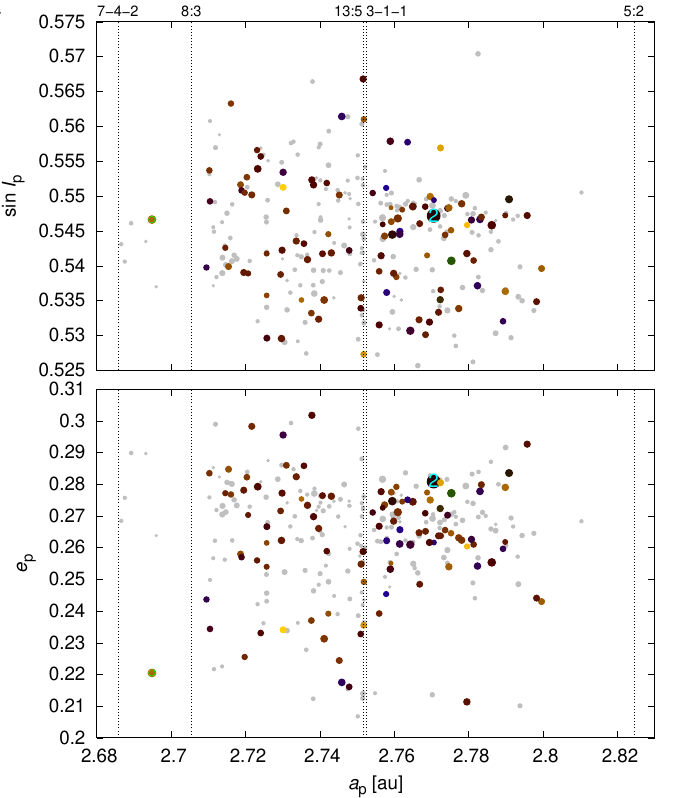} &
\\[0.1cm]
\kern0.5cm {\bf Polana (CI)} &
\kern0.5cm Sylvia (P) &
\kern0.5cm Tina (M) &
\\
\includegraphics[width=5.9cm]{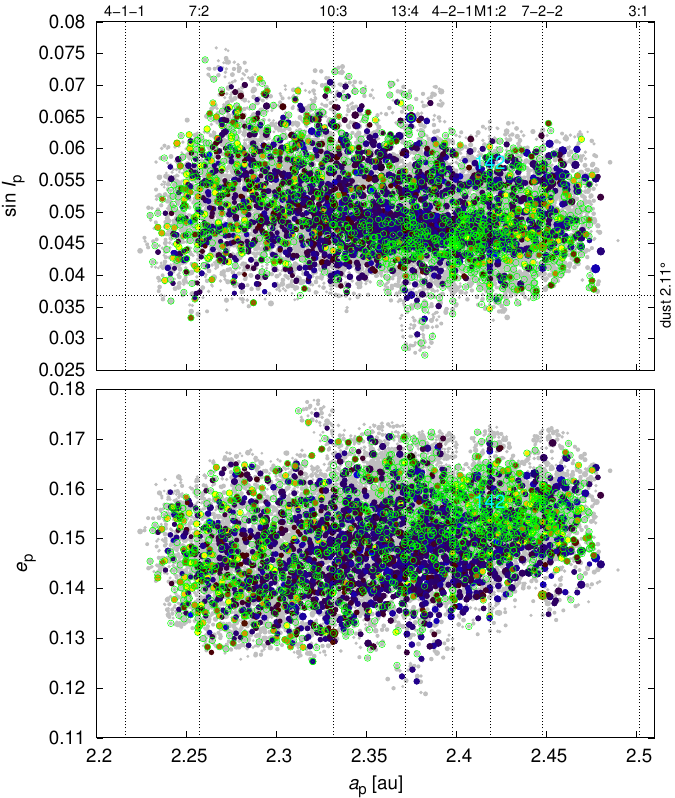} &
\includegraphics[width=5.9cm]{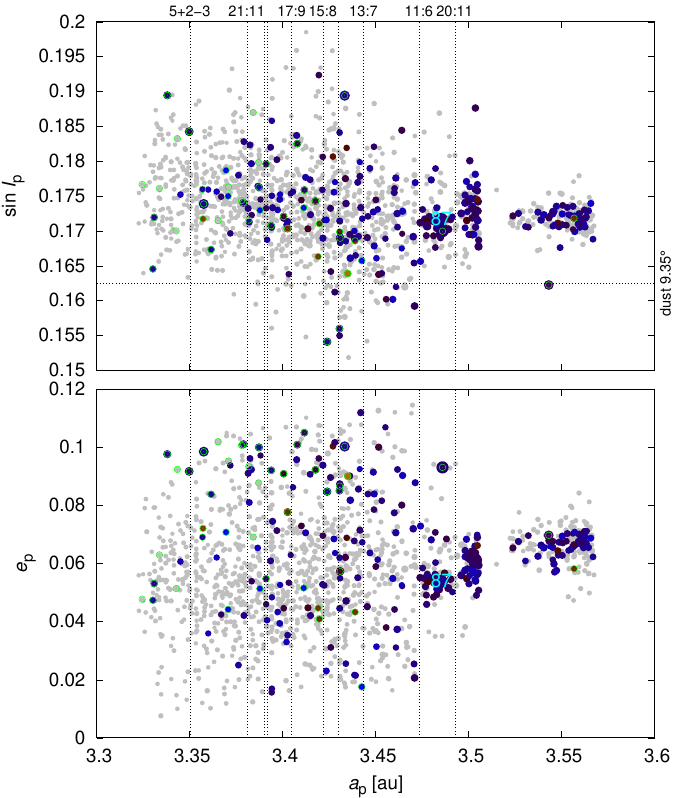} &
\includegraphics[width=5.9cm]{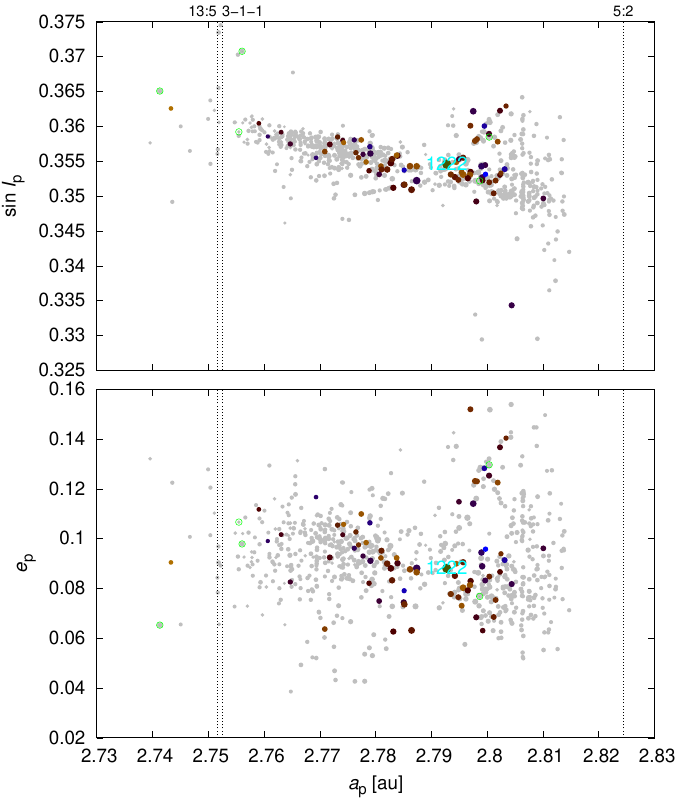} &
\\[0.1cm]
\kern0.5cm Themis (CI) &
\kern0.5cm Theobalda (CM) &
\kern0.5cm Ursula (CI) &
\\
\includegraphics[width=5.9cm]{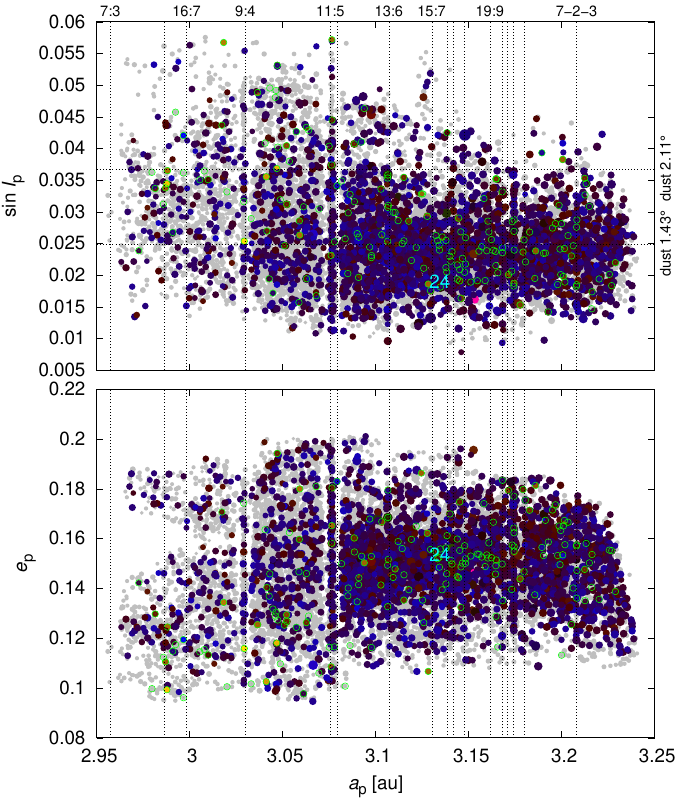} &
\includegraphics[width=5.9cm]{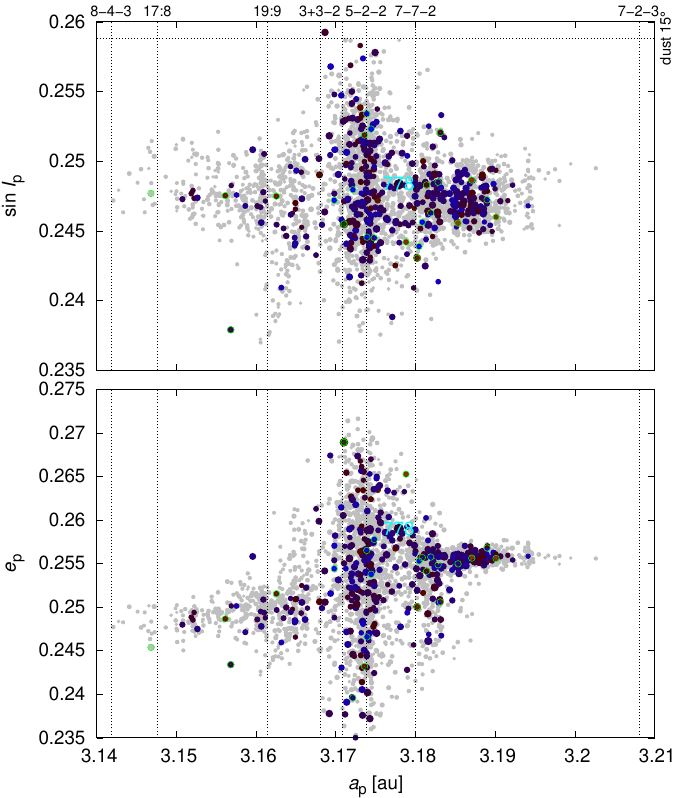} &
\includegraphics[width=5.9cm]{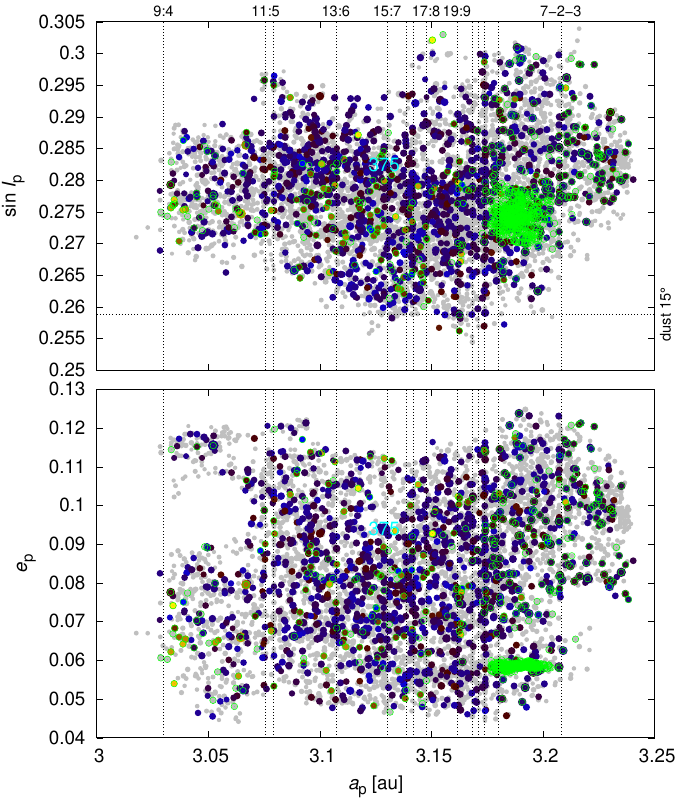} &
\\
\end{tabular}
\caption{continued.}
\end{figure*}

\addtocounter{figure}{-1}
\begin{figure*}
\centering
\begin{tabular}{c@{\kern0.1cm}c@{\kern0.1cm}c@{}c}
\kern0.5cm {\bf Veritas (CM)} &
\kern0.5cm Vibilia (CM) &
\kern0.5cm Watsonia (CO/CV/CK) &
\\
\includegraphics[width=5.9cm]{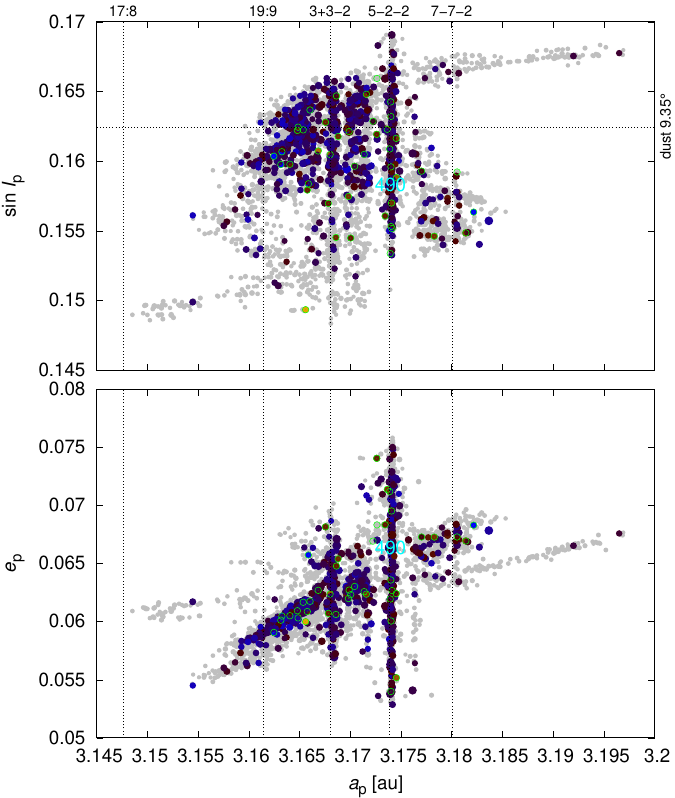} &
\includegraphics[width=5.9cm]{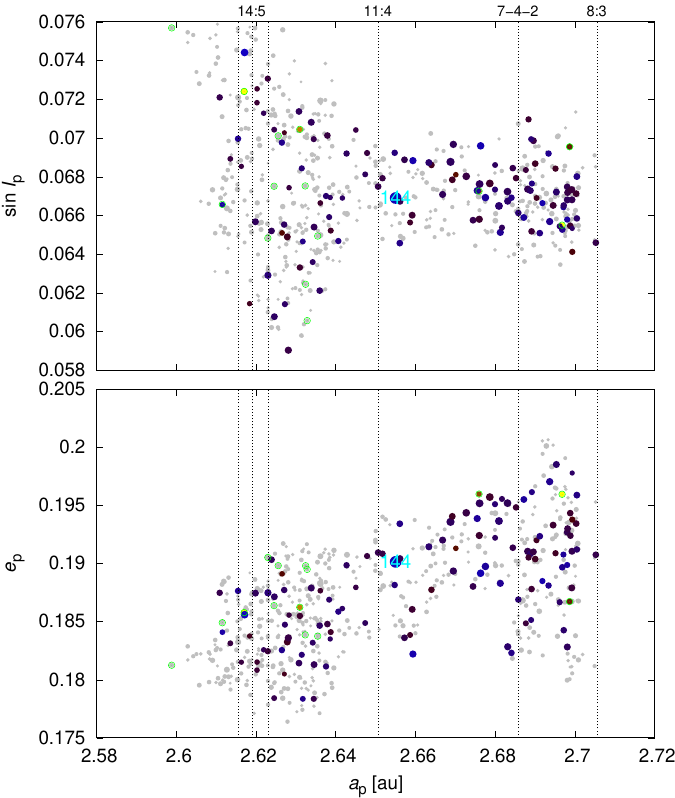} &
\includegraphics[width=5.9cm]{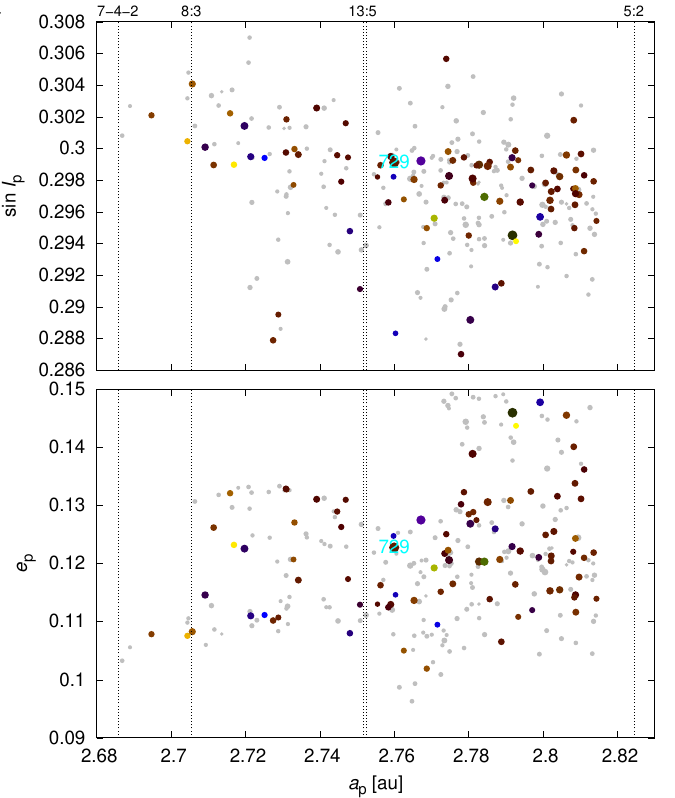} &
\\[0.1cm]
\kern0.5cm Witt (?) &
\\
\includegraphics[width=5.9cm]{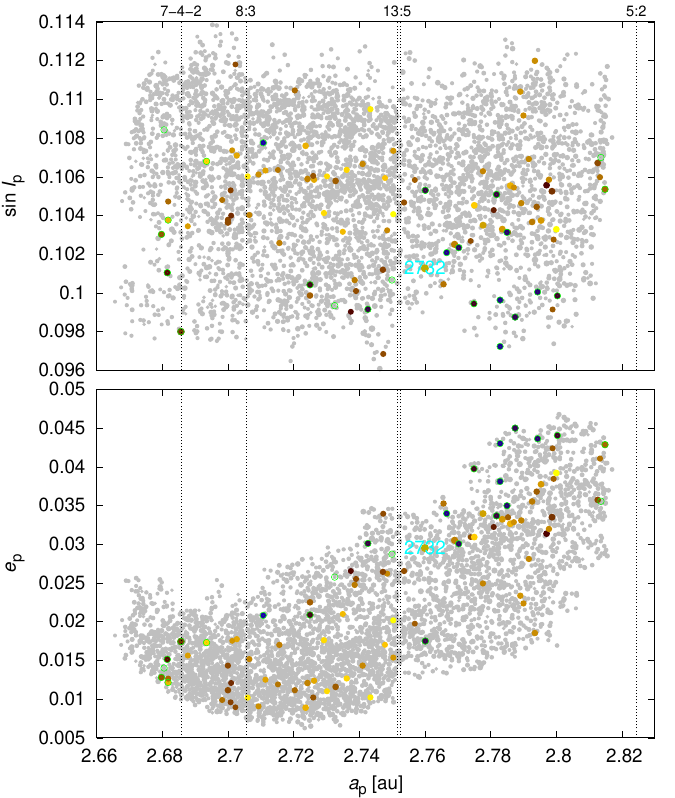} &
\\[0.1cm]
\end{tabular}
\caption{continued.}
\end{figure*}

\begin{figure*}
\begin{tabular}{c@{}c@{}c@{}c@{}c@{}c}
\kern0.0cm Adeona (CM) &
\kern0.0cm {\bf Aeolia (?)} &
\kern0.0cm A\"eria (M) &
\kern0.0cm Agnia (H) &
\kern0.0cm Alauda (CI) &
\\
\includegraphics[width=3.4cm]{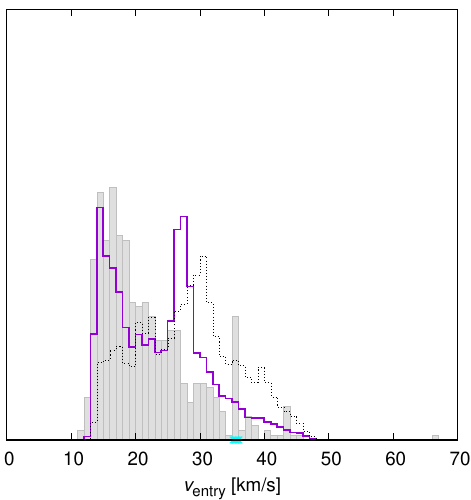} &
\includegraphics[width=3.4cm]{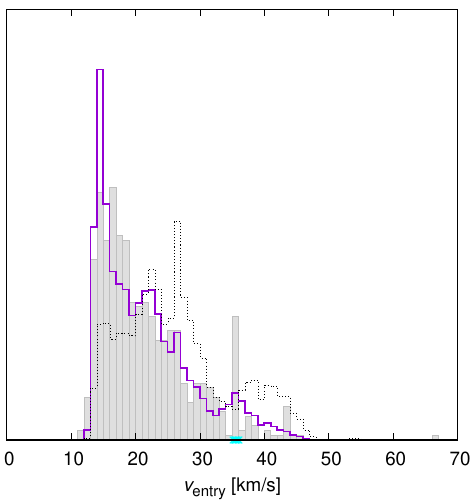} &
\includegraphics[width=3.4cm]{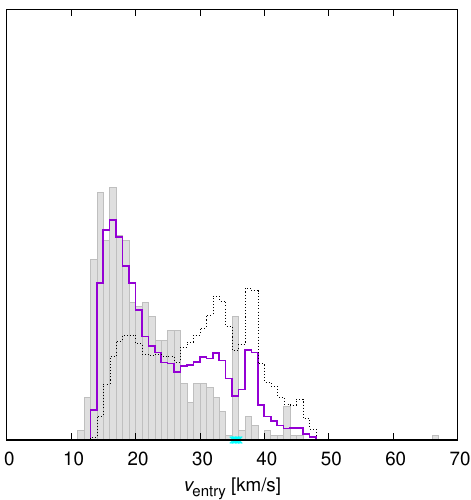} &
\includegraphics[width=3.4cm]{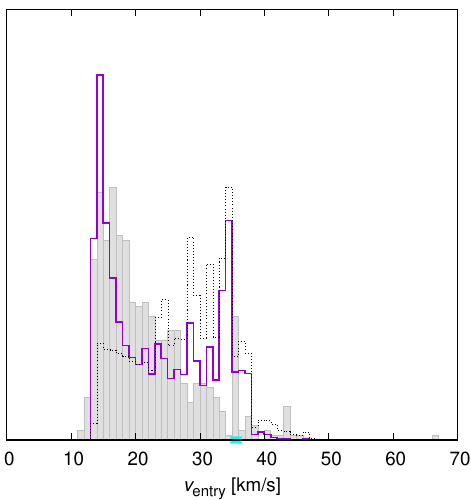} &
\includegraphics[width=3.4cm]{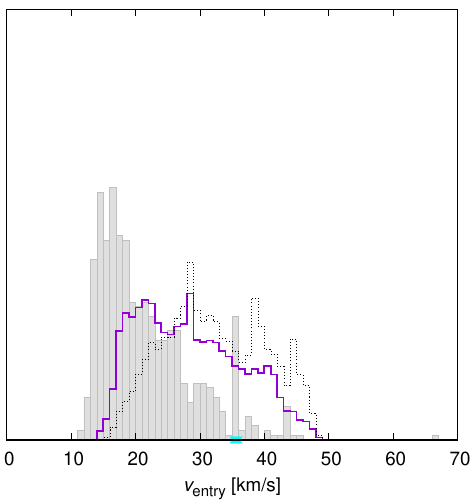} &
\\
\kern0.0cm Astrid (CM) &
\kern0.0cm {\bf Baptistina (?)} &
\kern0.0cm Beagle (CI) &
\kern0.0cm {\bf Brang\"ane (M)} &
\kern0.0cm Brasilia (M) &
\\
\includegraphics[width=3.4cm]{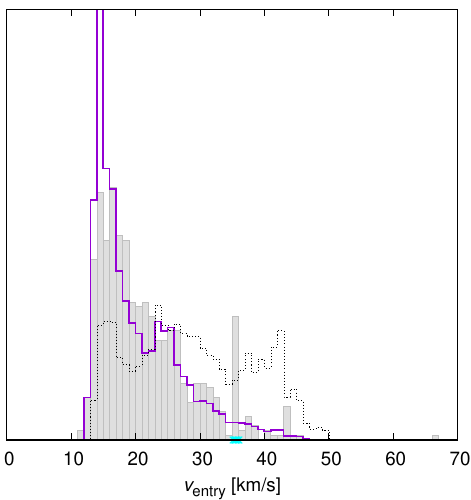} &
\includegraphics[width=3.4cm]{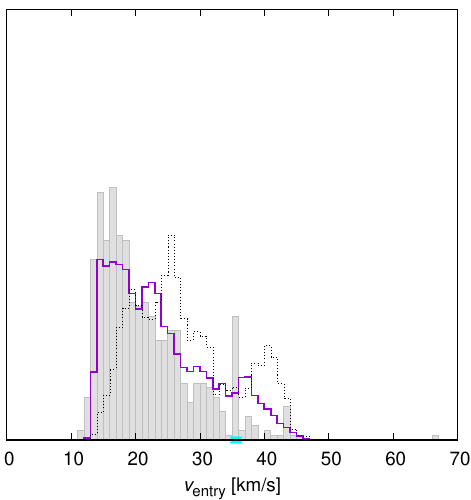} &
\includegraphics[width=3.4cm]{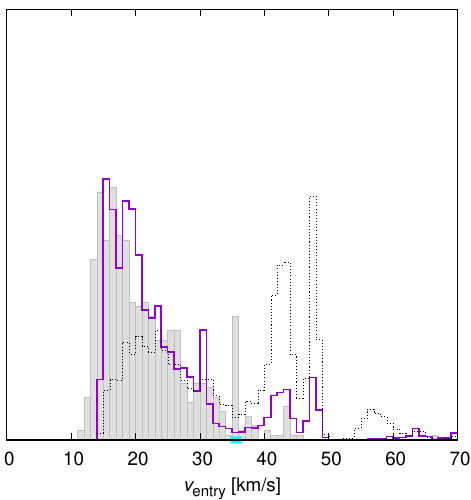} &
\includegraphics[width=3.4cm]{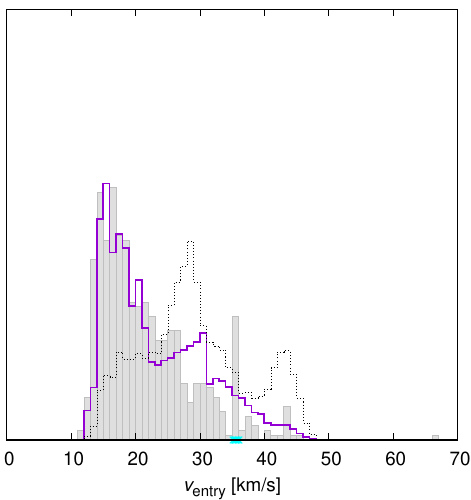} &
\includegraphics[width=3.4cm]{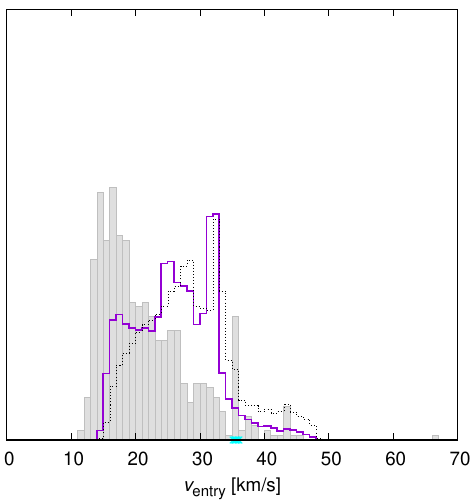} &
\\
\kern0.0cm Brucato (?) &
\kern0.0cm Chloris (CM) &
\kern0.0cm Clarissa (CI) &
\kern0.0cm Dora (CM) &
\kern0.0cm Elfriede (CI) &
\\
\includegraphics[width=3.4cm]{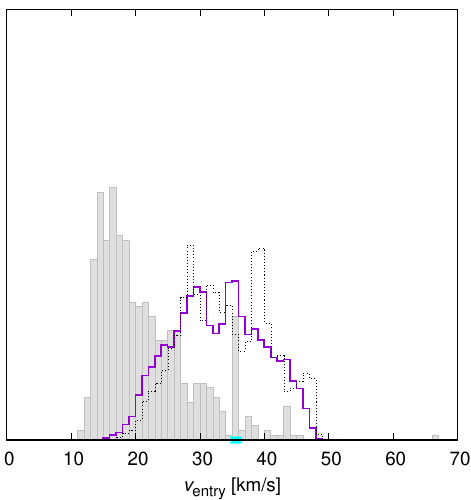} &
\includegraphics[width=3.4cm]{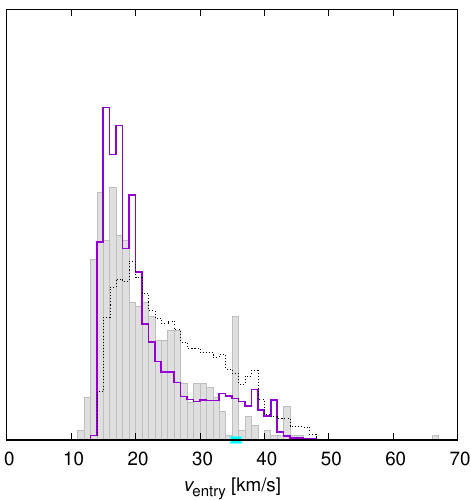} &
\includegraphics[width=3.4cm]{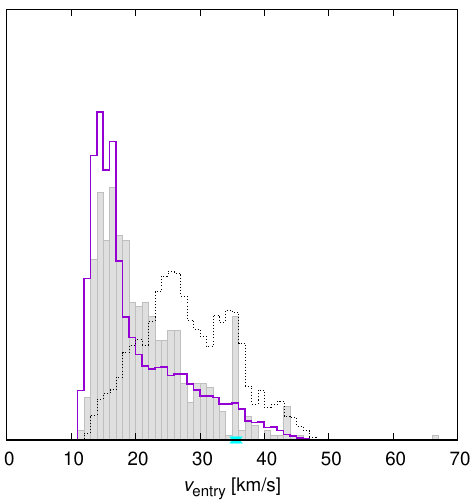} &
\includegraphics[width=3.4cm]{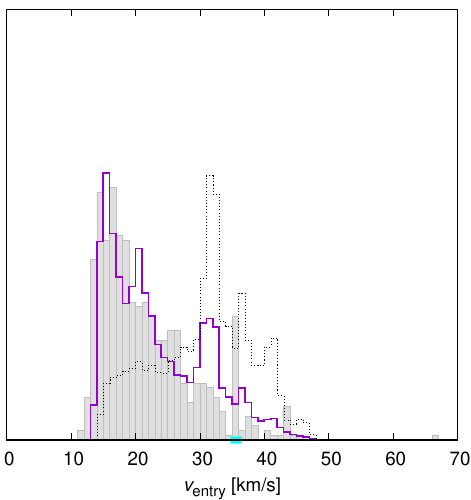} &
\includegraphics[width=3.4cm]{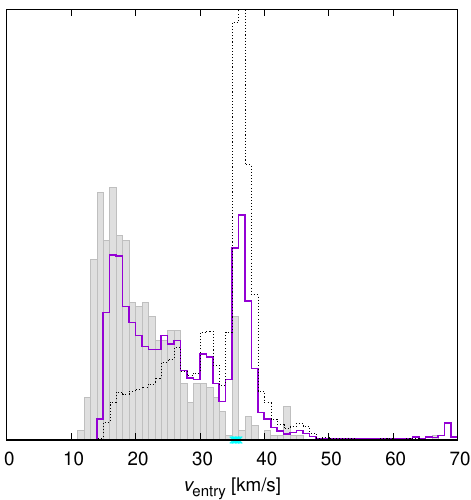} &
\\
\kern0.0cm Emma (IDP) &
\kern0.0cm Eos (CO/CV/CK) &
\kern0.0cm Erigone (CM) &
\kern0.0cm Eunomia (LL) &
\kern0.0cm {\bf Euphrosyne (CI)} &
\\
\includegraphics[width=3.4cm]{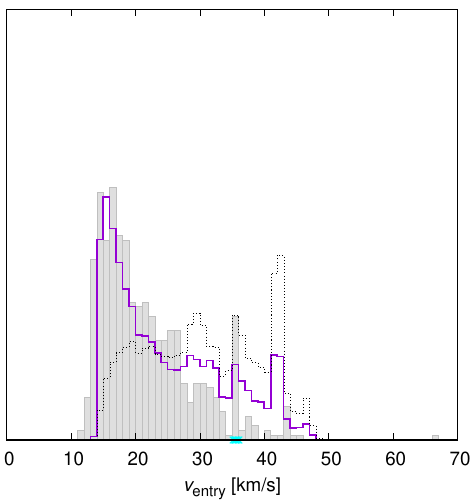} &
\includegraphics[width=3.4cm]{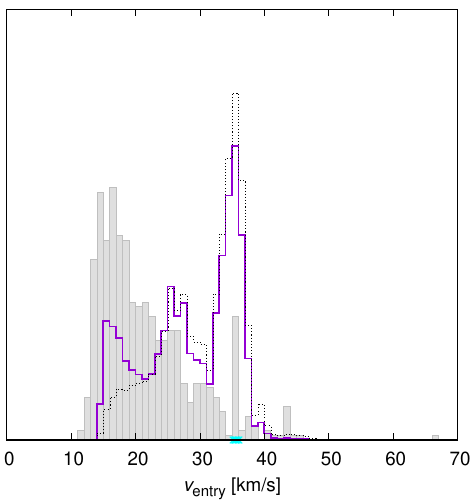} &
\includegraphics[width=3.4cm]{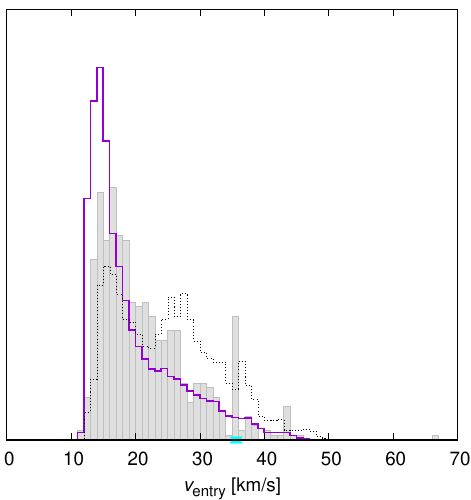} &
\includegraphics[width=3.4cm]{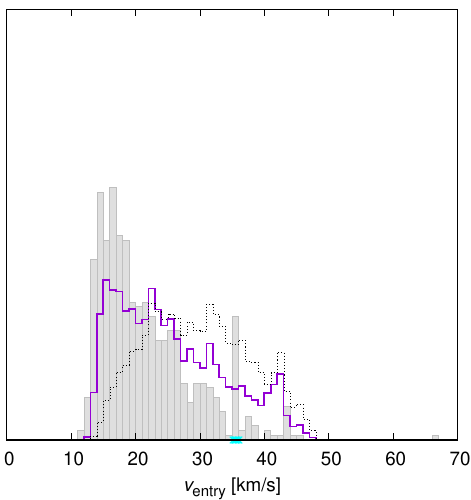} &
\includegraphics[width=3.4cm]{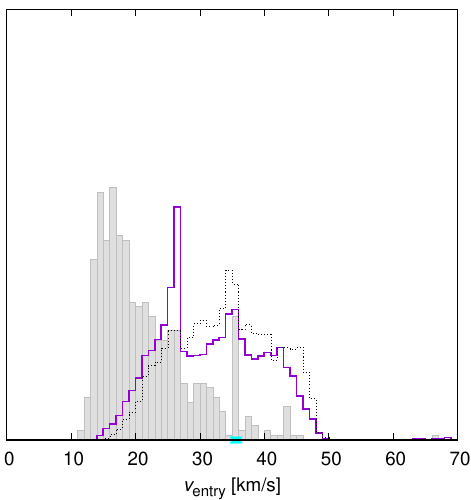} &
\\
\kern0.0cm Flora (LL) &
\kern0.0cm Gefion (L) &
\kern0.0cm Hoffmeister (CI) &
\kern0.0cm Hungaria (E) &
\kern0.0cm Hygiea (CI) &
\\
\includegraphics[width=3.4cm]{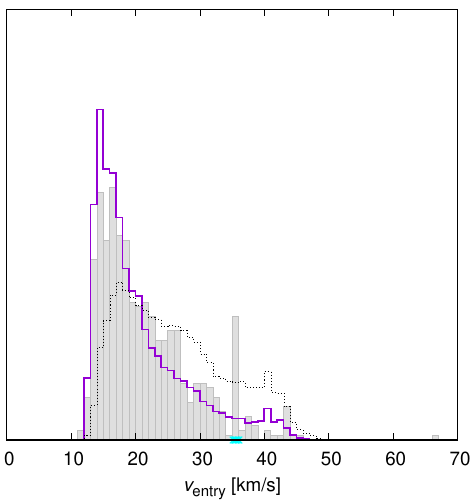} &
\includegraphics[width=3.4cm]{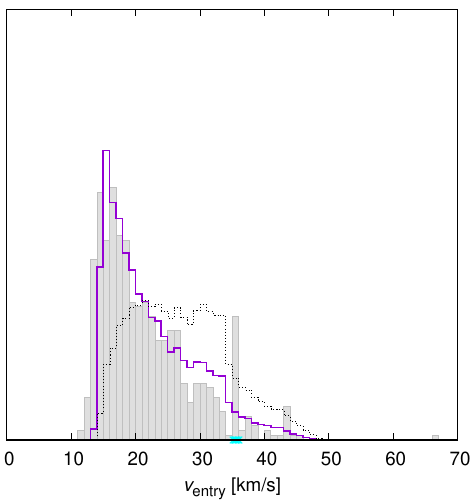} &
\includegraphics[width=3.4cm]{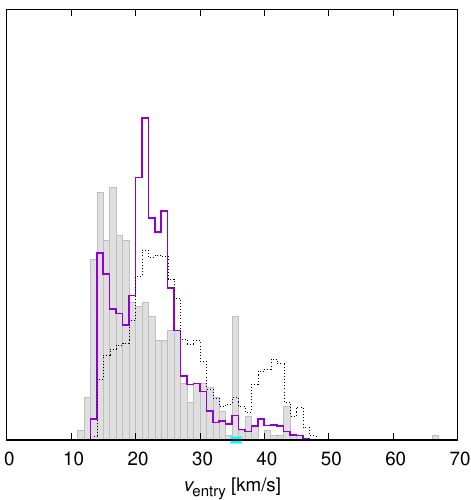} &
\includegraphics[width=3.4cm]{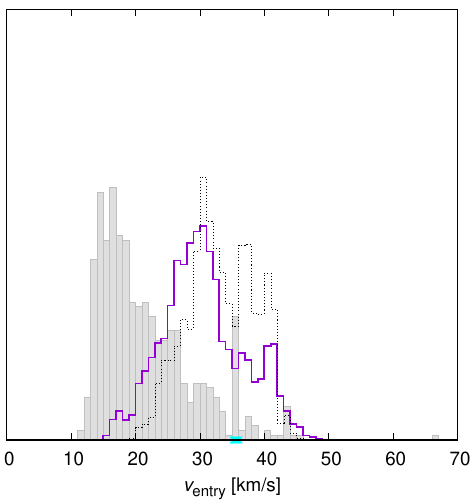} &
\includegraphics[width=3.4cm]{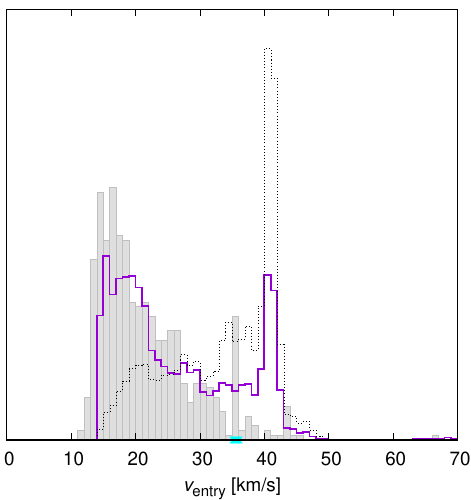} &
\\
\kern0.0cm Iannini (Aca/Lod) &
\kern0.0cm Juno (L) &
\kern0.0cm Kalliope (M) &
\kern0.0cm Karin (H) &
\kern0.0cm {\bf K\"onig (CM)} &
\\
\includegraphics[width=3.4cm]{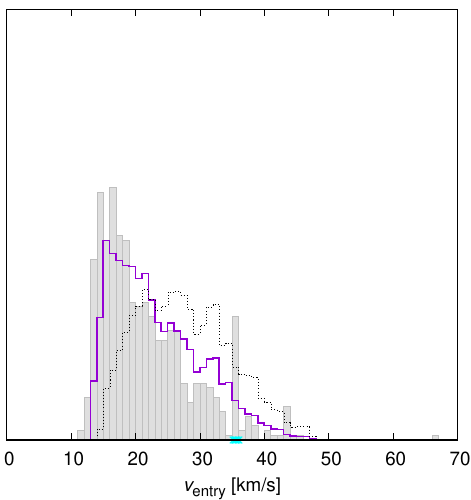} &
\includegraphics[width=3.4cm]{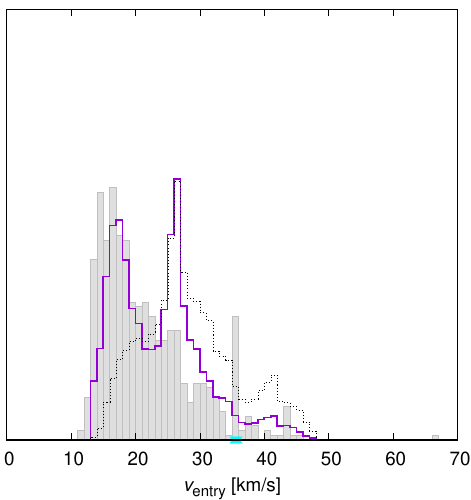} &
\includegraphics[width=3.4cm]{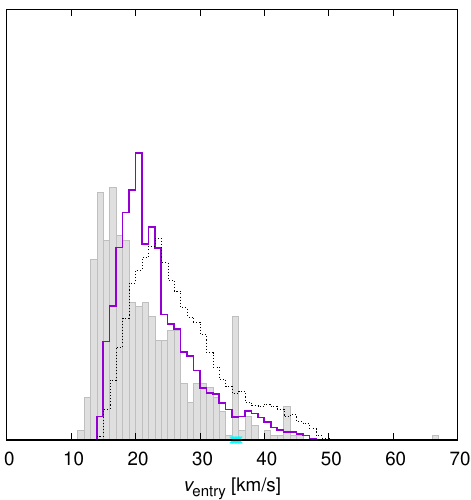} &
\includegraphics[width=3.4cm]{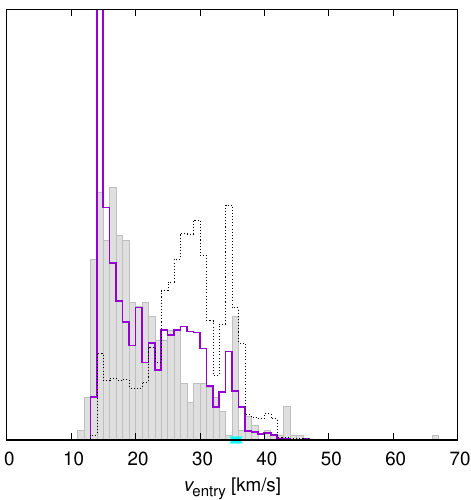} &
\includegraphics[width=3.4cm]{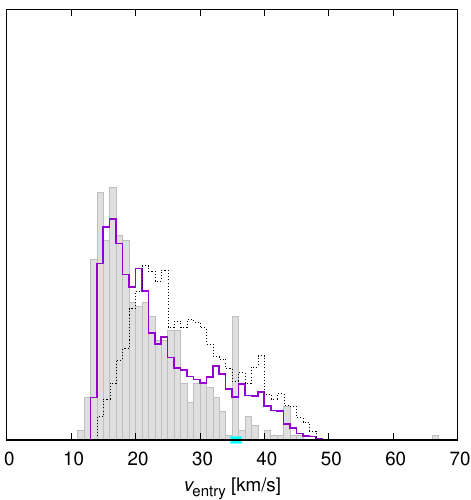} &
\\
\end{tabular}
\caption{
Same as Fig.~\ref{impvel_nooutliers} for individual families.
All histograms were renormalized; the respective weighting is discussed in Sec.~\ref{fripon}.
Since the numbers of particles were limited, not all peaks are statistically significant.
}
\label{impvel}
\end{figure*}

\addtocounter{figure}{-1}
\begin{figure*}
\begin{tabular}{c@{}c@{}c@{}c@{}c@{}c}
\kern0.0cm Koronis (H) &
\kern0.0cm Lixiauhoua (CI) &
\kern0.0cm Maria (H) &
\kern0.0cm Massalia (L) &
\kern0.0cm Merxia (H) &
\\
\includegraphics[width=3.4cm]{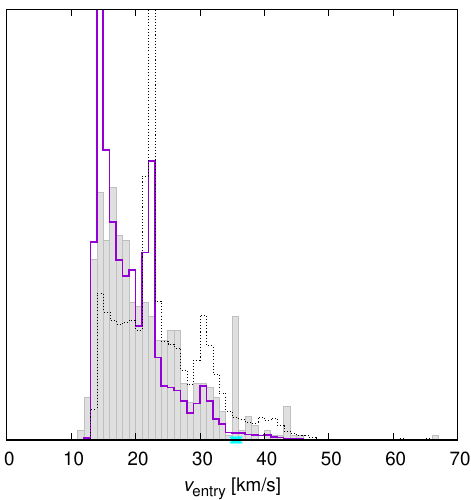} &
\includegraphics[width=3.4cm]{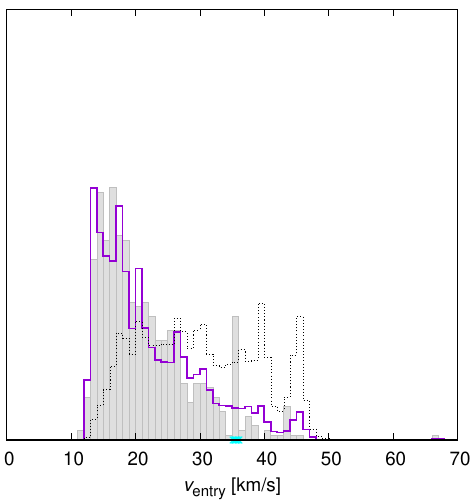} &
\includegraphics[width=3.4cm]{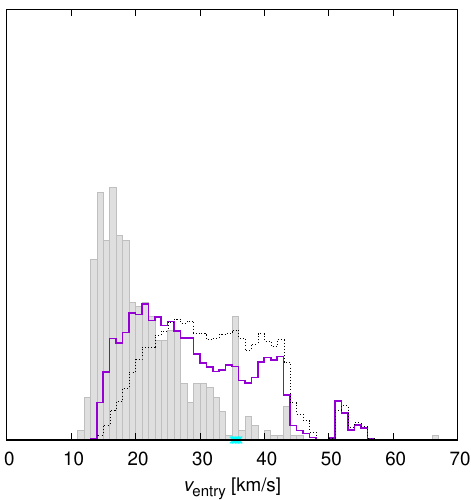} &
\includegraphics[width=3.4cm]{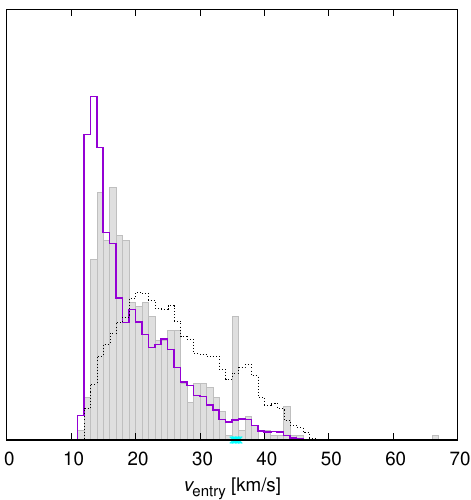} &
\includegraphics[width=3.4cm]{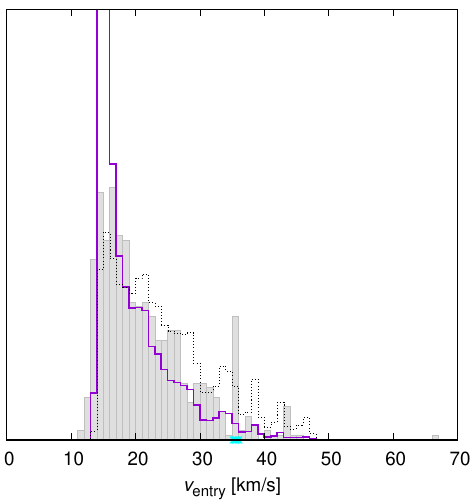} &
\\
\kern0.0cm Misa (CI) &
\kern0.0cm Naema (CI) &
\kern0.0cm Nemesis (CI) &
\kern0.0cm Nysa (LL) &
\kern0.0cm Padua (IDP) &
\\
\includegraphics[width=3.4cm]{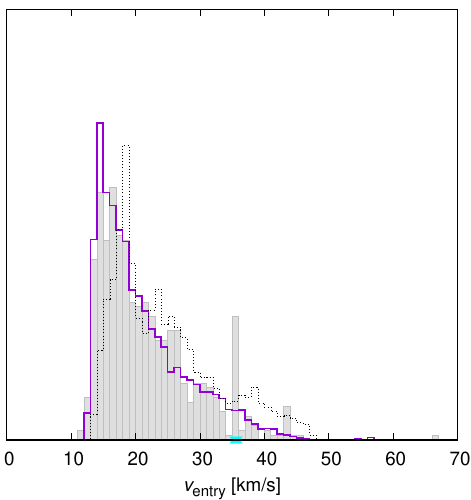} &
\includegraphics[width=3.4cm]{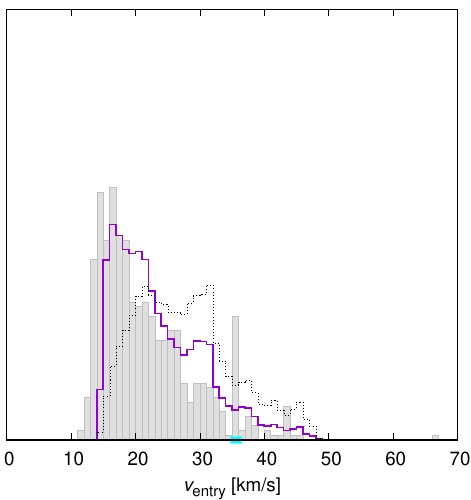} &
\includegraphics[width=3.4cm]{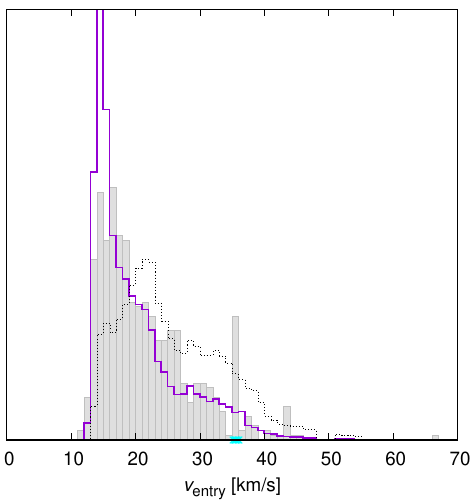} &
\includegraphics[width=3.4cm]{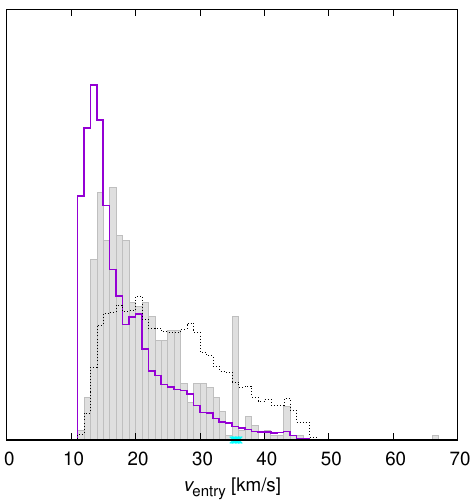} &
\includegraphics[width=3.4cm]{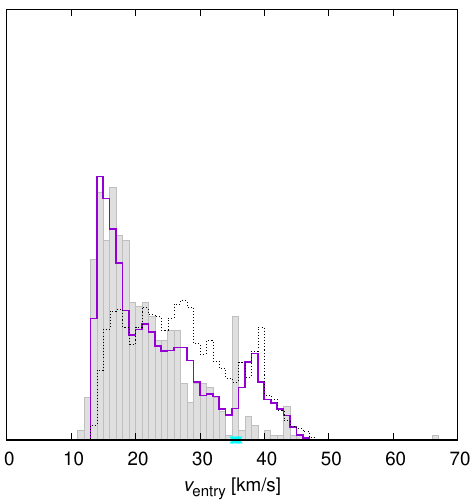} &
\\
\kern0.0cm Pallas (B) &
\kern0.0cm Phocaea (H) &
\kern0.0cm {\bf Polana (CI)} &
\kern0.0cm Sylvia (P) &
\kern0.0cm Themis (CI) &
\\
\includegraphics[width=3.4cm]{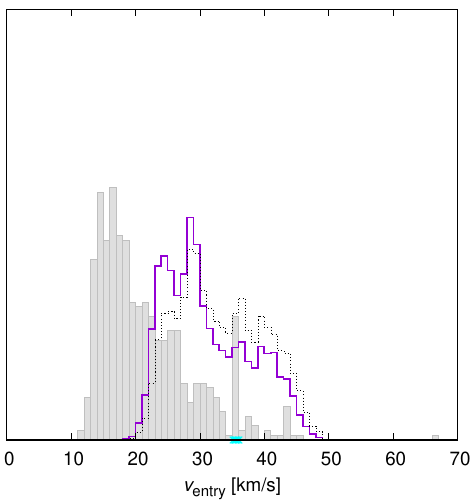} &
\includegraphics[width=3.4cm]{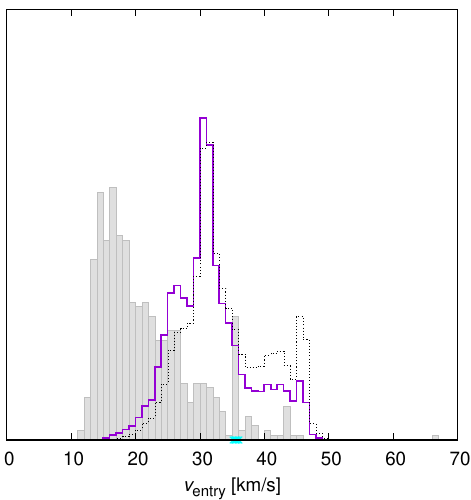} &
\includegraphics[width=3.4cm]{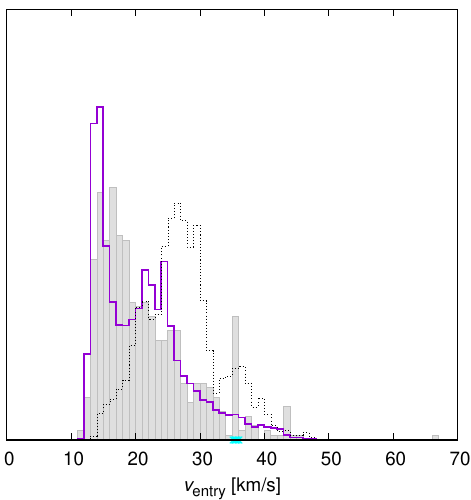} &
\includegraphics[width=3.4cm]{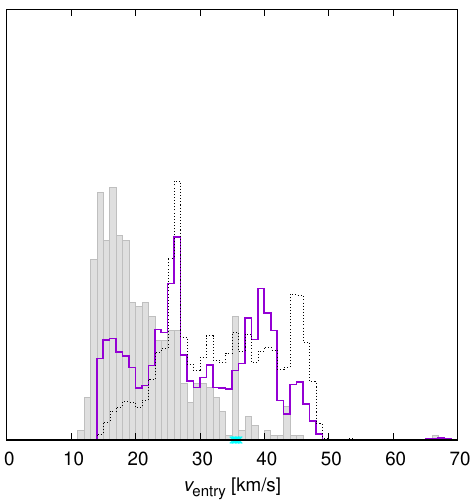} &
\includegraphics[width=3.4cm]{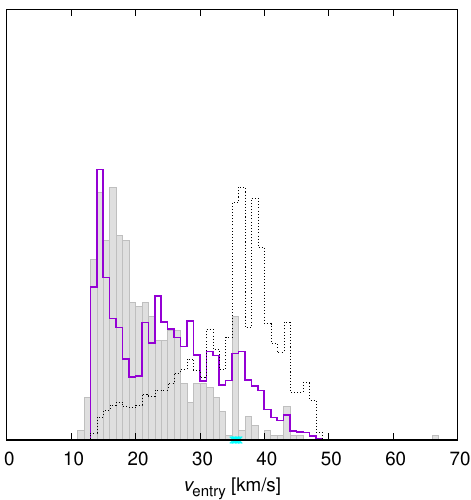} &
\\
\kern0.0cm Theobalda (CM) &
\kern0.0cm Ursula (CI) &
\kern0.0cm {\bf Veritas (CM)} &
\kern0.0cm Vesta (HED) &
\kern0.0cm Vibilia (CM) &
\\
\includegraphics[width=3.4cm]{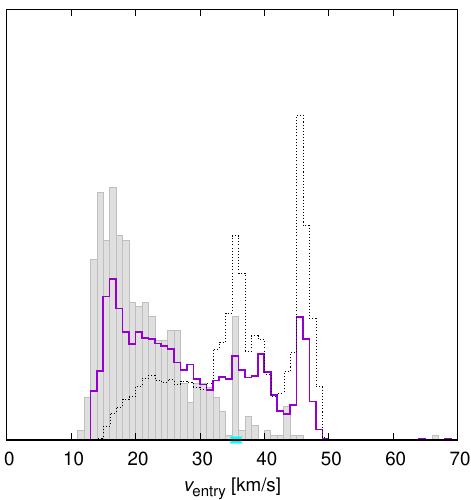} &
\includegraphics[width=3.4cm]{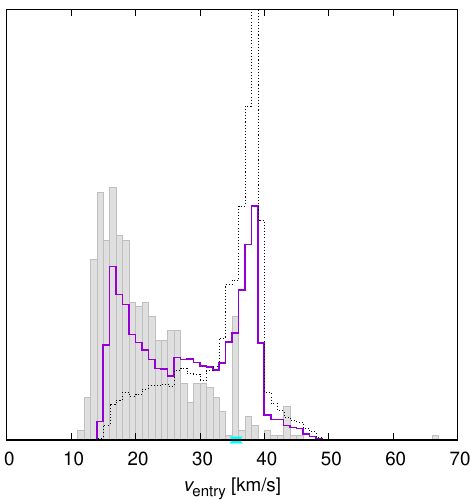} &
\includegraphics[width=3.4cm]{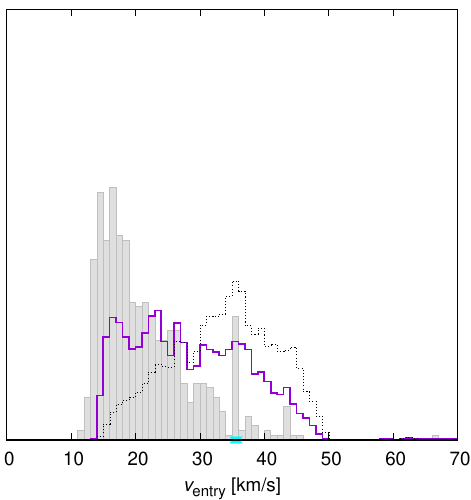} &
\includegraphics[width=3.4cm]{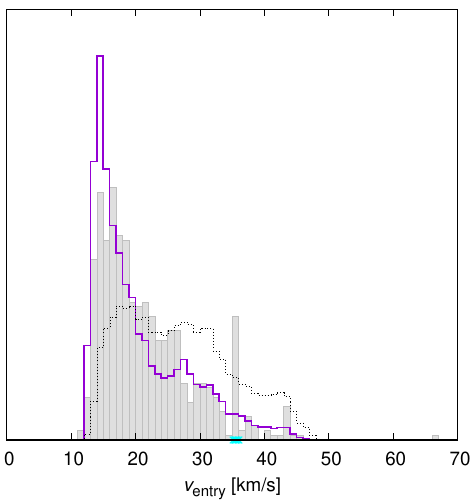} &
\includegraphics[width=3.4cm]{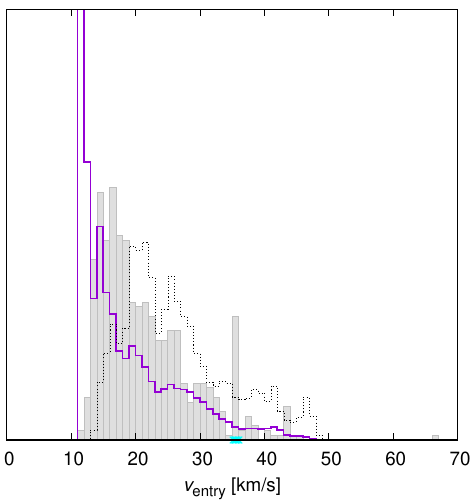} &
\\
\kern0.0cm Watsonia (CO/CV/CK) &
\kern0.0cm Phaethon (CY?) &
\\
\includegraphics[width=3.4cm]{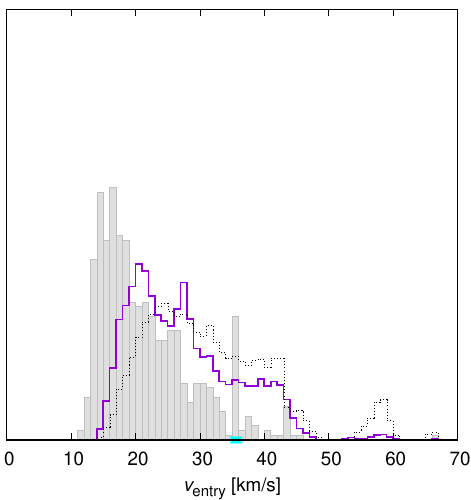} &
\includegraphics[width=3.4cm]{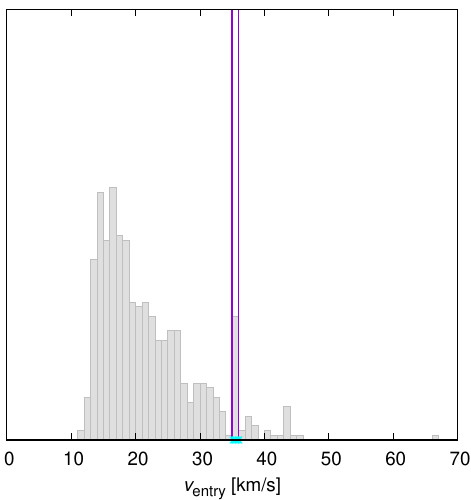} &
\end{tabular}
\caption{
continued.
}
\end{figure*}


\clearpage

\section{Supplementary tables}

We report properties of the observed SFDs in Tab.~\ref{tab:q},
parameters of synthetic families in Tab.~\ref{tab:synthetic},
and intrinsic collisional probabilities in
Tabs.~\ref{tabc1},
\ref{tabc2}.

\begin{table}[h]
\caption{Power-law slopes of the observed SFDs of the C-type families.}
\label{tab:q}
\centering
\begin{tabular}{lrrr}
\vrule width 0pt depth 4pt
family & $q_1$ & $q_2$ & $q_3$ \\
\hline
\vrule width 0pt height 9pt
Adeona (CM)             & $-2.8$ & $-1.7$ & $    $ \\
{\bf Aeolia (?)}        & $-4.0$ & $-3.3$!& $    $ \\
A\"eria (M)             & $-3.6$ & $-2.6$ & $    $ \\
Alauda (CI)             & $-2.8$ & $-2.2$ & $-1.3$ \\
Astrid (CM)             & $-2.8$ & $-3.3$!& $    $ \\
{\bf Baptistina (?)}    & $-3.4$ & $-2.4$ & $-1.8$ \\
Beagle (CI)             & $-3.0$!&        & $    $ \\
{\bf Brang\"ane (M)}    & $-3.9$ & $-2.8$!& $    $ \\
Brasilia (M)            & $-3.3$ & $-3.1$!& $    $ \\
Brucato (?)             & $-3.7$ & $-1.5$ & $    $ \\
Chloris (CM)            & $-2.7$ & $-1.7$ & $    $ \\
Clarissa (CI)           & $-3.8$ & $-3.0$!& $    $ \\
Dora (CM)               & $-3.8$ & $-2.1$ & $-1.6$ \\
Elfriede (CI)           & $-5.4$ & $-3.9$!& $    $ \\
Emma (IDP)              & $-3.2$ & $-2.1$ & $    $ \\
Eos (CO/CV/CK)          & $-3.3$ & $-2.3$ & $-1.9$ \\
Erigone (CM)            & $-3.6$ & $-1.9$ & $    $ \\
{\bf Euphrosyne (CI)}   & $-4.4$ & $-2.4$ & $    $ \\
Hoffmeister (CI)        & $-2.9$ & $-2.8$!& $    $ \\
Hygiea (CI)             & $-3.7$ & $-1.8$ & $    $ \\
Iannini (Aca/Lod)       & $-3.2$!&        & $    $ \\
{\bf K\"onig (CM)}      & $-3.4$ & $-2.6$ & $    $ \\
Lixiaohua (CI)          & $-3.6$ & $-2.4$ & $    $ \\
Misa (CI)               & $-4.1$ & $-2.8$!& $    $ \\
Naema (CI)              & $-4.5$ & $-2.1$ & $    $ \\
Nemesis (CI)            & $-4.3$ & $-3.1$ & $-2.6$ \\
Padua (IDP)             & $-3.2$ & $-2.2$ & $-1.4$ \\
Pallas (B)              & $-2.8$ & $-1.0$ & $    $ \\
{\bf Polana (CI)}       & $-2.7$ & $-2.3$ & $-1.8$ \\
Sylvia (P)              & $-3.7$ & $-3.2$!&        \\
Themis (CI)             & $-2.5$ & $-1.3$ & $-1.7$ \\
Theobalda (CM)          & $-3.9$ & $-3.0$!& $    $ \\
Tina (M)                & $-4.8$ & $-3.3$ & $    $ \\
Ursula (CI)             & $-4.8$ & $-3.5$ & $-2.3$ \\
{\bf Veritas (CM)}      & $-6.7$ & $-2.7$!& $    $ \\
Vesta (HED)             & $-4.6$ & $-3.3$ & $-1.5$ \\
Vibilia (CM)            & $-3.0$ & $-1.6$ & $    $ \\
Watsonia (CO/CV/CK)     & $-1.3$ & $-2.4$ & $-1.1$ \\
Witt (?)                & $-3.4$ & $-2.9$!& $    $ \\
\hline
\end{tabular}
\end{table}

\begin{table}
\caption{
Parameters of synthetic families.
}
\label{tab:synthetic}
\begin{tabular}{lrrrrrr}
family & $f$ & $\omega$ & $v_5$ & $\alpha$ & $D_{\rm pb}$ & $\rho$ \\
-- & deg & deg & ${\rm m}\,{\rm s}^{-1}$ & 1 & km & ${\rm g}\,{\rm cm}^{-3}$ \\
\hline
\vrule width 0pt height 9pt
{\bf Aeolia (CO/CV/CK)} & 105 &  15 &  5   & $-0.5$ &  48 & 1.7 \\
A\"eria (M)             &  90 &  15 & 25   & $-0.5$ &  70 & 3.5 \\
Astrid (CM)             &  90 &  30 & 15   & $-0.5$ &  63 & 1.7 \\
Beagle (CI)             & 105 &   0 &  8   & $-0.5$ &  38 & 1.3 \\
{\bf Brang\"ane (M)}    & 105 &   0 &  6   & $-0.5$ &  42 & 3.1 \\
Brasilia (M)            &  75 & 105 & 20   & $-0.5$ &  49 & 3.5 \\
Brucato (?)             &  90 &  90 & 50   & $-0.5$ &  90 & 1.7 \\
Chloris (CM)            &  90 &  30 & 25   & $-0.5$ & 118 & 1.7 \\
Clarissa (CI)           &  90 &  30 &  7   & $-0.5$ &  40 & 1.7 \\
Elfriede (CI)           &  90 &  90 & 15   & $-0.5$ & 122 & 1.7 \\
Emma (IDP)              &  75 &  45 & 15   & $-0.5$ & 128 & 1.7 \\
Eos (CO/CV/CK)          & 150 &  30 & 93   & $-0.5$ & 280 & 3.1 \\
Erigone (CM)            &  90 &  90 & 20   & $-0.5$ &  92 & 1.7 \\
Hoffmeister (CI)        &  90 &  15 & 40   & $-0.5$ &  99 & 1.7 \\
Iannini (Aca/Lod)       & 120 & 135 &  4.5 & $-0.5$ &  23 & 2.5!\\
Kalliope (M)            & 100 & 330 & 24   & $-1.0$ & 116 & 4.4 \\
{\bf K\"onig (CM)}      &  85 &   0 & 10   & $-0.5$ &  50 & 1.7 \\
Lixiaohua (CI)          &  90 &  30 & 30   & $-0.5$ & 116 & 1.7 \\
Misa (CI)               &  90 &  90 & 20   & $-0.5$ &  72 & 1.7 \\
Naema (CI)              &  90 &  30 & 20   & $-0.5$ &  90 & 1.7 \\
Nemesis (CI)            &  90 &  90 & 20   & $-0.5$ &  64 & 3.5 \\
Padua (IDP)             &  60 &  15 & 20   & $-0.5$ & 124 & 1.7 \\
Pallas (B)              & 135 &   0 & 20   & $-0.5$ & 512 & 2.9 \\
{\bf Polana (CI)}       &  75 &  60 & 40   & $-0.5$ & 100 & 1.7 \\
Theobalda (CM)          &  45 &   0 & 20   & $-0.5$ &  79 & 1.7 \\
Tina                    & 180 &  60 & 12   & $-0.5$ &  37 & 2.5!\\
{\bf Veritas (CM)}      &  20 & 160 & 10   & $-0.5$ & 126 & 2.5!\\
Vibilia (CM)            &  75 &   0 & 25   & $-0.5$ & 150 & 1.7 \\
\hline
\end{tabular}
\tablefoot{
$f$ denotes the true anomaly,
$\omega$, the argument of pericentre,
$v_5$, the speed at $D = 5\,{\rm km}$,
$\alpha$, the slope of $v(D)$,
$D_{\rm pb}$, the parent body size,
$\rho$, the bulk density.
}
\end{table}

\begin{table}
\caption{Intrinsic collisional probability and the mean collisional velocity for various main belt populations.}
\label{tabc1}
\centering
\begin{tabular}{lrr}
\vrule width 0pt depth 4pt
populations & $p$ & $v$ \\
--          & $10^{-18}\,{\rm km}^{-2}\,{\rm y}^{-1}$ & ${\rm km}\,{\rm s}^{-1}$ \\
\hline
\vrule width 0pt height 9pt
MB--MB                    &  2.971 &  5.741 \\
MB--Adeona                &  3.318 &  5.709 \\
MB--Aegle                 &  2.286 &  6.597 \\
MB--Aeolia                &  4.242 &  5.072 \\
MB--A\"eria               &  3.296 &  5.387 \\
MB--Alauda                &  2.115 &  7.665 \\
MB--Astrid                &  4.707 &  4.388 \\
MB--Baptistina            &  2.934 &  5.578 \\
MB--Beagle                &  3.397 &  4.786 \\
MB--Brang\"ane            &  3.489 &  5.570 \\
MB--Brasilia              &  3.017 &  6.016 \\
MB--Brucato               &  2.469 &  9.744 \\
MB--Clarissa              &  3.556 &  5.082 \\
MB--Chloris               &  3.240 &  6.151 \\
MB--Dora                  &  3.460 &  5.482 \\
MB--Elfriede              &  2.293 &  5.753 \\
MB--Emma                  &  2.999 &  4.934 \\
MB--Eos                   &  3.144 &  4.905 \\
MB--Erigone               &  3.824 &  5.614 \\
MB--Euphrosyne            &  1.823 &  9.157 \\
MB--Hoffmeister           &  4.130 &  4.482 \\
MB--Hungaria              &  0.750 &  9.209 \\
MB--Hygiea                &  2.945 &  4.688 \\
MB--Iannini               &  3.053 &  6.591 \\
MB--Kalliope              &  3.320 &  5.409 \\
MB--K\"onig               &  3.497 &  5.275 \\
MB--Lixiaohua             &  2.470 &  5.651 \\
MB--Misa                  &  4.619 &  5.167 \\
MB--Naema                 &  3.464 &  5.177 \\
MB--Nemesis               &  3.980 &  4.665 \\
MB--Padua                 &  3.944 &  4.577 \\
MB--Pallas                &  2.035 & 11.321 \\
MB--Polana                &  4.176 &  5.167 \\
MB--Sylvia                &  1.440 &  4.660 \\
MB--Themis                &  3.531 &  4.764 \\
MB--Theobalda             &  2.016 &  6.816 \\
MB--Tina                  &  2.463 &  7.834 \\
MB--Ursula                &  2.444 &  5.929 \\
MB--Veritas               &  2.623 &  4.669 \\
MB--Vibilia               &  4.261 &  5.258 \\
MB--Watsonia              &  2.953 &  6.614 \\
MB--Witt                  &  3.860 &  4.610 \\
\hline
\end{tabular}
\end{table}

\addtocounter{table}{-1}
\begin{table}
\caption{continued.}
\centering
\begin{tabular}{lrr}
\vrule width 0pt depth 4pt
populations & $p$ & $v$ \\
--          & $10^{-18}\,{\rm km}^{-2}\,{\rm y}^{-1}$ & ${\rm km}\,{\rm s}^{-1}$ \\
\hline
\vrule width 0pt height 9pt
Adeona--Adeona            &  5.978 &  5.512 \\
Aegle--Aegle              &  3.050 &  6.631 \\
Aeolia--Aeolia            &  9.911 &  4.072 \\
A\"eria--A\"eria          &  9.605 &  4.467 \\
Alauda--Alauda            &  2.816 &  7.853 \\
Astrid--Astrid            & 23.313 &  1.448 \\
Baptistina--Baptistina    & 14.429 &  4.349 \\
Beagle--Beagle            & 10.644 &  3.393 \\
Brasilia--Brasilia        &  4.877 &  5.797 \\
Brang\"ane--Brang\"ane    &  6.961 &  5.322 \\
Brucato--Brucato          &  4.198 & 10.964 \\
Clarissa--Clarissa        & 18.207 &  2.995 \\
Chloris--Chloris          &  4.993 &  6.505 \\
Dora--Dora                &  5.700 &  5.087 \\
Elfriede--Elfriede        &  5.423 &  5.580 \\
Emma--Emma                &  4.591 &  3.904 \\
Eos--Eos                  &  5.274 &  3.876 \\
Erigone--Erigone          & 12.913 &  5.400 \\
Euphrosyne--Euphrosyne    &  2.072 &  9.646 \\
Hoffmeister--Hoffmeister  & 11.422 &  2.086 \\
Hungaria--Hungaria        & 23.695 &  9.850 \\
Hygiea--Hygiea            &  4.921 &  3.234 \\
Iannini--Iannini          &  4.976 &  7.481 \\
Kalliope--Kalliope        &  5.757 &  4.771 \\
K\"onig--K\"onig          &  8.079 &  4.470 \\
Lixiaohua--Lixiaohua      &  3.263 &  5.207 \\
Misa--Misa                & 14.373 &  4.382 \\
Naema--Naema              &  8.624 &  4.395 \\
Nemesis--Nemesis          &  9.297 &  2.821 \\
Padua--Padua              & 11.946 &  2.287 \\
Pallas--Pallas            &  2.638 & 13.249 \\
Polana--Polana            & 17.305 &  3.989 \\
Sylvia--Sylvia            &  3.703 &  3.522 \\
Themis--Themis            & 10.256 &  3.373 \\
Theobalda--Theobalda      &  2.454 &  6.940 \\
Tina--Tina                &  3.267 &  8.538 \\
Ursula--Ursula            &  3.531 &  4.764 \\
Veritas--Veritas          &  5.501 &  3.409 \\
Vibilia--Vibilia          & 10.020 &  4.608 \\
Watsonia--Watsonia        &  4.980 &  6.628 \\
Witt--Witt                & 12.690 &  2.454 \\
\hline
\end{tabular}
\end{table}

\begin{table}
\caption{Same as Tab.~\ref{tabc1} for the Earth and meteoroids in the NEO space.}
\label{tabc2}
\centering
\begin{tabular}{lrr}
\vrule width 0pt depth 4pt
populations & $p$ & $v$ \\
-- & $10^{-18}\,{\rm km}^{-2}\,{\rm y}^{-1}$ & ${\rm km}\,{\rm s}^{-1}$ \\
\hline
\vrule width 0pt height 9pt
Earth--Adeona      & 24.907 & 28.822 \\
Earth--Aeolia      & 33.466 & 27.246 \\
Earth--A\"eria     & 18.136 & 29.839 \\
Earth--Alauda      &  7.735 & 32.346 \\
Earth--Astrid      & 54.689 & 27.557 \\
Earth--Baptistina  & 31.289 & 28.464 \\
Earth--Beagle      &  9.215 & 35.075 \\
Earth--Brang\"ane  & 24.446 & 28.317 \\
Earth--Brasilia    & 13.910 & 28.873 \\
Earth--Brucato     &  8.900 & 34.705 \\
Earth--Chloris     & 22.566 & 25.873 \\
Earth--Clarissa    & 41.365 & 26.673 \\
Earth--Dora        & 21.475 & 28.184 \\
Earth--Elfriede    &  7.285 & 32.525 \\
Earth--Emma        & 11.913 & 30.569 \\
Earth--Eos         & 16.049 & 30.298 \\
Earth--Erigone     & 46.200 & 25.920 \\
Earth--Euphrosyne  &  7.679 & 34.731 \\
Earth--Hoffmeister & 38.050 & 26.869 \\
Earth--Hungaria    & 12.800 & 33.037 \\
Earth--Hygiea      &  7.218 & 33.474 \\
Earth--Iannini     & 13.355 & 27.607 \\ 
Earth--Kalliope    & 16.135 & 27.643 \\
Earth--K\"onig     & 21.540 & 29.461 \\
Earth--Lixiaohua   & 15.461 & 30.019 \\
Earth--Misa        & 52.399 & 26.743 \\
Earth--Naema       & 10.208 & 27.823 \\
Earth--Nemesis     & 24.246 & 25.767 \\
Earth--Nysa        & 79.017 & 25.823 \\
Earth--Padua       & 25.296 & 29.583 \\
Earth--Pallas      &  7.174 & 33.273 \\
Earth--Phaethon    & 31.706 & 35.407 \\
Earth--Polana      & 48.218 & 26.951 \\
Earth--Sylvia      &  9.136 & 33.395 \\
Earth--Themis      & 21.244 & 34.213 \\
Earth--Theobalda   &  8.853 & 34.606 \\
Earth--Tina        & 11.526 & 33.009 \\
Earth--Ursula      &  8.610 & 33.213 \\
Earth--Veritas     &  8.736 & 33.585 \\
Earth--Vibilia     & 54.686 & 27.195 \\
Earth--Watsonia    & 11.984 & 31.053 \\
Earth--Witt        & 29.088 & 29.137 \\
\hline
\end{tabular}
\end{table}


\clearpage

\section{Classification of NEOs}

Several NEOs from MITHNEOS \citep{Binzel_2019Icar..324...41B,Marsset_2022AJ....163..165M} were taxonomically reclassified in this work. In Table\,\ref{tabd1}, we provide our new classification of these objects within the Bus-DeMeo system \citep{DeMeo_2009Icar..202..160D}. We introduce a new class, "298", which stands for bodies spectrally analog to (298)~Baptistina and (396)~Aeolia. Those bodies exhibit a single wide 1.00-$\mu{\rm m}$ absorption band but no 2-$\mu{\rm m}$ band. "U" stands for unclassified/unknown.

In Table\,\ref{tabd2}, we classify B, C and P-type NEOs into the two main groups of carbonaceous chondrites: CI (concave-up) and CM (concave-down). NEOs that could not be classified due to ambiguous spectra (e.g., low SNR) are not included in the table.

\begin{table}[h!]
\caption{Reclassification of NEOs.}
\label{tabd1}
\centering
\begin{tabular}{l cc l}
\vrule width 0pt depth 4pt
NEO & MITHNEOS & This work \\
\hline
\vrule width 0pt height 9pt
(3199) Nefertiti                & K    & A    \\ 
(4581) Asclepius 1989 FC 		& L    & 298 \\
(5496) 1973 NA                  & C    & Xe   \\ 
(5620) Jasonwheeler	1990 OA 	& Sq   & 298 \\
(8037) 1993 HO1 				& L    & 298 \\
(13553) Masaakikoyama 1992 JE 	& L    & 298 \\
(14402) 1991 DB 				& Xk   & 298 \\
(15817) Lucianotesi	1994 QC     & Xk   & Xn   \\
(29075) 1950 DA                 & L    & C \\
(106589) 2000 WN107             & C,X  & Xn   \\
(144411) 2004 EW9 				& L    & 298 \\
(153591) 2001 SN263 			& Xn   & C,X  \\
(162635) 2000 SS164             & C,X  & K    \\
(163191) 2002 EQ9 				& L    & 298 \\
(185851) 2000 DP107 			& L    & 298 \\
(190208) 2006 AQ 				& S    & 298 \\
(220839) 2004 VA 				& L    & 298 \\
(230979) 2005 AT42              & U    & K    \\
(250706) 2005 RR6               & S,Sr & K    \\
(257744) 2000 AD205 			& L    & 298  \\
(354952) 2006 FJ9               & C,X  & Xn    \\ 
(360191) 1988 TA 				& S    & 298  \\
(363790) 2005 JE46 				& L    & 298  \\
(370307) 2002 RH52              & C,X  & L     \\ 
(371660) 2007 CN26 				& L    & 298  \\
(380929) 2006 HU30 				& K    & 298  \\
(388838) 2008 EZ5 				& S,Sr & 298 \\
(410088) 2007 EJ 				& K    & 298  \\
(411165) 2010 DF1               & Xk   & K     \\
(418797) 2008 VF 				& K    & 298  \\
(420187) 2011 GA55 				& L    & 298  \\
(420738) 2012 TS 				& S\_comp & 298 \\
(433953) 1997 XR2 				& L    & 298  \\
(453707) 2010 XY72              & U    & K     \\
(481532) 2007 LE 				& L    & 298  \\
(518507) 2006 EE1               & L    & 298  \\
(523631) 2009 SX1 				& Xk   & 298  \\
(551685) 2013 HT15 				& L    & 298  \\
2008 QS11                       & L    & 298  \\
2011 TN9                        & S\_comp & 298  \\
2011 WK15                       & Xk   & 298  \\
2015 XB379                      & L    & K     \\
2018 JA                         & S,Sr & 298  \\
2019 SH6                        & K    & 298  \\
2020 WU5                        & L    & 298  \\
\hline
\end{tabular}
\end{table}

\begin{table}[h!]
\caption{Classification of CC-like NEOs}
\label{tabd2}
\centering
\begin{tabular}{lr}
\vrule width 0pt depth 4pt
NEO & This work \\
\hline
\vrule width 0pt height 9pt
 (3200)             Phaethon    1983 TB &    CI \\
 (4015)    Wilson-Harrington    1979 VA &    CI \\
 (5645)                      1990 SP &    CM \\
 (7092)               Cadmus    1992 LC &    CM \\
 (7753)                      1988 XB &    CI \\
(20086)                      1994 LW &    CI \\
(26760)                    2001 KP41 &    CM \\
(29075)                      1950 DA &    CM \\
(38071)                     1999 GU3 &    CI \\
(53319)                     1999 JM8 &    CI \\
(65679)                      1989 UQ &    CI \\
(65690)                      1991 DG &    CM \\
(65996)                     1998 MX5 &    CM \\
(68278)                     2001 FC7 &    CM \\
(85275)                      1994 LY &    CM \\
(90367)                     2003 LC5 &    CM \\
(90416)                   2003 YK118 &    CI \\
(96189)            Pygmalion   1991 NT3 &    CI \\
(108519)                      2001 LF &    CI \\
(136874)                    1998 FH74 &    CI \\
(138175)                   2000 EE104 &    CM \\
(139345)                    2001 KA67 &    CM \\
(141424)                      2002 CD &    CM \\
(153201)                   2000 WO107 &    CM \\
(153219)                    2000 YM29 &    CM \\
(153591)                   2001 SN263 &    CI \\
(154007)                      2002 BY &    CM \\
(159504)                    2000 WO67 &    CM \\
(162004)                      1991 VE &    CM \\
(162173)                Ryugu   1999 JU3 &    CI \\
(162566)                    2000 RJ34 &    CI \\
(162581)                    2000 SA10 &    CM \\
(163348)                     2002 NN4 &    CI \\
(164206)                    2004 FN18 &    CM \\
(170502)                     2003 WM7 &    CI \\
(175706)                     1996 FG3 &    CM \\
(234061)                     1999 HE1 &    CI \\
(242643)                     2005 NZ6 &    CI \\
(256412)                     2007 BT2 &    CI \\
(265962)                      2006 CG &    CM \\
(267494)                     2002 JB9 &    CM \\
(269690)                     1996 RG3 &    CI \\
(275611)                   1999 XX262 &    CI \\
(279744)                     1998 KM3 &    CI \\
(285263)                     1998 QE2 &    CM \\
(289315)                    2005 AN26 &    CM \\
(307005)                     2001 XP1 &    CM \\
(308635)                    2005 YU55 &    CM \\
(312473)                   2008 SX245 &    CI \\
(329291)                     2000 JB6 &    CI \\
(333358)                     2001 WN1 &    CM \\
(354030)                    2001 RB18 &    CI \\
(355256)                     2007 KN4 &    CM \\
(370307)                    2002 RH52 &    CM \\
(382503)                     2001 RE8 &    CM \\
(388945)                     2008 TZ3 &    CI \\
(409204)                    2003 WX25 &    CM \\
(410650)                     2008 SQ1 &    CM \\
(410777)                      2009 FD &    CM \\
(410778)                    2009 FG19 &    CI \\
(411280)                    2010 SL13 &    CM \\
(414287)                     2008 OB9 &    CM \\
(437316)                     2013 OS3 &    CM \\
(438105)                    2005 GO22 &    CI \\
\hline
\end{tabular}
\end{table}

\addtocounter{table}{-1}
\begin{table}[h!]
\caption{continued.}
\label{tabd3}
\centering
\begin{tabular}{lr}
\vrule width 0pt depth 4pt
NEO & CC group \\
\hline
\vrule width 0pt height 9pt
(438429)                     2006 WN1 &    CI \\
(438902)                   2009 WF104 &    CI \\
(440212)                      2004 OB &    CM \\
(443103)                    2013 WT67 &    CM \\
(448972)                    2011 YV15 &    CI \\
(451124)                     2009 KC3 &    CM \\
(452334)                      2001 LB &    CM \\
(452561)                      2005 AB &    CM \\
(455322)                    2002 NX18 &    CM \\
(461912)                     2006 RG2 &    CI \\
(464798)                    2004 JX20 &    CM \\
(465749)                     2009 WO6 &    CI \\
(467460)                    2006 JF42 &    CM \\
(475665)                    2006 VY13 &    CI \\
(481032)                    2004 YZ23 &    CI \\
(489486)                     2007 GS3 &    CM \\
(492143)                      2013 OE &    CM \\
(505657)                   2014 SR339 &    CI \\
(518735)                     2009 JL1 &    CM \\
(523654)                     2011 SR5 &    CM \\
(523811)                     2008 TQ2 &    CM \\
(525552)                      2005 JB &    CM \\
(558307)                    2015 AS45 &    CM \\
                         2002 GZ8 &    CM \\
                         2002 LY1 &    CM \\
                        2003 AF23 &    CI \\
                        2003 EG16 &    CM \\
                         2004 QD3 &    CI \\
                        2005 GR33 &    CM \\
                          2006 RZ &    CI \\
                       2006 UN216 &    CM \\
                        2007 PF28 &    CM \\
                        2008 SV11 &    CM \\
                         2008 US4 &    CI \\
                         2011 PT1 &    CM \\
                        2011 YS62 &    CI \\
                        2012 AA11 &    CI \\
                        2012 KT42 &    CI \\
                         2012 LZ1 &    CM \\
                        2012 XQ55 &    CM \\
                        2013 FY13 &    CI \\
                         2013 PY6 &    CM \\
                        2014 LW21 &    CI \\
                        2014 PR62 &    CM \\
                       2014 UF206 &    CM \\
                         2014 VH2 &    CM \\
                       2014 WF201 &    CM \\
                          2015 BC &    CM \\
                         2015 SV2 &    CM \\
                          2015 SZ &    CM \\
                         2016 AZ8 &    CI \\
                       2016 CO247 &    CI \\
                        2016 ED85 &    CM \\
                        2016 LX48 &    CI \\
                         2016 PR8 &    CM \\
                         2016 XH1 &    CM \\
                       2017 BM123 &    CM \\
                        2017 CR32 &    CM \\
                         2018 WX1 &    CI \\
                          2019 FU &    CI \\
                          2019 UC &    CM \\
                         2019 YH2 &    CM \\
                         2020 RB6 &    CM \\
                          2020 SN &    CM \\
                         2020 WK3 &    CM \\
\hline
\end{tabular}
\end{table}

\end{document}